\documentclass[apj,numberedappendix]{emulateapj}
\usepackage{amsmath}

\PassOptionsToPackage{breaklinks=true}{hyperref}

\usepackage[z]{tmsfnts}

\newcommand{\ergcm}{erg~s$^{-1}$~cm$^{-2}$}
\newcommand{\ergs}{erg~s$^{-1}$}
\renewcommand{\deg}{\ensuremath{^\circ}}
\newcommand{\BS}{Bright SHARC}
\newcommand{\SSH}{SHARC South}
\newcommand{\fmin}{f_{\text{min}}}
\newcommand{\Psel}{P_{\text{sel}}}
\newcommand{\Aeff}{A_{\text{eff}}}
\newcommand{\Veff}{V_{\text{eff}}}
\newcommand{\texp}{t_{\text{exp}}}
\def\400d{400d}
\def\pI{\phantom{1}}
\def\pII{\phantom{1}\phantom{1}}

\journalinfo{Submitted to ApJS; astro-ph/0610739}
\slugcomment{Submitted to the Astrophysical Journal Supplement Series
  October 25, 2006. \emph{astro-ph/0610739}}

\shorttitle{400 SQUARE DEGREES CLUSTER SURVEY}
\shortauthors{BURENIN ET AL.}

\begin{document}

\title{The 400 square degrees \emph{ROSAT} PSPC galaxy cluster survey:\\
  Catalog and statistical calibration}

\author{R.\ A.\ Burenin\altaffilmark{1}, A.~Vikhlinin\altaffilmark{2,1},
  A.\ Hornstrup\altaffilmark{3}, H.\ Ebeling\altaffilmark{4},
  H.~Quintana\altaffilmark{5}, A.\ Mescheryakov\altaffilmark{1}}

\email{burenin@hea.iki.rssi.ru}
\altaffiltext{1}{Space Research Institute (IKI), Profsoyuznaya 84/32,
                 Moscow, Russia}
\altaffiltext{2}{Harvard-Smithsonian Center for Astrophysics, 
                 60 Garden Street, Cambridge, MA 02138, USA}
\altaffiltext{3}{Danish National Space Center, Juliane Maries Vej 30, 
                 Copenhagen 0, DK-2100, Denmark}
\altaffiltext{4}{Institute for Astronomy, University of Hawaii, 2680
Woodlawn Drive, Honolulu, HI 96822, USA}
\altaffiltext{5}{Departamento de Astronomia y Astrofisica, 
                 Pontificia Universidad Catolica de Chile, Casilla 306, 
                 Santiago, 22, Chile}

\begin{abstract}
  
  We present a catalog of galaxy clusters detected in a new \emph{ROSAT}
  PSPC survey. The survey is optimized to sample, at high redshifts, the
  mass range corresponding to $T>5$\,keV clusters at $z=0$. Technically,
  our survey is the extension of the 160 square degrees survey
  \citep[160d,][]{vikhl98a,mullis03}.  We use the same detection
  algorithm, thus preserving high quality of the resulting sample; the
  main difference is a significant increase in sky coverage.  The new
  survey covers 397 square degrees and is based on 1610 high Galactic
  latitude \emph{ROSAT} PSPC pointings, virtually all pointed
  \emph{ROSAT} data suitable for the detection of distant clusters.  The
  search volume for X-ray luminous clusters within $z<1$ exceeds that of
  the entire local Universe ($z<0.1$). We detected 287 extended X-ray
  sources with fluxes $f>1.4\times 10^{-13}\,$erg$\,$s$^{-1}\,$cm$^{-2}$
  in the 0.5--2~keV energy band, of which 266 (93\%) are optically
  confirmed as galaxy clusters, groups or individual elliptical
  galaxies. This paper provides a description of the input data, the
  statistical calibration of the survey via Monte-Carlo simulations, and
  the catalog of detected clusters.  We also compare the basic results
  to those from previous, smaller area surveys and find good agreement
  for the $\log N$--$\log S$ distribution and the local X-ray luminosity
  function. Our sample clearly shows a decrease in the number density
  for the most luminous clusters at $z>0.3$. The comparison of our
  \emph{ROSAT}-derived fluxes with the accurate \emph{Chandra}
  measurements for a subset of high-redshift clusters demonstrates the
  validity of the 400 square degree survey's statistical calibration.
  
\end{abstract}

\section{Introduction} 
\label{sec:intro} 

Observations of galaxy clusters over a range of redshifts is an
attractive way to probe fundamental cosmological parameters. Cluster
data have been extensively used in the past for this purpose
\citep[e.g.,][]{evrard89,white93,oukbirblanchard92,vianaliddle96,
  vianaliddle99,eke98,henry97,henry00,viana03,vikhl03,voevvikhl04,
  2003A&A...398..867S}, primarily to constrain the cosmological density
parameter. Clusters also can be used as an independent and complementary
Dark Energy probe \citep{starobinky98,wangsteinhardt98,hutererturner01,
  haiman01,battyeweller03,molnar04,2006PhRvD..73f7301H}. 

These cosmological applications rely on the existence of large,
unbiased, and statistically complete cluster samples. Using the X-ray
emission of the hot intracluster medium (ICM) is one of the best methods
of finding distant clusters \citep{gioia90a,rosati95,vikhl98a}. A
comparison of the efficiency of detecting clusters in X-rays versus
other methods can be found, for example, in \cite{rosati02}. The
presently available X-ray selected samples are those from the
\emph{Einstein} Extended Medium Sensitivity Survey
\citep[EMSS;][]{gioia90,stocke91} and various samples derived from the
\emph{ROSAT} PSPC observations --- the 160d survey
\citep{vikhl98a,mullis03}, Bright SHARC \citep{romer00}, WARPS
\citep{scharf97,perlman02}, \SSH{} \citep{burke03}, NEP
\citep{henry01,gioia03}, and RDCS \citep{rosati98}.  There is an ongoing
survey based on \emph{ROSAT} HRI data \citep[BMW,][]{moretti04}, and
also a survey sampling very X-ray luminous clusters using the data from
the \emph{ROSAT} All-Sky Survey \citep[MACS]{ebeling01a}. 

The largest sample published to date comes from the 160d survey. In that
survey, clusters were serendipitously selected as extended X-ray sources
in the inner region of the \emph{ROSAT} PSPC field of view where the
angular resolution is sufficient to spatially resolve clusters even at
high redshifts. The 160d catalog includes 203 clusters, 43 of them at
$z>0.4$.  MACS is expected to find a similar number of distant clusters,
all with much higher X-ray luminosities. 

None of the previous \emph{ROSAT} surveys based on pointed observations
made use of all data suitable for finding distant clusters.  Our new
survey does exactly that. It is obtained by applying the 160d cluster
detection algorithm to virtually all suitable \emph{ROSAT} PSPC fields
(1610 in total), resulting in a sky coverage of 397~deg$^2$.  Hereafter,
we call it the \400d{} survey. The \400d{} sample includes only objects
with an observed X-ray flux above $1.4\times10^{-13}$~\ergcm{} in the
0.5--2 keV band. This flux limit corresponds to an X-ray luminosity of
$1.1\times 10^{44}$~\ergs{} at $z=0.5$\footnote{We compute all
  distance-dependent quantities assuming $\Omega_M=0.3$,
  $\Omega_\Lambda=0.7$, $h=0.71$. The luminosities are in the 0.5--2~keV
  band (source rest frame).}  or temperature $T\approx 5$~keV through
the $L_x-T$ relation \citep{markevitch98}. Because of the relatively
high flux threshold, our catalog does not include low-luminosity systems
at $z>0.3$, nor any clusters at very high redshifts ($z\gtrsim1$). 
Instead, it provides a representative snapshot of the population of
``typical'' clusters at $z=0.3-0.8$. In this Paper, we present the
\400d{} cluster catalog and describe the calibration of the survey's
effective area and volume through extensive Monte-Carlo simulations. We
also provide updated measurements of the cluster $\log N$--$\log S$
relation and the X-ray luminosity function. 

\section{X-Ray Data and Source Detection}
\label{sec:data} 

The \400d{} survey is based on the \emph{ROSAT} PSPC pointed
observations selected from the archive by the following criteria: 1)
Galactic latitude $|b|>25\deg$, 2) Galactic absorption
$N_H<10^{21}$~cm$^{-2}$, 3) total clean exposure
$\texp>1000$~s, and 4) targeted at least $10\deg$ away from LMC
and SMC.  Several pointings were discarded because of the large optical
extent of the \emph{ROSAT} targets.  Compared to the 160d survey, we
used pointings at lower Galactic latitudes and also with shorter
exposures and higher $N_H$. We also included pointed observations of
extended targets such as star clusters, normal galaxies, and galaxy
clusters at moderate and high redshifts, if the target emission did not
affect more than 50\% of the area in the inner 17.5\arcmin{} region; no
such pointings were used in the 160d survey.  The overall sample quality
is not degraded by these additional data because the final catalog uses
a relatively high flux threshold, $1.4\times10^{-13}\,$\ergcm. 

S.~Snowden's software \citep{snowden94} was used to clean the PSPC data
from high background intervals and to generate exposure maps.  Cluster
detection is performed in the hard energy band, 0.6--2\,keV
\citep[justified in][]{vikhl98a}.  Images from multiple observations of
the same target were merged, and fields with a total merged exposure
time of $\texp<1000$~s were discarded.  The detection threshold for
clusters in the $\texp=1000$~s fields is $\simeq
3\times10^{-13}$~\ergcm.

\begin{figure}
  \vspace*{-1.5mm}
  \centerline{
    \includegraphics[width=0.95\linewidth,bb=20 170 567 685]{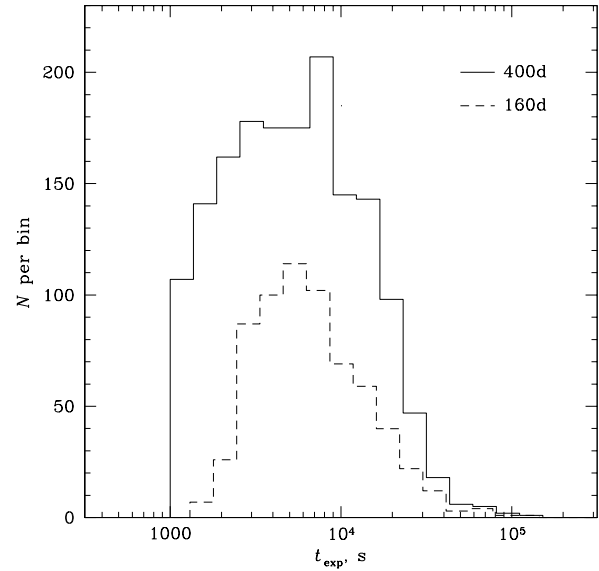}
    }
    \vspace*{-0.5mm}
    \caption{Distribution of exposure times in the \400d{} and 160d
      surveys.} 
  \label{fig:exp}
  \vspace*{-3mm}
\end{figure}

The final set contains 1610 fields, including all data from the 160d
survey (646 fields\footnote{The only exception is the pointing towards
  Arcturus which was included in the 160d survey (it contains the
  extended source 1415$+$1906) but discarded here because of the target's
  optical brightness.}) plus 964 additional fields. The exposure time
distribution in the \400d{} and 160d surveys is shown in
Fig.\,\ref{fig:exp}. The field center coordinates and exposure times are
listed in Table~\ref{tab:fields}. For each field, we defined a region
affected by either X-ray or optical emission from the target (also given
in Table~\ref{tab:fields}) and thus unsuitable for serendipitous cluster
detection. These target regions are typically 1\arcmin{} circles for
on-axis, point-like targets. Larger regions were used for extended
targets. No such regions were defined for pointings without a declared
target --- those classified under ``extragalactic survey'' or used to
complete the All-Sky Survey (\emph{ROSAT} sequences rp190xxx). 

\label{sec:det}

Extended X-ray sources in the central 17.5\arcmin{} of the field of view
were detected with the 160d analysis pipeline.  This algorithm
\citep[fully described in][]{vikhl98a} is a three-step procedure which
includes identification of the candidate sources through the wavelet
transform, Maximum Likelihood fitting of the selected sources, and final
selection based mainly on the significances of source existence and
extent. The only modification we made to the 160d pipeline is to drop
the requirement that the cluster core-radius exceeds $1/4$ of the PSF
FWHM. A closer examination showed that this selection significantly
decreased the detection efficiency for clusters with small angular size
while most of the associated false detections can be easily identified
optically. For the 160d fields, the effect of removing this criterion is
to add three clusters (0209$-$5116, 1338$+$3851, 1514$+$3636) and three
false detections (0522$-$3628, 1007$+$3502, 1428$+$0106). 

We detected 287 extended X-ray sources with fluxes above
$1.4\times10^{-13}$~\ergcm, compared to 116 such sources in the 160d
sample. For each source we measure its location and total X-ray flux,
and also derive uncertainties on these quantities as described in
\citet{vikhl98a}.

\begin{deluxetable}{ccrl}
  \tablecaption{List of \emph{ROSAT} pointings\label{tab:fields}}
  \tablewidth{0.9\linewidth}
  \tablehead{
      \colhead{$\alpha$} &
      \colhead{$\delta$} &
      \colhead{$T_{\rm exp}$,} &
      \colhead{Target region\tablenotemark{a}}\\
      \multicolumn{2}{c}{(J2000)} &
      \colhead{ks} &
      \colhead{$\Delta\alpha$, $\Delta\delta$, $r$}
    }
  \startdata
00 00 07.2 & $+$29 57 01 &   3.5\phd & \nodata\\
00 02 28.8 & $+$31 28 47 &  14.1\phd & \nodata\\
00 03 19.2 & $-$35 57 00 &   8.7\phd &   115\arcsec,   80\arcsec, 355\arcsec \\
00 03 21.5 & $-$26 03 36 &  38.2\phd &    30\arcsec,   15\arcsec,  50\arcsec \\
00 05 19.2 & $+$05 23 59 &   7.4\phd &   $-$15\arcsec,   15\arcsec,  80\arcsec
\enddata
\tablenotetext{a}{Defined by offsets from the field center and radius.} 
\tablecomments{The complete version of this table is in the electronic
  edition of the Journal. The printed edition contains only a sample.} 
\end{deluxetable}

\section{Optical observations}
\label{sec:optical}

The X-ray analysis was followed by an extensive optical program whose
purpose was to confirm the cluster identifications of our extended X-ray
sources, and to measure redshifts of previously unknown clusters. The
identification was based on an examination of the optical images of the
X-ray candidates. The images were obtained mainly with the
Russian-Turkish 1.5-m telescope in the North and with the Danish 1.54-m
telescope in the South. The images obtained were sufficiently deep to
detect cluster member galaxies out to $z\approx1$.  We considered the
X-ray source to be confirmed as a cluster if 1) there was an obvious
associated excess in the galaxy number density, or 2) there was a bright
elliptical galaxy at the X-ray centroid, even if it was seemingly
isolated \citep[this is a signature of so-called fossil groups,
see][]{ponman94,vikhl99b,jones03}. Our cluster identifications are
sufficiently reliable even though they do not use spectroscopic
redshifts, because extended X-ray emission is by itself a strong
indication of a cluster. The cluster identification had to be later
revised only for one object (see below). We identified 266 of 287 X-ray
candidates as galaxy clusters, groups, or X-ray luminous isolated
ellipticals.  An additional 5 objects are legitimate extended X-ray
sources (e.g., nearby spiral galaxies).  Only 16 objects (5\% of the
sample) remained unidentified; they are most likely false detections
(see below). All clusters from the 160d survey with fluxes above our
flux limit (116 objects) were re-confirmed in our catalog. 

Spectroscopic redshifts for a significant fraction of the \400d{}
clusters were previously known (Table~\ref{tab:id}). Redshifts for 88
clusters were measured in the 160d survey and an additional 27 clusters
had known redshifts from other \emph{ROSAT} and \emph{Einstein} surveys:
EMSS \citep[][8 objects]{stocke91}, WARPS \citep[][6
objects]{perlman02,ebeling01b}, Bright SHARC \citep[][8
objects]{romer00}, \SSH{} \citep[][1 object]{burke03}, NEP \citep[][2
objects]{gioia03}, and NORAS \citep[][2 objects]{boeringer00}.  The
redshifts for 62 low-$z$ clusters were available from the literature and
various public catalogs. The redshifts of the remaining 89 clusters were
measured by us with the Keck II, ESO 3.6-m, NTT, Magellan, FLWO 1.5-m,
Nordic Optical Telescope, and Danish 1.54-m telescopes (Hornstrup et
al., in preparation). 

\begin{deluxetable}{p{7cm}r}
  \tablecaption{Summary of optical identifications \label{tab:id}}
  \tablehead{
    \multicolumn{1}{l}{Description} &
    \colhead{Objects}
  }
  \tablewidth{0.99\linewidth}
  \startdata
  Detected extended X-ray sources\dotfill & 287 \\ 
  Confirmed clusters, groups, and galaxies\dotfill  & 266 \\
  Other extended X-ray sources\dotfill     & 5 \\
  False detections\dotfill & 16 \\
  Clusters at target $z$\dotfill & 24 \\
  Clusters in main sample\dotfill & 242 \\
  Previously known redshifts\dotfill & 177 \\
   \hline\\[-4pt]
   \multicolumn{2}{c}{Objects present in other catalogs}\\[2pt]
   \hline\\[-6pt]
  160d\dotfill & 116 \\
  EMSS\dotfill & 13 \\
  WARPS\dotfill & 15 \\
  Bright SHARC\dotfill & 22 \\
  \SSH\dotfill & 12 \\
  NEP\dotfill & 4 \\
  Abell clusters\dotfill & 43 \\
  NGC galaxies\dotfill & 10
  \enddata
\end{deluxetable}

\begin{figure}
  \vspace*{-1mm}
  \centerline{\includegraphics[width=0.95\linewidth,bb=20 170 567 685]{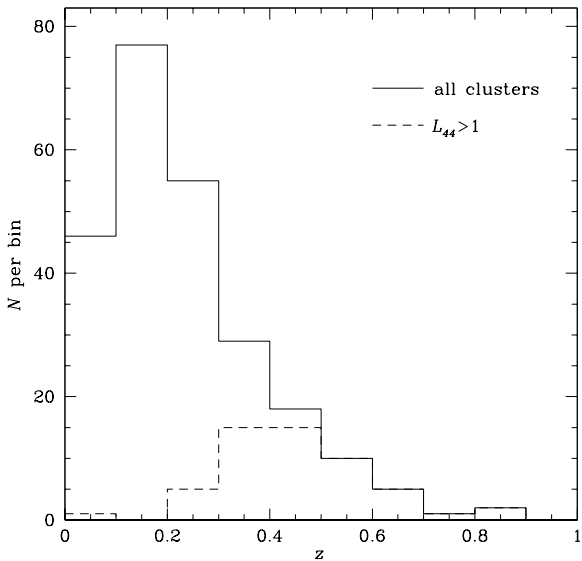}}
  \vspace*{-0.5mm}
    \caption{Redshift distribution of clusters in the \400d{}
      catalog. Dashed histogram shows the higher luminosity clusters,
      $L_x>10^{44}$~erg~s$^{-1}$.} 
  \label{fig:dsz}
\end{figure}

The redshift distribution of the \400d{} sample is shown in
Fig.\,\ref{fig:dsz}. The median redshift is relatively low, $z=0.20$, as
expected, since all X-ray flux limited samples are dominated by
low-luminosity systems at low redshifts. Clusters with higher X-ray
luminosities are at higher redshifts on average. For example, the median
redshift of clusters with $L>10^{44}$~\ergs, is $z=0.46$ (dashed
histogram in Fig.\,\ref{fig:dsz}). The most distant \400d{} cluster is
ClJ1226$+$3332 at $z=0.888$, a system previously discovered in the WARPS
survey \citep{ebeling01b}. 

Sixteen extended X-ray sources did not have any obvious counterparts in
the deep optical images. In principle, we cannot exclude that these are
very distant clusters ($z>1$). However, we note that a similar number of
\emph{false detections} is expected in our sample because of point
source confusion (\S~\ref{sec:false}) and so the unidentified sources
are most likely not clusters. To be conservative, however, one should
use an upper redshift boundary of $z\approx1$ for the \400d{} catalog
within which our sample should be essentially complete and clean.

\section{The catalog}
\label{sec:cat}

The \400d{} object catalog is presented in
Tables~\ref{tab:cat}--\ref{tab:false}.  The main cluster list is given
in Table~\ref{tab:cat}. The clusters within $|\Delta z|<0.01$ of the
\emph{ROSAT} target redshift are listed separately in Table~\ref{tab:tz}
because they are not entirely serendipitous. Extended non-cluster
sources and likely false X-ray detections are listed in
Tables~\ref{tab:notcl} and \ref{tab:false}, respectively. For each
source we provide the coordinates of the X-ray centroid (columns 2--3; a
typical positional uncertainty is 10\arcsec\ -- 30\arcsec), the total
unabsorbed flux in the 0.5--2~keV band (column 4), the redshift (column
5) with reference (column 6), the total X-ray luminosity in the
0.5--2~keV band (column 7), and notes on the optical IDs (column 8). 
The X-ray luminosity was computed as described in
Appendix~\ref{sec:cluster:popul:models}. 

\begin{figure}
  \vspace*{1.5mm}
  \centering \includegraphics[width=0.85\linewidth]{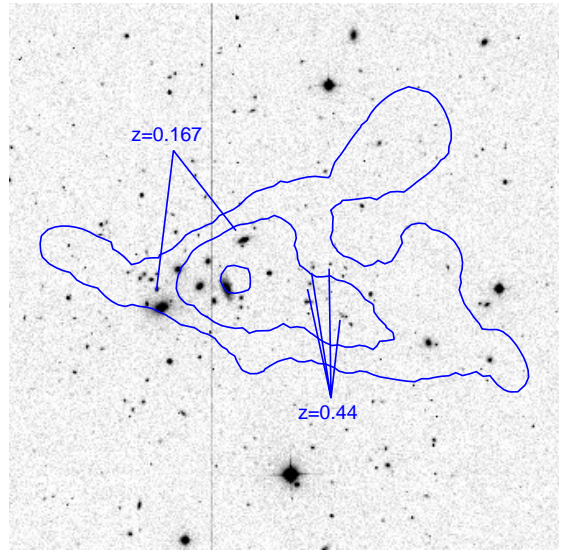}
  \caption{R-band image of 0141$-$3034 with overlaid \emph{ROSAT}
    contours. North is up and East is to the left.} 
  \label{fig:cl0141}
\end{figure}

\subsection{Notes on Individual Objects}
\label{sec:notes}

0106$+$3209 --- This source was detected in the EMSS and identified as a
QSO at $z=2.03$. However, this identification is ambiguous
\citep{stocke84} because the QSO is projected on a foreground elliptical
galaxy. \emph{ROSAT} PSPC data clearly show that the X-ray source is
extended, and that a point source at its center cannot contribute more than
10--30\% of the total flux.  Therefore we identify this object as a
galaxy group, a conclusion that is confirmed by a \emph{Chandra} observation of this
field \citep{hardcastle02}. 

0141$-$3034 --- A $z=0.44$ cluster near the X-ray centroid is projected
on the galaxy group AM\,0139--305 ($z=0.17$). Each object is associated
with a separate X-ray peak (Fig.~\ref{fig:cl0141}). The fluxes of these
systems cannot be deblended using the \emph{ROSAT} data.  We estimate
that the foreground group can contribute up to 30\% of the total X-ray
flux. 

0350$-$3801 --- The aspect solution for this \emph{ROSAT} observation
(sequence rp190505n00) was incorrect (systematically shifted
$\approx5.4\arcmin$ to the South--West). We reconstructed the aspect
solution by cross-correlating locations of bright X-ray sources in this
field with their locations in the overlapping pointings.  After this
correction, the extended X-ray source is unambiguously identified as a
galaxy cluster at $z=0.36$. 

0809$+$2811 --- This object was detected in the EMSS (MS\,0806.6+2820)
and classified as an AGN at $z=0.30$.  The optical AGN is located
0.6\arcmin\ from the \emph{ROSAT} centroid. The source is significantly
extended in the \emph{ROSAT} PSPC image, and there is no point source
near the AGN location. We instead identify this source as a cluster at
$z=0.399$. 

1002$+$6858 --- This object was identified as a QSO in the EMSS
(MS\,0958.4+6913, $z=0.93$). In the \emph{ROSAT} data, we detect both
the point source associated with the QSO and extended X-ray emission
centered 0.9\arcmin{} off the QSO. This object is therefore
classified as a cluster.  The QSO flux was correctly subtracted by our
automatic detection software. 

1142$+$1027 --- We revise the Bright SHARC identification of this object
as A\,1356 \citep{romer00}. A\,1356 is not detected in the
\emph{ROSAT} image. Instead, the extended X-ray source is associated
with a more distant galaxy group.

\begin{figure}
  \vspace*{1.5mm}
  \centering \includegraphics[height=0.85\linewidth]{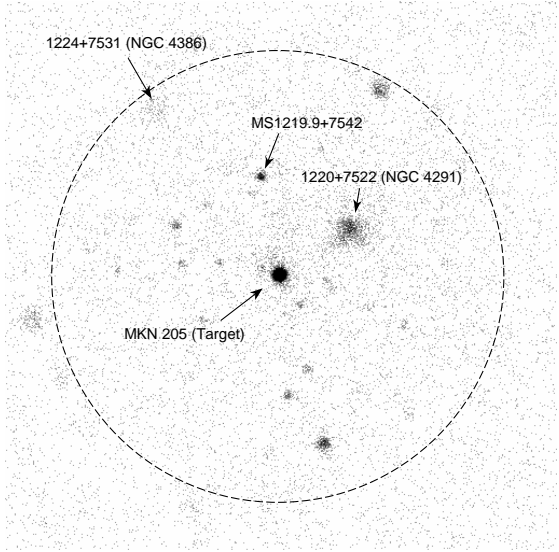}
  \caption{\emph{ROSAT} image containing the EMSS source MS1219.9+7542
    (= \BS{} source RXJ1222.1+7526). The source spatial extent is
    comparable to that of the \emph{ROSAT} target, MKN~205 (point-like
    AGN). Therefore, MS1219.9+7542 is clearly dominated by emission from
    a point source.} 
  \label{fig:rxj1222}
\end{figure}

1338$+$3851 --- The galaxy near the center of this source is a radio
source, 3C\,288. The X-ray source extent is significant but we cannot
exclude the possibility of considerable contamination by
AGN emission.

1500$+$2244 --- This object is a false detection, based on a recent
\emph{Chandra} observation. 

\begin{figure}
  \vspace*{1.55mm}
  \centerline{\includegraphics[height=0.85\linewidth]{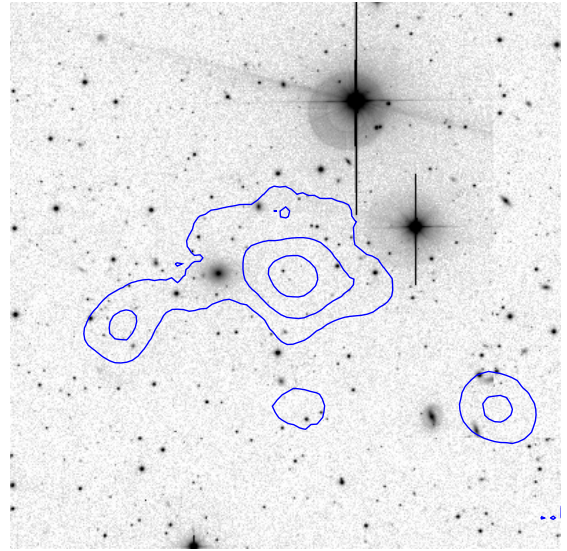}}
  \caption{R-band image of the \400d{} cluster 1313$-$3250 (\BS{} object
    RXJ1313.6$-$3250) with overlaid \emph{ROSAT} contours.} 
  \label{fig:cl1313}
\end{figure}

\section{Comparison with other X-ray surveys}
\label{sec:other}

The \400d{} fields overlap with those covered by several earlier surveys. A comparison of
our catalog with those from these previous studies, provided below, helps to assess any
systematic errors in the X-ray cluster selection, optical
identifications, and X-ray flux measurements.

\paragraph{EMSS}

The angular resolution of the \emph{Einstein} IPC is $\sim1\arcmin$
\citep{giacconi79}, which is larger than the angular core-radius of most
of the distant clusters. The inability to rely on the X-ray source
extent sometimes leads to incorrect identifications.  \emph{ROSAT} data
clearly show that, of 16 clusters in common between the \400d{} and EMSS
samples, 3 were incorrectly identified with AGNs (0106$+$3209,
0809$+$2811, and 1002$+$6858, see \S\,\ref{sec:notes}), and the fossil
group 1159$+$5531 \citep{vikhl99b} was classified as a galaxy with weak
emission lines.  The area covered by the \400d{} survey contains 13 EMSS
clusters, of which we detect 10.  MS1019.0+5139, MS1209.0+3917, and
MS1219.9+7542 are not included in our catalog because these sources were
not recognized as extended.  Examination of the \emph{ROSAT} data
confirms that their fluxes are dominated by emission from point X-ray
sources (see, e.g., Fig.~\ref{fig:rxj1222} for MS1219.9+7542).

\paragraph{WARPS}

The detection algorithm used in the \emph{WARPS} survey \citep{scharf97}
essentially selects X-ray candidates by peak surface brightness, with
only a weak reliance on the angular extent.  This detection algorithm is
very different from ours and therefore the comparison with \emph{WARPS}
is particularly useful for assessing the systematics of the X-ray
selection.  Our survey includes 74 of 80 \emph{ROSAT} PSPC fields used
in \emph{WARPS}.  In this area, the \emph{WARPS} catalog includes 14
clusters and one normal galaxy above our flux limit. We detected all
these objects. For two objects, we measure a lower X-ray flux
($<1.4\times10^{-13}$~\ergcm) so they are not included in our main
catalog. All 15 of our clusters in the overlapping area were also
detected by WARPS. We conclude that there is no difference in the source
lists and possibly a small difference in the flux measurements, which
are not statistical since virtually the same data were used, as is the
case also with \BS{} and \SSH.

\paragraph{Bright SHARC}

\begin{figure}
  \vspace*{-1mm}
  \centerline{\includegraphics[width=0.95\linewidth,bb=20 170 567 685]{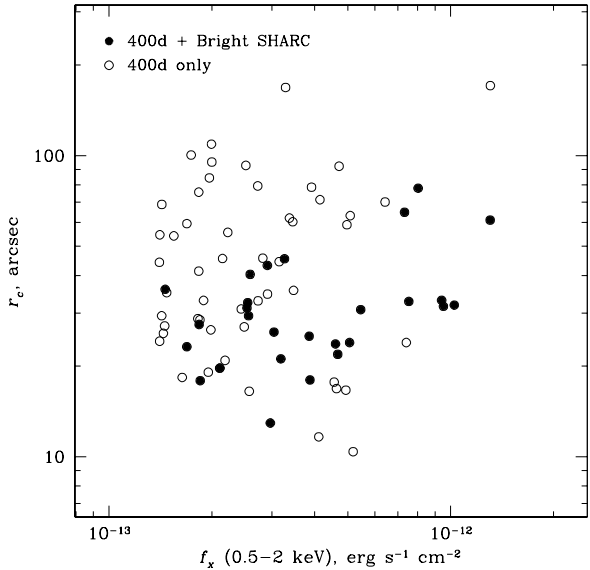}}
  \caption{The distribution of fluxes and core-radii for the clusters
    detected in \400d{} and \BS{} surveys in the overlapping area.} 
  \label{fig:bs}
\end{figure}

The detection algorithm used in the \BS{} survey \citep{romer00} is
based on the wavelet analysis at a single angular scale.  The \BS{}
catalog contains 32 clusters in the area covered by \400d{}, all above
our flux limit.  Our catalog includes 26 of these objects. Of the
remaining 6 objects, RXJ0209.4$-$1008, RXJ0415.7$-$5535,
RXJ0416.1$-$5546, RXJ1250.4$+$2530, and RXJ1349.2$-$0712 were in fact
the observation targets, and RXJ1222.1$+$7526 ($=$MS1219.9+7542) is not
extended (Fig.~\ref{fig:rxj1222}). The \400d{} catalog contains 78
clusters in the overlapping area, of which only 26 are listed in \BS. 
The \400d{} cluster 1313$-$3250 was detected (RXJ1313.6$-$3250) but
classified as a ``blend''. \emph{ROSAT} and optical images, however,
support our cluster identification (Fig.~\ref{fig:cl1313}).  Obviously,
there are large differences between the \BS{} and \400d{} samples in the
overlapping data. We attribute this mostly to the X-ray detection
algorithm used in \BS{}. This is well illustrated by the comparison of
the distributions of cluster fluxes and core-radii in the two surveys
(Fig.\,\ref{fig:bs}).  Obviously, \BS{} tends to miss clusters with
either large ($r_c\gtrsim 50''$) or small ($r_c\lesssim 20''$) angular
extent.

\begin{figure}
  \vspace*{0.82mm}
  \centering \includegraphics[height=0.85\linewidth]{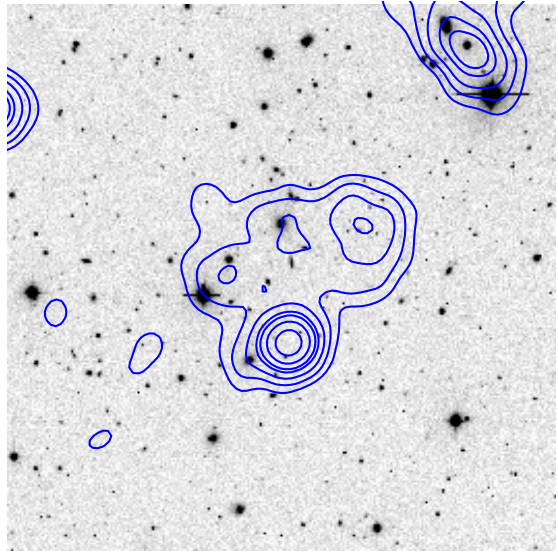}
  \caption{R-band image of the \400d{} cluster 0506$-$2840 (identified
    as ``multiple point sources'' in \SSH) with overlaid \emph{ROSAT}
    contours. The point source to the South was correctly removed from
    the cluster emission.} 
  \label{fig:cl0506m2840}
\end{figure}

\paragraph{\SSH} Our survey used 61 of 66 \emph{ROSAT} pointings used in
the \SSH{} survey \citep{burke03}. In these fields, we detect 15
clusters with off-axis angles $>5\arcmin$ (the inner radius used by
Burke~et~al.). All 15 objects were detected in \SSH{} but 3 were not
optically identified as clusters: 1252$-$2920 and 2305$-$3545 were
listed as unidentified, and third object, 0506$-$2840, was listed as
``multiple point sources''.  Inspection of the \emph{ROSAT} image
(Fig.~\ref{fig:cl0506m2840}) shows that there is both a point source
(correctly detected in \400d) and an extended X-ray source, clearly
associated with a galaxy group. All of the \SSH{} clusters above our
flux limit are included in the \400d{} sample. To summarize, \SSH{} and
our survey have nearly identical X-ray source lists and the difference
is in the optical identifications.

\begin{figure*}
  \vspace*{1mm}
  \setlength{\fboxsep}{0pt}
  \centering
  \framebox{\includegraphics[width=0.406\linewidth]{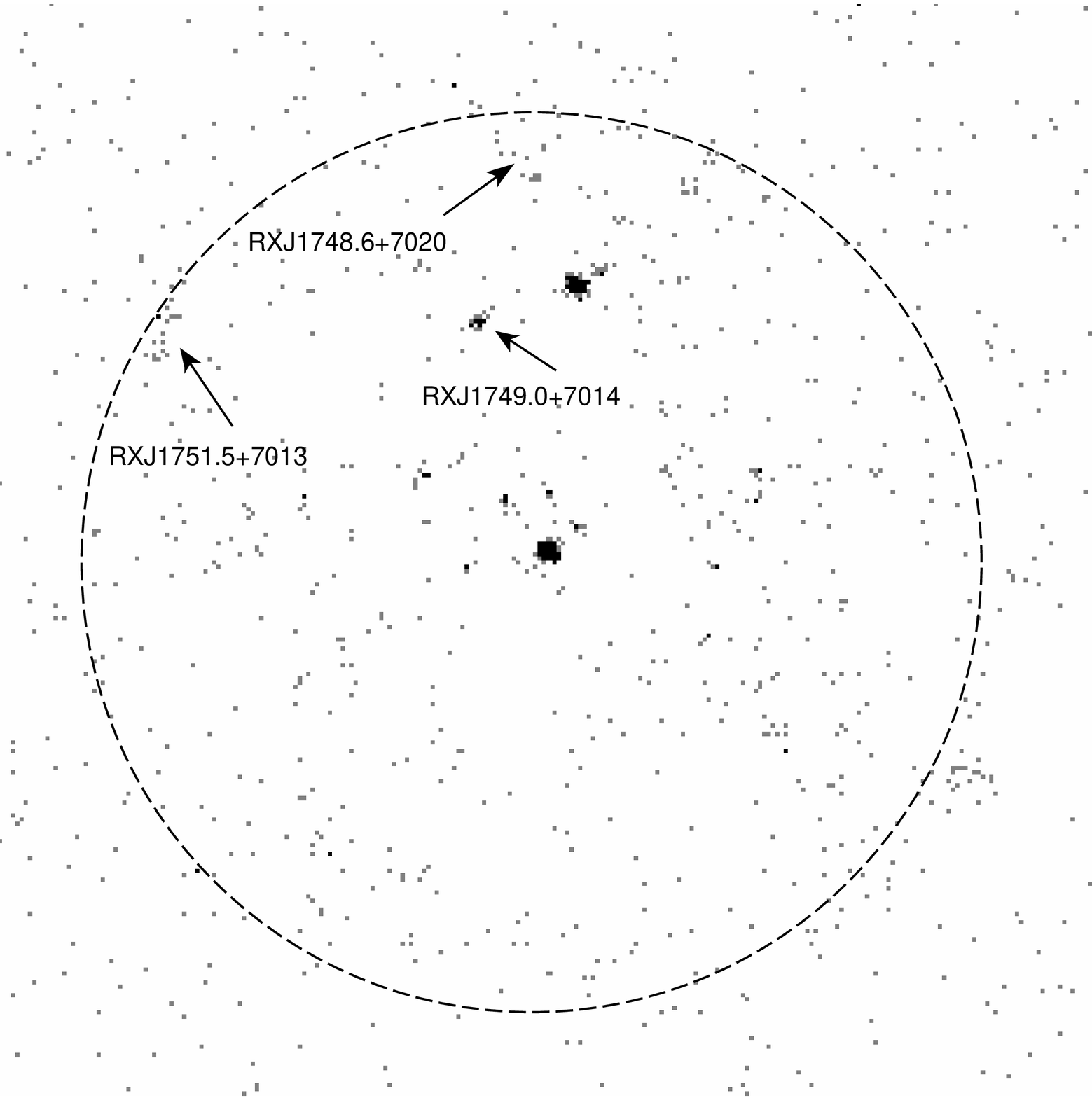}}
  ~
  \framebox{\includegraphics[width=0.406\linewidth]{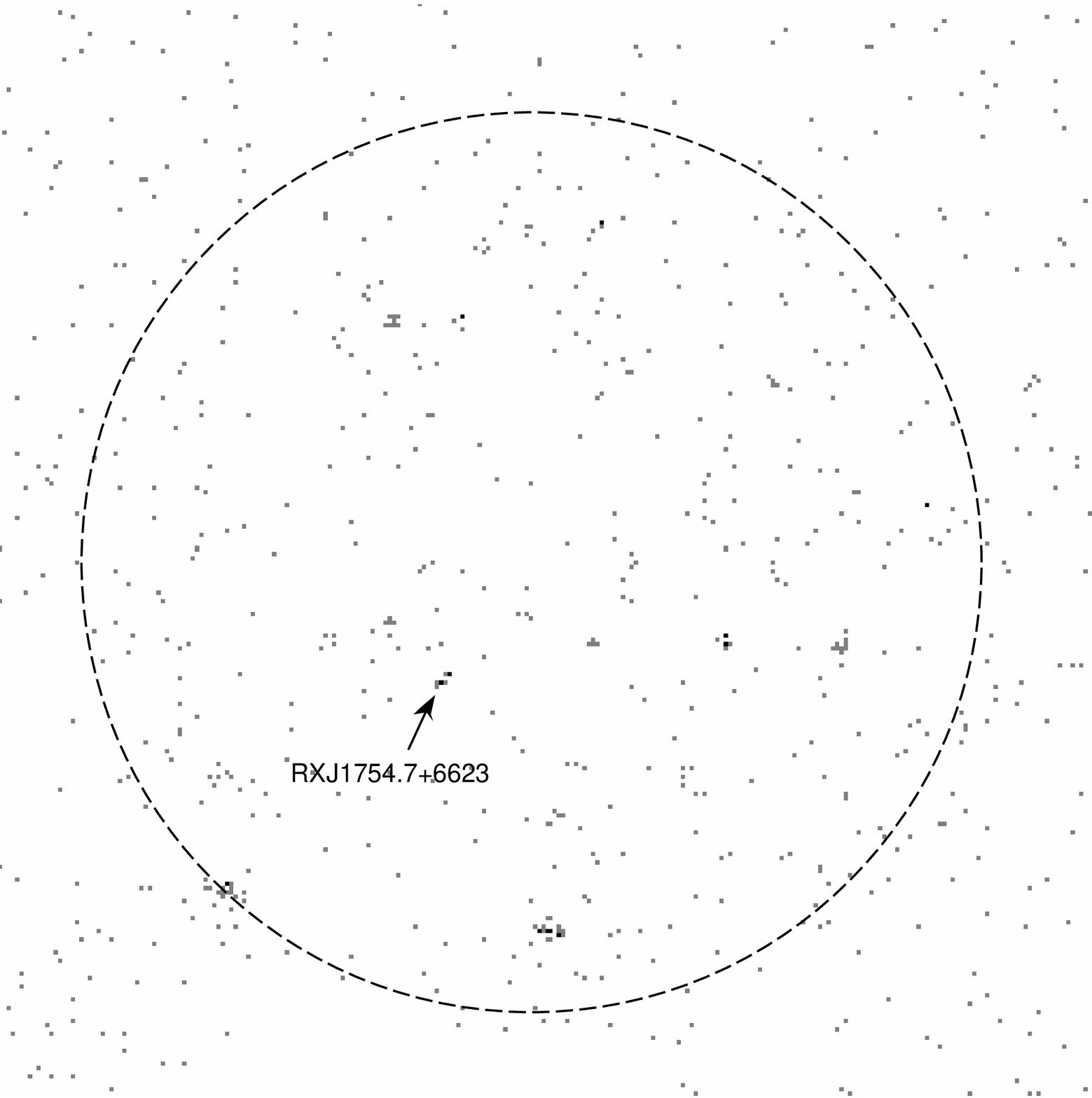}}

  \vspace*{1.7mm}

  \framebox{\includegraphics[width=0.406\linewidth]{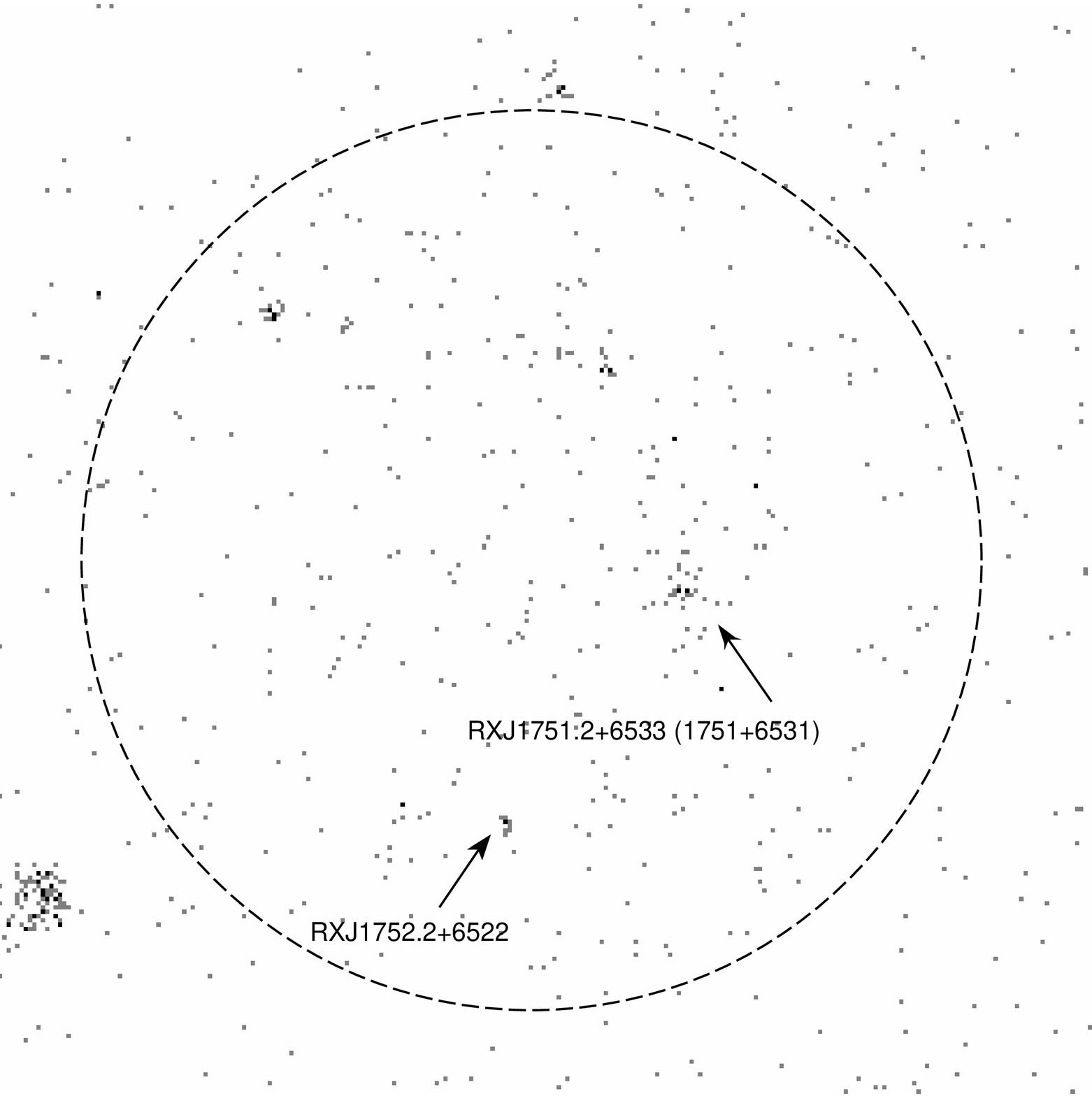}}
  ~
  \framebox{\includegraphics[width=0.406\linewidth]{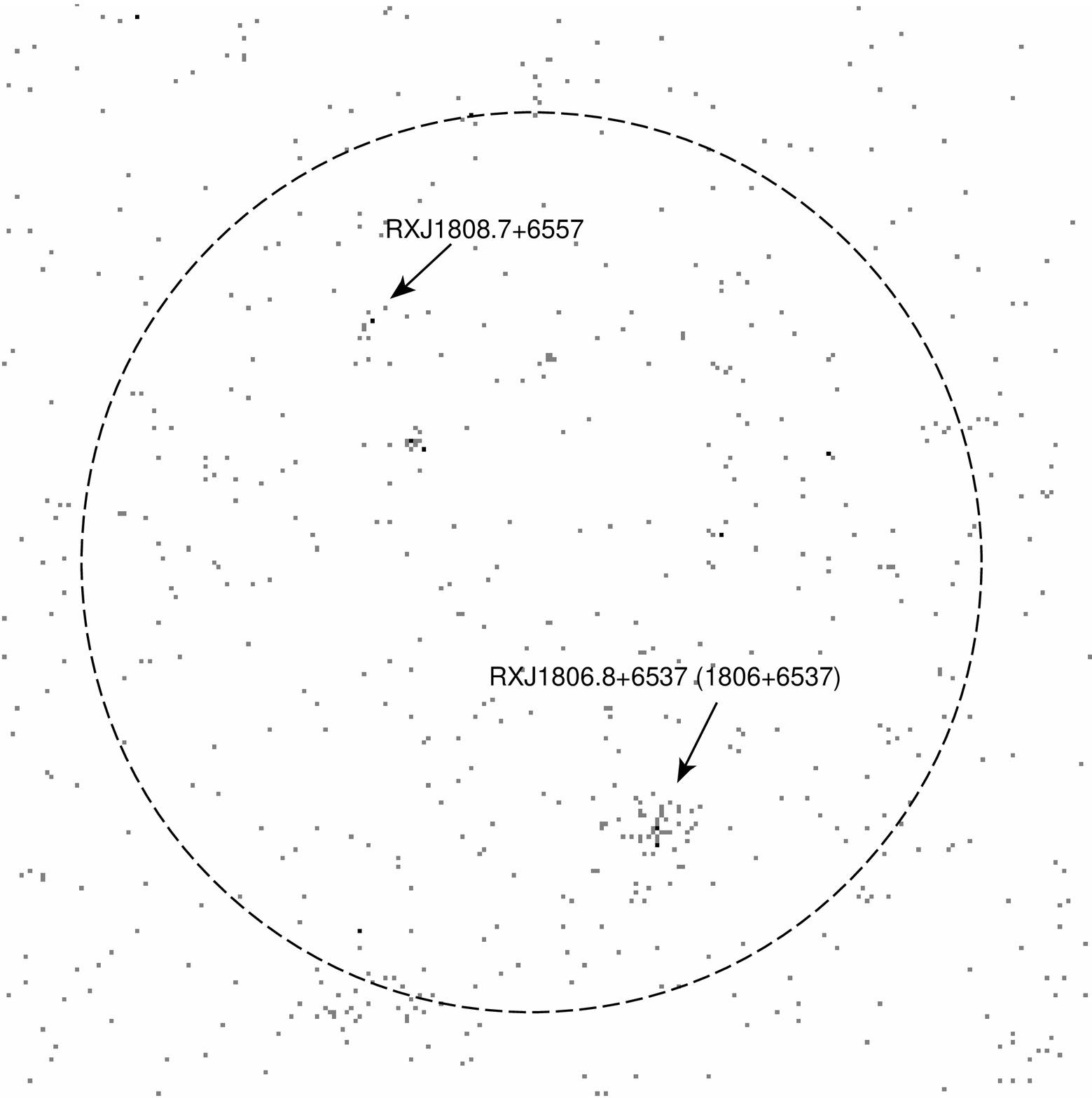}}

  \vspace*{0.5mm}
  \caption{Pointed \emph{ROSAT} PSPC observations containing NEP sources
    RXJ1749.0+7014, RXJ1751.5+7013, RXJ1752.2+6522, RXJ1754.7+6623, and
    RXJ1808.7+6557 which appear point-like. Also marked are the NEP
    sources RXJ1748.6+7020 (too faint to be detected in the pointed
    observation), and RXJ1751.2+6533 and RXJ1806.8+6537 that correspond
    to our clusters 1751+6531 and 1806+6537. Dashed circles show the
    central 17.5\arcmin\ of the FOV.} 
  \label{fig:nep_point_sources}
\end{figure*}

\paragraph{NEP} The North Ecliptic Pole (NEP) survey
\citep{henry01,gioia03} includes sources in the high-exposure region of
the \emph{ROSAT} All-Sky Survey. The nominal detection threshold of the
NEP survey is below our flux limit.  The advantage of NEP is a large
contiguous area (81~deg$^2$) with fairly uniform X-ray data. The
disadvantage for cluster detection is the poor angular resolution,
$\approx2\arcmin$ \citep{boese00}, which complicated the determination
of the spatial extent of detected sources. The total overlap between the
\400d{} and NEP surveys is $\approx5.8$~deg$^2$, where we detect 6
clusters of which 4 are also listed in the NEP catalog. Our clusters
1746$+$6848 and 1807$+$6946 were missed by the NEP probably because of
the presence of bright point X-ray sources in their vicinity. There are
13 NEP clusters in the overlapping region, of which we do not detect~8:
five appear point-like (Fig.~\ref{fig:nep_point_sources}), while 3 have
fluxes which are too faint for the pointed observations. 

\medskip

Comparison of our cluster list with the overlapping data from other
X-ray surveys generally shows good agreement between the catalogs and
thus demonstrates robustness of the X-ray cluster selection.  In cases
where a disagreement is found the cause is not related to the X-ray or
optical analysis in the \400d{} sample. Most of the discrepancies can be
traced to errors in the optical identifications, and therefore the role
of optical data in the cluster selection should be minimized. The
misclassification rate in the NEP and EMSS surveys is relatively high by
modern standards. This underscores the need for X-ray telescopes with
adequate angular resolution in implementing efficient distant cluster
surveys. Robust X-ray analysis techniques applied to such surveys can
provide high-quality cluster samples. 

\section{Statistical calibration of the X-ray detection algorithm}
\label{sec:sim}

The X-ray detection procedure of the \400d{} survey was extensively
calibrated. Our approach is similar to that used in the 160d survey
\citep{vikhl98a}. It is based on Monte-Carlo simulations which provide
the probability of detecting clusters with given X-ray flux and size, the
bias and scatter in the X-ray flux measurements, and the expected number
of false detections. We also studied more subtle effects, e.g., how the
cluster detectability is affected by substructure in the ICM, or by the
presence of central X-ray surface brightness peaks as well as
intracluster point sources.

\label{sec:simruns}

The cluster detectability depends on the observation exposure, the
object off-axis distance, the proximity to other sources, and more
weakly on secondary effects such as the Galactic absorption and the
level of diffuse X-ray background in the field. To properly treat all
these effects, we performed simulations in which clusters with different
input parameters were placed at random in the real \emph{ROSAT} images. 
The simulated data were then run through the complete X-ray analysis
pipeline. All statistical properties of the \400d{} survey are obtained
from comparison of the measured properties and input parameters for the
simulated clusters. 

The input clusters in the majority of the simulations were represented
by the elliptical $\beta$-models (see
Appendix~\ref{sec:cluster:popul:models} for details) but more
complicated cases were also considered (\S\,\ref{sec:areasys}). Input values for fluxes and core radii were selected on a grid spanning the range
$10^{-14}<f_x<3\times10^{-12}$~erg~s$^{-1}$~cm$^{-2}$ and
$5''<r_c<300''$. Simulated clusters were placed in randomly chosen
survey fields and randomly positioned within the central 18.5\arcmin{} of the
field. This radius is larger than the maximum off-axis angle for
detected clusters, $17.5'$. This allows us to properly treat edge
effects but results in detection probabilities $<1$ even for very bright
objects.  In total, we simulated 1,500,000 clusters.

\subsection{Detection Probability}
\label{sec:eff}

\begin{figure}
  \vspace*{-1mm}
  \centering 
  \includegraphics[width=0.95\linewidth,bb=20 170 567 685]{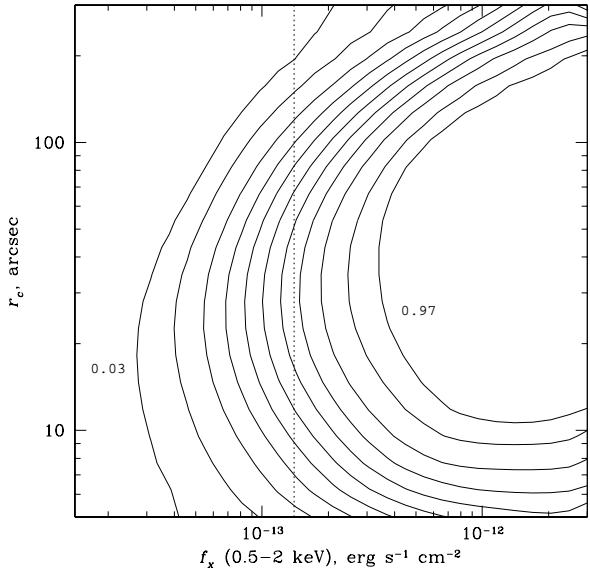}
  \caption{Cluster detection probability as a function of input flux and
    angular core-radius. The contours represent probabilities of 0.03,
    0.1, 0.2, 0.3, 0.4, \ldots, 0.9, and 0.97. Note that this
    probability does not include the effect of imposing the minimum flux
    requirement ($\fmin=1.4\times 10^{-13}\,$\ergcm, dotted line) for
    the clusters to enter our final catalog.} 
  \label{fig:detprob}
\end{figure}

The derived probability of cluster detection as a function of flux and
core radius is shown in Fig.\,\ref{fig:detprob}. The probability is
normalized to the ratio of input and nominal survey areas (see above) to
remove the trivial geometric effects. As expected, the detection
probability is nearly 1 for high-flux clusters and decreases at lower
$f_x$ primarily because faint clusters are not detected in lower
exposure fields. The detection probability for $f_x=1.4\times
10^{-13}$~erg$~$s$^{-1}\,$cm$^{-2}$, the \400d{} catalog flux limit,
exceeds 0.5 in a large range of angular core-radii.

At a fixed flux, the detection probability peaks in the core-radius
range $r_c=10\arcsec-100\arcsec$. The probability decreases at large
$r_c$ because very extended clusters have a lower ratio of the source
and background flux and hence a lower detection significance. The
detection probability is small for compact clusters, $r_c<10''$, because
such clusters are more difficult to distinguish from the point sources
in the off-axis regions of the field of view (the PSF size changes from
25\arcsec{} FVWM on-axis to 60\arcsec{} at an off-axis distance of
17.5\arcmin). We note that the detection probability is a much stronger
function of X-ray flux than of angular size. In fact, the probability
is non-negligible even for very compact clusters, $r_c\approx5\arcsec$. 
Such objects can still be identified as extended sources because their
surface brightness profiles have power-law wings unlike the PSF. 

The core radius range within which our detection algorithm is sensitive,
10\arcsec--100\arcsec, matches well the typical sizes of high-redshift
clusters. For example, this corresponds to a range of 60--600~kpc at
$z=0.5$. This range compares well with the observed core radii for X-ray
luminous clusters which have the median $r_c=180$\,kpc and are
distributed in the range 70--350\,kpc at both $z\approx0$ \citep{jf99}
and $z\approx0.5$ \citep{vikhl98b}. Even at $z=1$, an angular size of 10\arcsec{} corresponds to a physical radius of 80~kpc, which is near
the lower boundary of the observed core-radius distribution.

\subsection{Bias and Scatter in the Flux Measurements}
\label{sec:me}

\begin{figure}
  \vspace*{-1mm}
  \centering 
  \includegraphics[bb=50 140 550 713,width=0.97\linewidth]{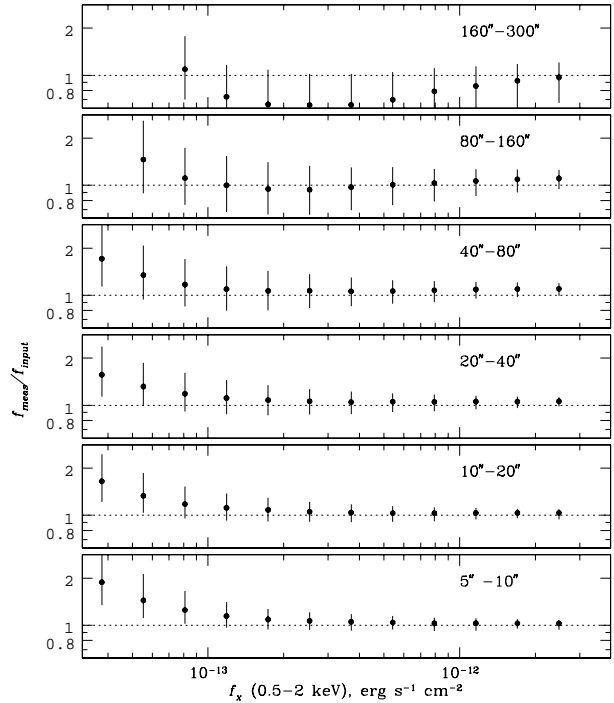}
  \vspace*{-1.3mm}
  \caption{Bias and scatter of the flux measurement as a function of
    input flux. The average value of $f_{\rm meas}/f_{\rm input}$ is
    shown by points and the rms scatter is indicated by error bars. Each
    panel corresponds to different ranges of input core-radii,
    $5''-10''$, $10''-20''$, \ldots, $160''-300''$.} 
  \label{fig:mefluxax}
\end{figure}

Figure\,\ref{fig:mefluxax} shows the average bias and scatter in the
flux measurements for detected clusters of different angular size. There
is no significant bias at fluxes $f_x>1.4\times10^{-13}\,$\ergcm, except
for $r_c>160''$ where the measured $f_x$ are biased low because the
local background is overestimated. This effect is strong only at low
redshifts. For example, the largest core radius in the \cite{jf99}
sample is $r_c=350$\,kpc, corresponding to an angular size of
$r_c<100\arcsec$ at $z>0.2$.  Fluxes of faint clusters are
systematically overestimated which is a typical example of Malmquist
bias. This problem is confined to the flux range below the limiting flux
of our catalog.

\begin{figure}
  \vspace*{-1mm}
  \centering 
  \includegraphics[width=0.95\linewidth,bb=20 170 567 685]{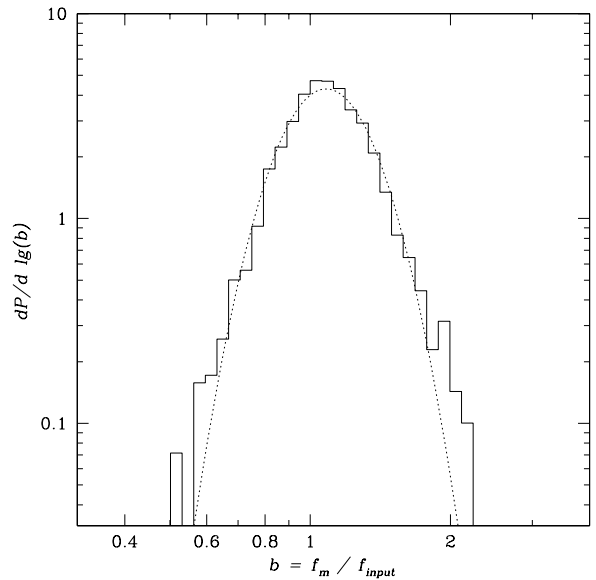}
  \vspace*{-0.3mm}
  \caption{The distribution of scatter in the flux
    measurements for simulated clusters with $f_x=(1.4-2) \times
    10^{-13}$\,\ergcm{} and $r_c=20\arcsec-35\arcsec$. The dotted line shows
    the log-normal approximation.} 
  \label{fig:dsfluxdev}
\end{figure}

Our simulations provide also the shape of the $f_{\rm meas}/f_{\rm
  input}$ distribution. The knowledge of this distribution is required,
e.g., for accurate calculations of the survey area or volume for
clusters near our flux limit. An example of the flux scatter
distribution derived from the simulations is shown in
Fig.\,\ref{fig:dsfluxdev}.  The scatter distribution can be approximated
by a log-normal function (dotted line). However, there is no need to use
this approximation because the distribution is sampled sufficiently
accurately except for the extreme $\sim 1\%$ upper and lower tails.

\subsection{False Detections}
\label{sec:false}

As mentioned above, the \400d{} catalog contains 16 unidentified
sources. They are most likely false X-ray detections arising from
confusion of point sources. We identified most such cases through deep
optical imaging (\S\,\ref{sec:optical}). However, it is useful to
estimate the false detection rate independently from the optical data,
to make sure that we are not missing a new interesting class of sources
(e.g.{} clusters at very high redshift).  Our approach is identical to that
used for the 160d survey. We simulated \emph{ROSAT} PSPC images
containing only point sources and the diffuse background.  The source
fluxes were derived from the observed $\log N$--$\log S$ relation
\citep{hasinger93} and their locations were chosen either randomly or
according to the angular correlation function from \cite{vikhl95a}. The
resulting images correctly reproduce the fluxes and spatial distribution
of detectable sources as well as the background fluctuations caused by
undetected sources. 

\begin{figure}
  \vspace*{-1mm}
  \centering 
  \includegraphics[width=0.95\linewidth,bb=20 170 567 685]{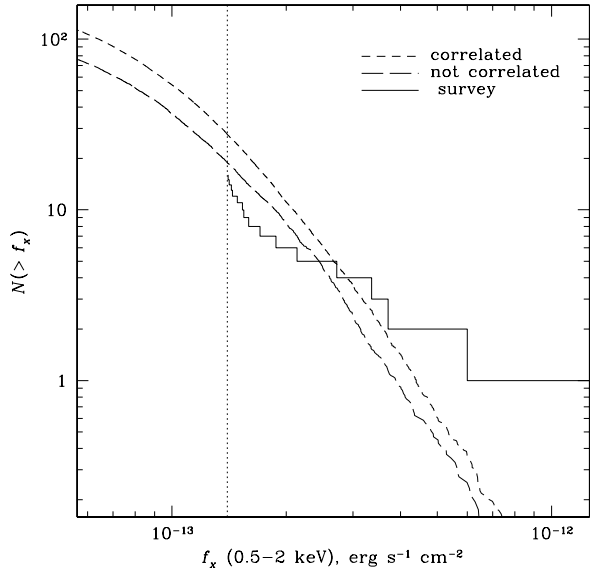}
  \vspace*{-0.34mm}
  \caption{The flux distributions for unidentified \400d{} sources
    (histogram) and false detections derived in the simulations with
    randomly distributed (long-dashed line) and correlated (short-dashed
    line) point sources.} 
  \label{fig:falsecounts}
\end{figure}

\begin{figure}
  \vspace*{-1mm}
  \centering 
  \includegraphics[width=0.95\linewidth,bb=20 170 567 685]{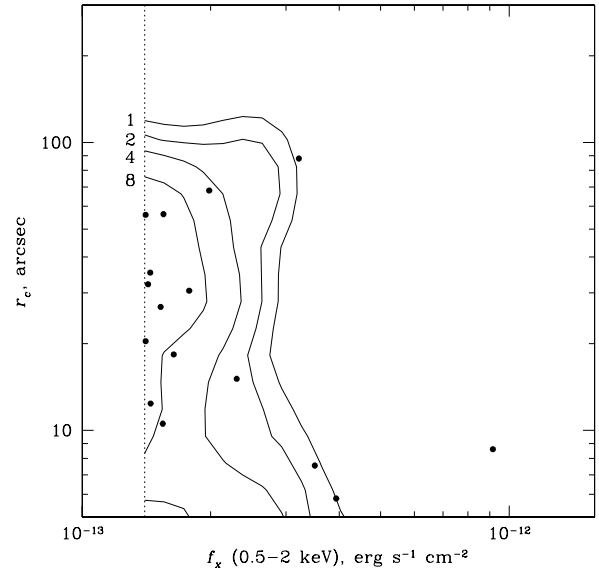}
  \vspace*{-0.34mm}
  \caption{The flux vs.{} size distribution of false detections. The
    isodensity levels are shown by contours labeled to indicate how many
    false sources are outside the contour.  The points show unidentified
    \400d{} sources.} 
  \label{fig:falsedist}
  \vspace{1mm}
\end{figure}

All extended sources detected in these simulations are false by design. 
The distributions of their fluxes and sizes are shown in
Fig.\,\ref{fig:falsecounts} and \ref{fig:falsedist}.  The total number
of unidentified sources in the \400d{} catalog (16) is consistent with
the number of false detections expected for randomly distributed point
sources (18.6).  It is smaller than, but marginally consistent with,
27.1 false detections expected for correlated point sources.  We note
that the angular correlation amplitude measured by \citet{vikhl95a} for
point sources with $f\gtrsim10^{-14}$\,\ergcm{} is not necessarily
directly applicable in the flux range of sources mostly responsible for
false detections in the \400d{} sample, $f\gtrsim10^{-13}$\,\ergcm. The
distributions of fluxes and radii for unidentified sources and false
detection also match very well (Fig.\,\ref{fig:falsedist}). Given the
good agreement in the number and properties of unidentified \400d{}
sources and false detections in the simulations, we conclude that the
unidentified sources are most likely not clusters. 

\subsection{Effects of Point Sources Near Cluster Centers}
\label{sec:point}

In some low-redshift clusters there are X-ray bright AGNs associated
with the central cluster galaxies \citep{2003MNRAS.339.1163C}. The X-ray
luminosity of the central AGN is typically much lower than that of the
host cluster. However, the AGN fraction as well as their typical X-ray
luminosities can increase at high redshift.  Our analysis software does
not attempt to subtract the contribution of point sources near the
center (within $\approx 1$ FWHM of the PSF) from the cluster flux. This
is a deliberate decision to avoid that the peaked X-ray emission
from the cluster cooling cores is misinterpreted as the signature of
a central AGN.  Also, deblending of the central AGNs is an ill-posed
problem in general because the angular size of high-redshift clusters is
close to the PSF width.  The effect of central point sources is twofold.  First,
they increase the source detectability but decrease our ability to
identify it as extended.  Second, the cluster flux is biased high if the
point source emission is not subtracted.  These effects were studied
through additional Monte-Carlo simulations. 

First, we note that our main simulations already include the effects
associated with the chance projection of background sources. Also,
individually detectable point sources are automatically removed from the
cluster flux if they are located outside the central region (e.g.,
Fig.\ref{fig:cl0506m2840}).  Therefore, we consider here only sources
located within the cluster core radius. The simulations proceed as usual
but we place point sources of various flux on top of input $\beta$-model
clusters.  Point sources were placed either at the center of the
$\beta$-model (to simulate AGNs in central galaxies of relaxed clusters)
or randomly distributed within a $1\,r_c$ circle (to simulate AGNs in
non-central galaxies or merging clusters). 

\begin{figure}
  \vspace*{-1mm}
  \centering 
  \includegraphics[width=0.95\linewidth,bb=20 170 567 685]{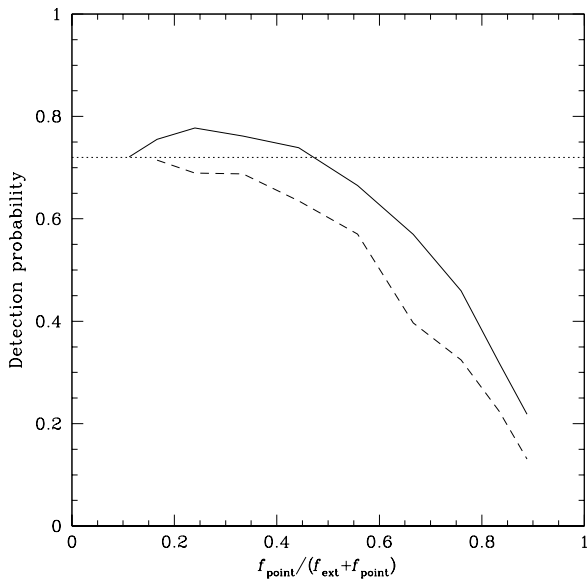}
  \caption{Detection probability for simulated clusters with point
    sources near the center, as a function of the point-source flux
    fraction. The extended component has $f=(1.4-3)\times
    10^{-13}$\,\ergcm{} and $r_c=10\arcsec-60\arcsec$. The horizontal line
    shows the mean detection probability for such clusters without the
    point sources.  Dashed and solid lines show the cases of point
    sources located within 15\arcsec{} and $1\,r_c$ from the cluster
    center, respectively.} 
  \label{fig:pointdetprob}
\end{figure}

\begin{figure}
  \vspace*{-1mm}
  \centering 
  \includegraphics[width=0.95\linewidth,bb=20 170 567 685]{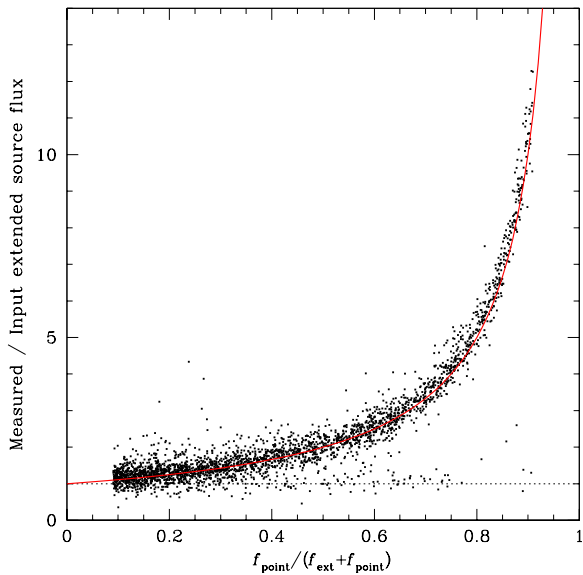}
  \caption{Bias in the flux measurements caused by point sources. The solid
    line shows the expected bias in the case when the point source flux is
    simply added to the cluster, $(f_{\rm point}+f_{\rm ext})/f_{\rm ext}$.} 
  \label{fig:pointflux}
\end{figure}

Figure~\ref{fig:pointdetprob} shows the effect of central point sources
on the detection probability for clusters with fluxes near our catalog
threshold ($f_{\rm ext}=(1.4-3)\times 10^{-13}$~\ergcm). The curves show
the detection probability as a function of the ratio $f_{\rm point}/(f_{\rm
  ext}+f_{\rm point})$. The average detection probability for clusters
without central sources is shown by the horizontal line. The detection
probability decreases significantly only for $f_{\rm point}>0.5$ (i.e.,
the point source is more luminous than the host cluster). Even though
the detection probability is affected weakly, the measured flux is
always strongly biased (Fig.\ref{fig:pointflux}). We measure essentially
$f_{\rm ext}+f_{\rm point}$, except in a small number of cases when our
algorithm enters the deblending mode and $f_{\rm ext}$ is correctly
recovered. 

To summarize, the presence of X-ray bright central AGNs should not affect
the catalog completeness unless the AGN luminosities exceed those of the
host clusters, $\sim 10^{44}$~\ergs{} at $z=0.5$. However, the \emph{ROSAT}
X-ray fluxes can be overestimated in such cases. \emph{Chandra} observations
of our distant clusters will help to assess the importance of this effect.

\section{Using the Survey Statistical Calibration}
\label{sec:areas:and:volumes}

The most general way to fit models of the cluster population to the
\400d{} data is through the detection probability and flux measurement
scatter functions derived above. The predicted ``response'' of the
\400d{} survey to clusters with a distribution of fluxes and sizes
$n(f,r_c,z)$ is
\begin{equation}
  \label{eq:nobs}
  n_{\mathrm{obs}}(f_m,z) = \iint P_m(f_m|f,r_c) P_d(f,r_c)
  \,n(f,r_c,z) \, d r_c\, df, 
\end{equation}
where $f_m$ is the measured flux, $P_d(f,r_c)$ is the detection
probability, and $P_m(f_m|f,r_c)$ is the scatter in the flux measurement as a
function of true flux and size. The function $n_{\mathrm{obs}}(f_m,z)$
is used to compute the likelihood given the number of actually observed
clusters. However, it is useful also to have simpler functions such as
the sky coverage or the survey volume as a function of cluster
luminosity.  Below, we demonstrate how such functions can be computed
and used for non-parametric derivations of the cluster $\log N - \log S$ distribution
or the X-ray luminosity function.

\begin{deluxetable}{p{3cm}cccc}
  \tablecaption{Effective sky coverage of the \400d{} catalog
    \label{tab:selprob}}
  \tablewidth{0.99\linewidth}
  \tablehead{
    \colhead{$f_x$,} & & \multicolumn{3}{c}{$A\,P_{\mathrm{sel}}(f,z)$}\\[1pt]
    \cline{3-5} \colhead{(\ergcm)} & \colhead{$A\,P_{\mathrm{sel}}(f)$} &
    \colhead{$z=0.3$} & \colhead{$z=0.5$\rlap{\large\phantom{$|$}}} &
    \colhead{$z=0.8$} } \startdata
  ~$4.0\times10^{-14}$\dotfill & \pII1.0  &\pII1.1 &\pII1.2 &\pII1.1\\
  ~$6.0\times10^{-14}$\dotfill & \pII4.5  &\pII4.8 &\pII4.7 &\pII4.7\\
  ~$8.0\times10^{-14}$\dotfill & \pI13.5  &\pI14.6 &\pI15.0 &\pI14.9\\
  ~$9.0\times10^{-14}$\dotfill & \pI20.3  &\pI21.7 &\pI21.6 &\pI21.8\\
  ~$1.0\times10^{-13}$\dotfill & \pI32.7  &\pI34.2 &\pI33.9 &\pI34.3\\
  ~$1.1\times10^{-13}$\dotfill & \pI51.0  &\pI52.2 &\pI51.4 &\pI52.2\\
  ~$1.2\times10^{-13}$\dotfill & \pI76.0  &\pI77.3 &\pI76.2 &\pI77.6\\
  ~$1.3\times10^{-13}$\dotfill &   108.6  &  109.0 &  108.1 &  109.8\\
  ~$1.4\times10^{-13}$\dotfill &   145.3  &  143.0 &  142.2 &  146.0\\
  ~$1.5\times10^{-13}$\dotfill &   177.6  &  172.5 &  173.2 &  177.9\\
  ~$1.6\times10^{-13}$\dotfill &   207.5  &  201.7 &  203.6 &  208.4\\
  ~$1.7\times10^{-13}$\dotfill &   230.1  &  222.6 &  226.8 &  231.7\\
  ~$1.8\times10^{-13}$\dotfill &   248.2  &  240.5 &  245.8 &  250.2\\
  ~$2.0\times10^{-13}$\dotfill &   278.8  &  270.7 &  278.1 &  281.1\\
  ~$2.4\times10^{-13}$\dotfill &   313.6  &  303.6 &  314.4 &  315.9\\
  ~$3.0\times10^{-13}$\dotfill &   340.6  &  333.4 &  342.7 &  342.7\\
  ~$4.0\times10^{-13}$\dotfill &   366.0  &  364.0 &  368.8 &  367.0\\
  ~$5.0\times10^{-13}$\dotfill &   378.9  &  379.7 &  382.5 &  380.1\\
  ~$7.0\times10^{-13}$\dotfill &   384.2  &  388.9 &  390.7 &  388.3\\
  ~$1.0\times10^{-12}$\dotfill &   388.9 & 395.4 & 394.9 & 392.4\relax
 \enddata
 \tablecomments{The table lists effective sky coverage (in units of
   deg$^2$) of the \400d{} catalog as a function of true cluster flux. 
   The coverage is the product of the selection probability
   (\S\,\ref{sec:Psel}) and geometric survey area, $446.3$~deg$^2$. 
   Column~2 gives the area averaged over all redshifts
   (eq.\ref{eq:prob:400d:f}) and columns 3--5 give more accurate
   averages for clusters at $z=0.3$, 0.5, and~0.8
   (eq.\ref{eq:prob:400d:f,z}).} 
\end{deluxetable}

\subsection{Efficiency of Selection to \400d{} Catalog}
\label{sec:Psel}

The \400d{} catalog includes only clusters with a measured flux above
$1.4\times10^{-13}$~\ergcm{}, i.e. we discard objects with fainter fluxes
even if they pass the detection significance criteria.  The probability
for a cluster to be selected in the catalog can be computed as
\begin{equation}\label{eq:prob:400d}
  \Psel(f,r_c) =  P_d(f,r_c) \int_{\fmin}^{\infty}
  P_m(f_m|f,r_c)\,d f_m
\end{equation}
where $\fmin$ is the minimum flux required for the cluster selection
($\fmin=1.4\times10^{-13}$~\ergcm{} for the main \400d{} catalog, or a
higher value if the probability is computed for a brighter
subsample).  Note that $\Psel(f,r_c)$ is different from the detection
efficiency $P_d(f,r_c)$. 

\begin{figure}
  \vspace*{-8.59mm}
  \centerline{\includegraphics[width=1.023\linewidth]{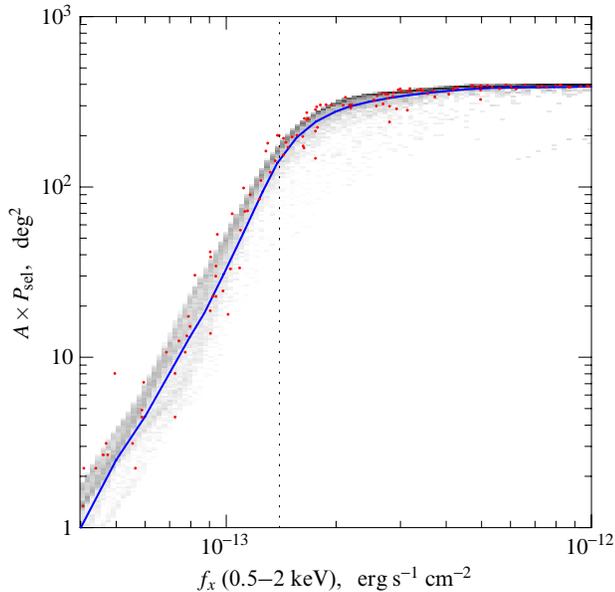}}
  \vspace*{-2mm}
  \caption{Selection efficiency as a function of flux (solid line;
    numerical values are given in the second column of
    Table~\ref{tab:selprob}).  Selection efficiencies for clusters with
    X-ray morphologies from the $z\sim0$ sample
    (\S\ref{sec:areasys}) are shown by points. They agree very well with
    the efficiencies for $\beta$-model clusters drawn from our reference
    population model (shown by shading). The mean of the distribution at
    each flux (represented by the line) is offset from the peak (densest
    shading) because of a long tail towards lower efficiencies. This
    tail is present both in the $\beta$-model clusters and those from
    the real population.} 
  \label{fig:selprob}
\end{figure}

In practice, $P_d(f,r_c)$ and $P_m(f_m|f,r_c)$ are weak functions of
$r_c$ in the plausible range of angular sizes of distant clusters.  This
allows us to eliminate the $r_c$-dependence by averaging
eq.(\ref{eq:prob:400d}) with a realistic distribution of core
radii\footnote{We use the measurements by \cite{jf99}, see
  Appendix\,\ref{sec:cluster:popul:models}.}, $n(f,r_c,z)$:
\begin{equation}\label{eq:prob:400d:f,z}
  \Psel(f,z) = \frac{\int \Psel(f,r_c) \, n(f,r_c,z) \, d r_c}
  {\int n(f,r_c,z) \, d r_c}. 
\end{equation}
The $z$-dependence in eq.(\ref{eq:prob:400d:f,z}) arises because the
X-ray analysis depends on the angular, not proper size. A further
simplification is to use a realistic model for the cluster distribution
as a function of flux and $z$\footnote{Such a distribution can be
  computed using a model for the X-ray luminosity function, see
  Appendix\,\ref{sec:cluster:popul:models}} and average over $z$:
\begin{equation}\label{eq:prob:400d:f}
  \Psel(f) = \frac
   {\iint \Psel(f,r_c) \, n(f,r_c,z) \, d r_c dz}
   {\iint n(f,r_c,z) \, dr_c\, dz}
\end{equation}
The probabilities in eq.(\ref{eq:prob:400d}),~(\ref{eq:prob:400d:f,z}),
and~(\ref{eq:prob:400d:f}), multiplied by the geometric area of the
\400d{} survey, give (in the decreasing order of accuracy) the effective
sky coverage of the \400d{} catalog as a function of \emph{true} flux. 
The calculations for our reference cluster population model
(Appendix\,\ref{sec:cluster:popul:models}) are shown in
Fig.\ref{fig:selprob} and tabulated in Table~\ref{tab:selprob}. 

\subsubsection{Sensitivity to the Cluster Population Models}
\label{sec:areasys}

The calculation of the cluster selection probability
(eq.~\ref{eq:prob:400d:f,z},~\ref{eq:prob:400d:f}) in principle depends
on the exact form of the core-radius distribution and also on the
luminosity function and its evolution. However, these dependencies are
weak as shown below. 

Figure~\ref{fig:darea} shows the relative deviations of $\Psel(f)$
(eq.\ref{eq:prob:400d:f}) calculated with different models of the
cluster X-ray luminosity function (XLF) and core-radius distribution. 
The top panel demonstrates the sensitivity to the assumed evolution of
the XLF (this affects the $z$-distribution of clusters with given X-ray
flux and hence their angular core-radii). $\Psel(f)$ changes by less
than $\pm 2\%$ in the full range of the XLF evolution models consistent
with the \citet{mullis04} measurements. $\Psel(f)$ is slightly more
sensitive to the assumed distribution of core radii (bottom panel in
Fig.\ref{fig:darea}). For example, if the average core radius for
distant clusters is scaled by factors 1.3 and 0.77 relative to the
non-evolving \cite{jf99} distribution, $\Psel(f)$ decreases by 3\%. Such
an evolution of the average core radius is inconsistent with
observations \citep{vikhl98b}.

\begin{figure}
  \vspace*{-1mm}
  \centering 
  \includegraphics[width=0.95\linewidth,bb=20 170 567 685]{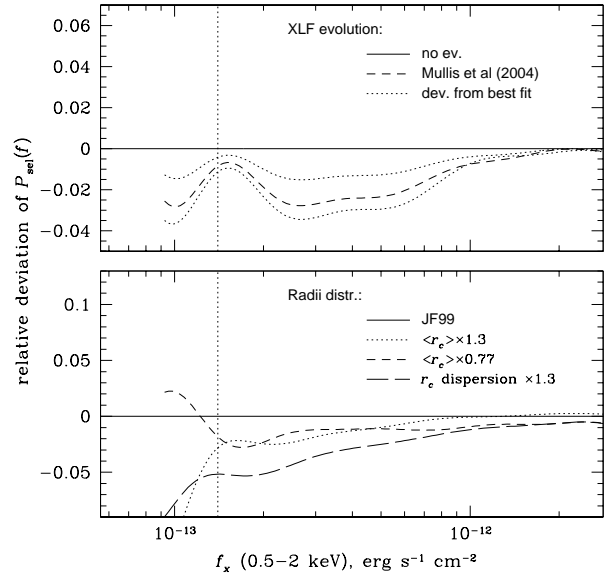}
  \caption{Sensitivity of the survey area calculations to the XLF
    evolution (top) and distributions of core-radii (bottom).  In the
    top panel, we compare the selection probabilities calculated
    assuming non-evolving XLF and the best-fit model from
    \citet[][dashed line]{mullis04}. The range of $\Psel(f)$
    corresponding to the measurement uncertainties in the Mullis et
    al.{} is shown by dashed lines. The bottom panel illustrates the
    sensitivity of $\Psel(f)$ to variations of the mean and width of the
    core radius distribution by $\pm30\%$.} 
  \label{fig:darea}
\end{figure}

\begin{figure*}
  \vspace*{1mm}
  \centering
\includegraphics[width=0.17\linewidth]{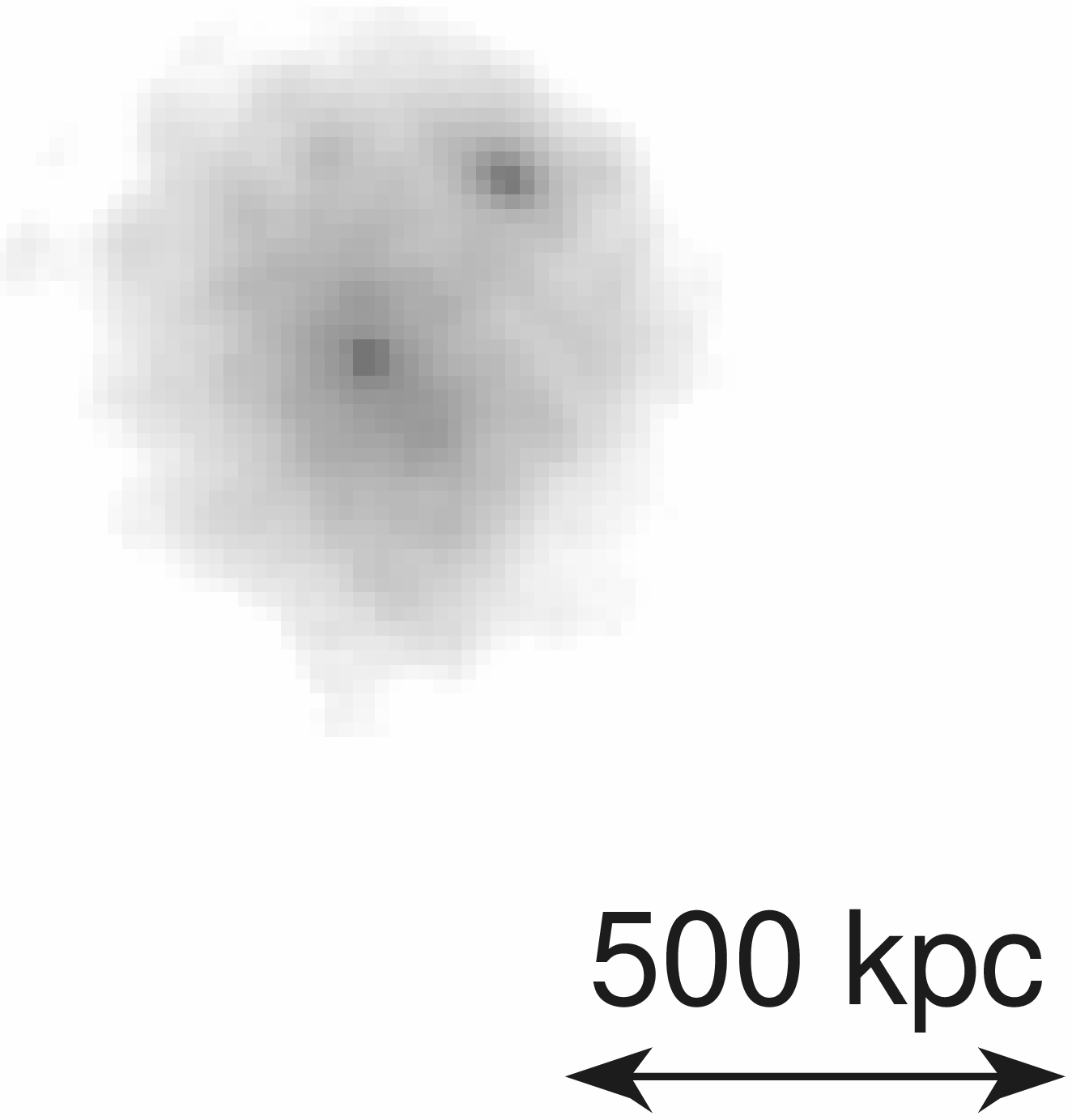}
\includegraphics[width=0.17\linewidth]{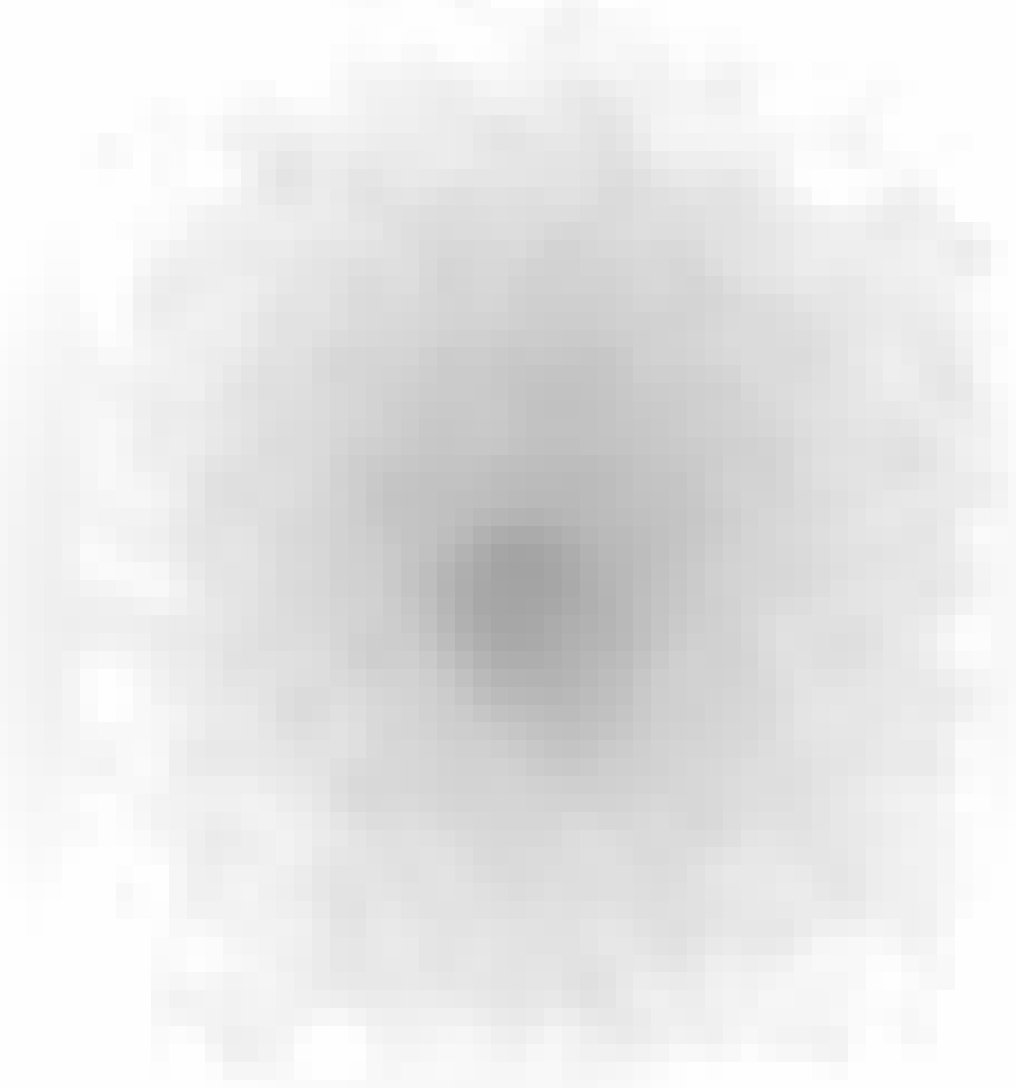}
\includegraphics[width=0.17\linewidth]{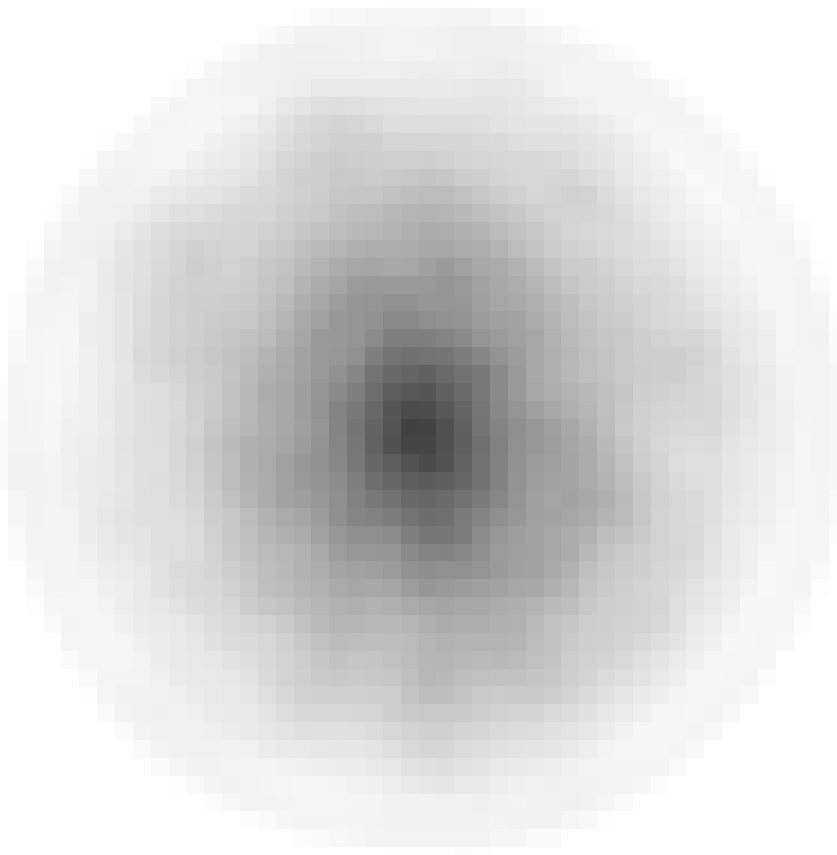}
\includegraphics[width=0.17\linewidth]{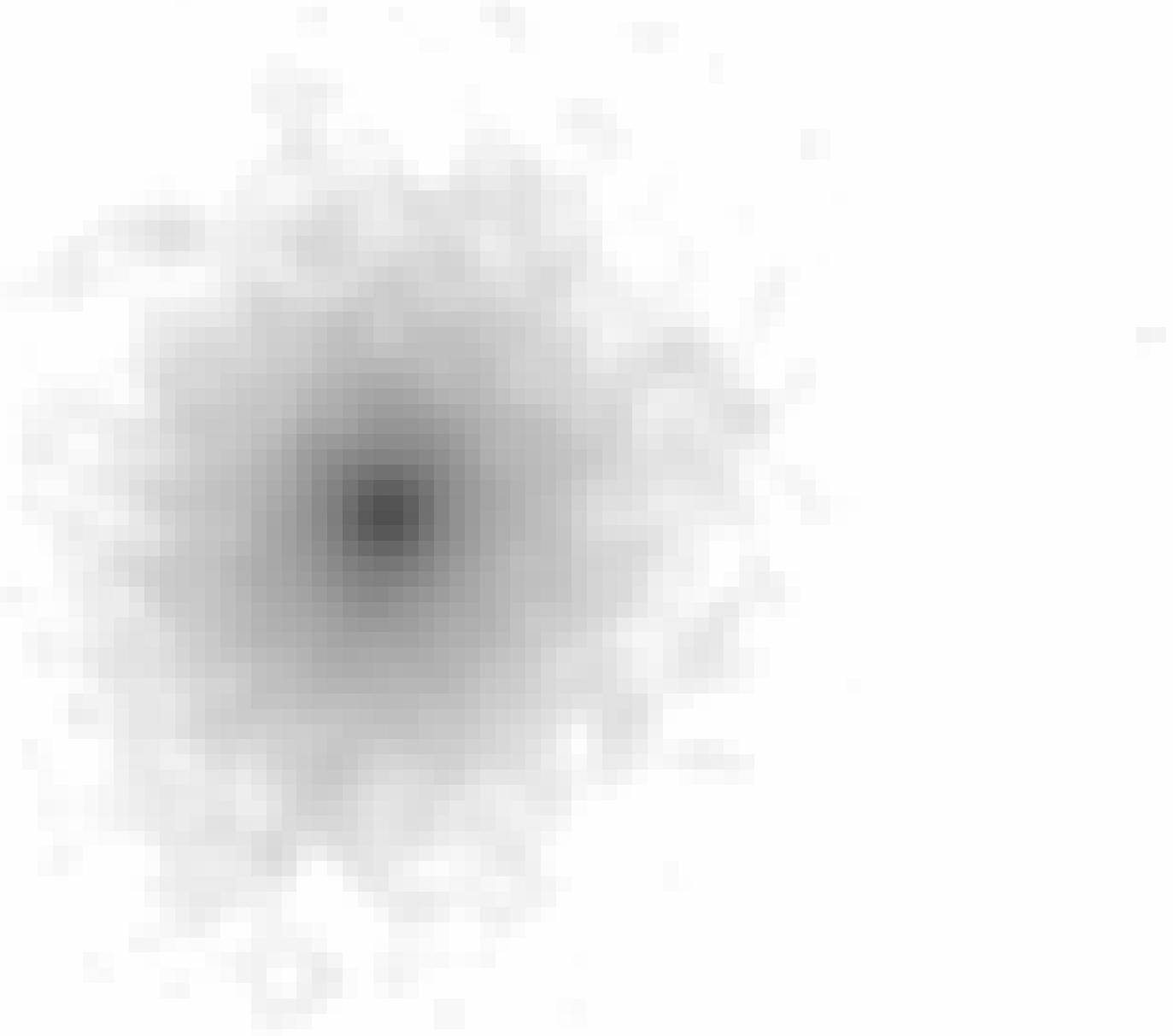}
\par                                           
\includegraphics[width=0.17\linewidth]{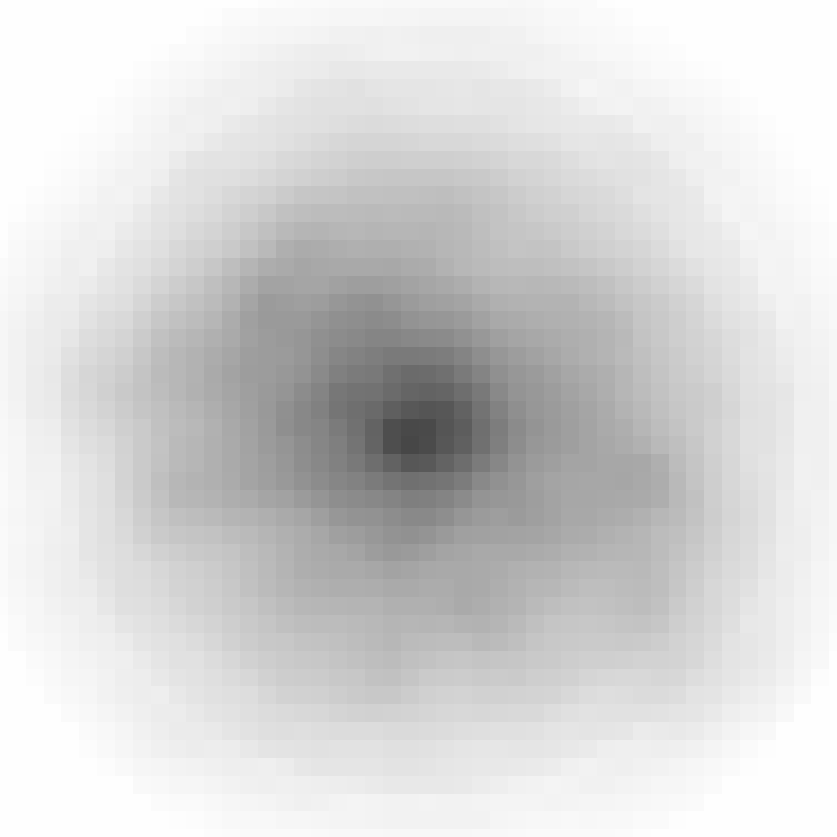}
\includegraphics[width=0.17\linewidth]{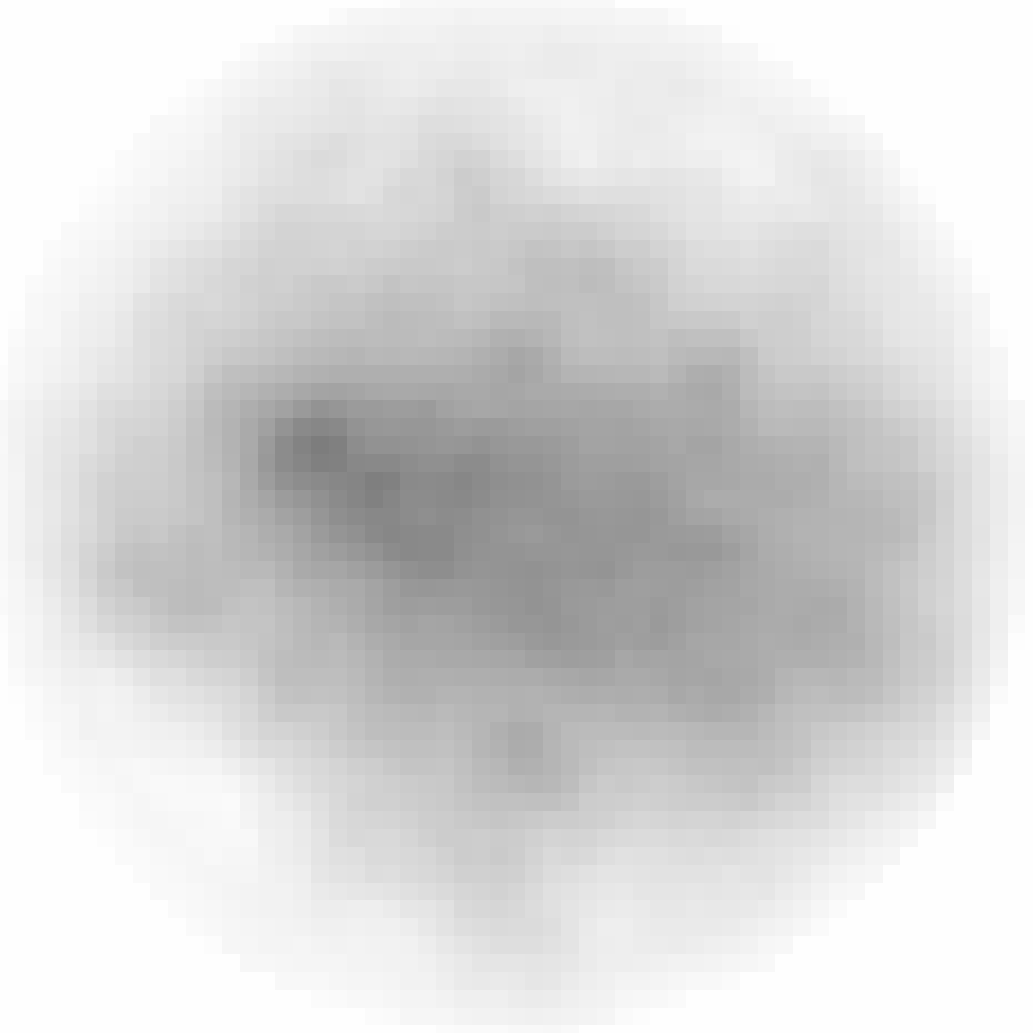}
\includegraphics[width=0.17\linewidth]{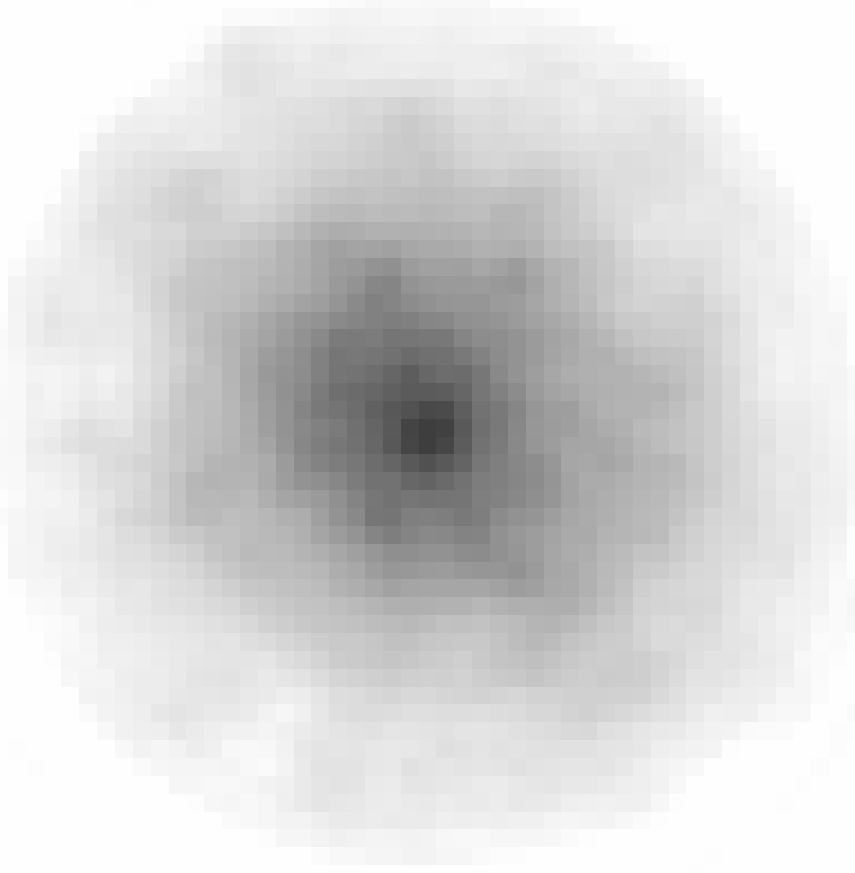}
\includegraphics[width=0.17\linewidth]{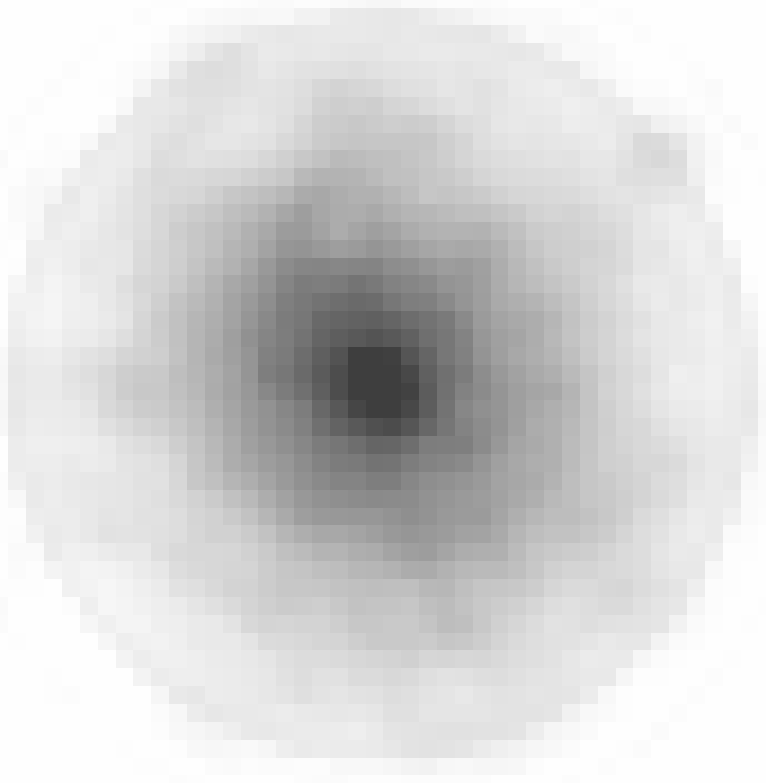}
\par                                           
\includegraphics[width=0.17\linewidth]{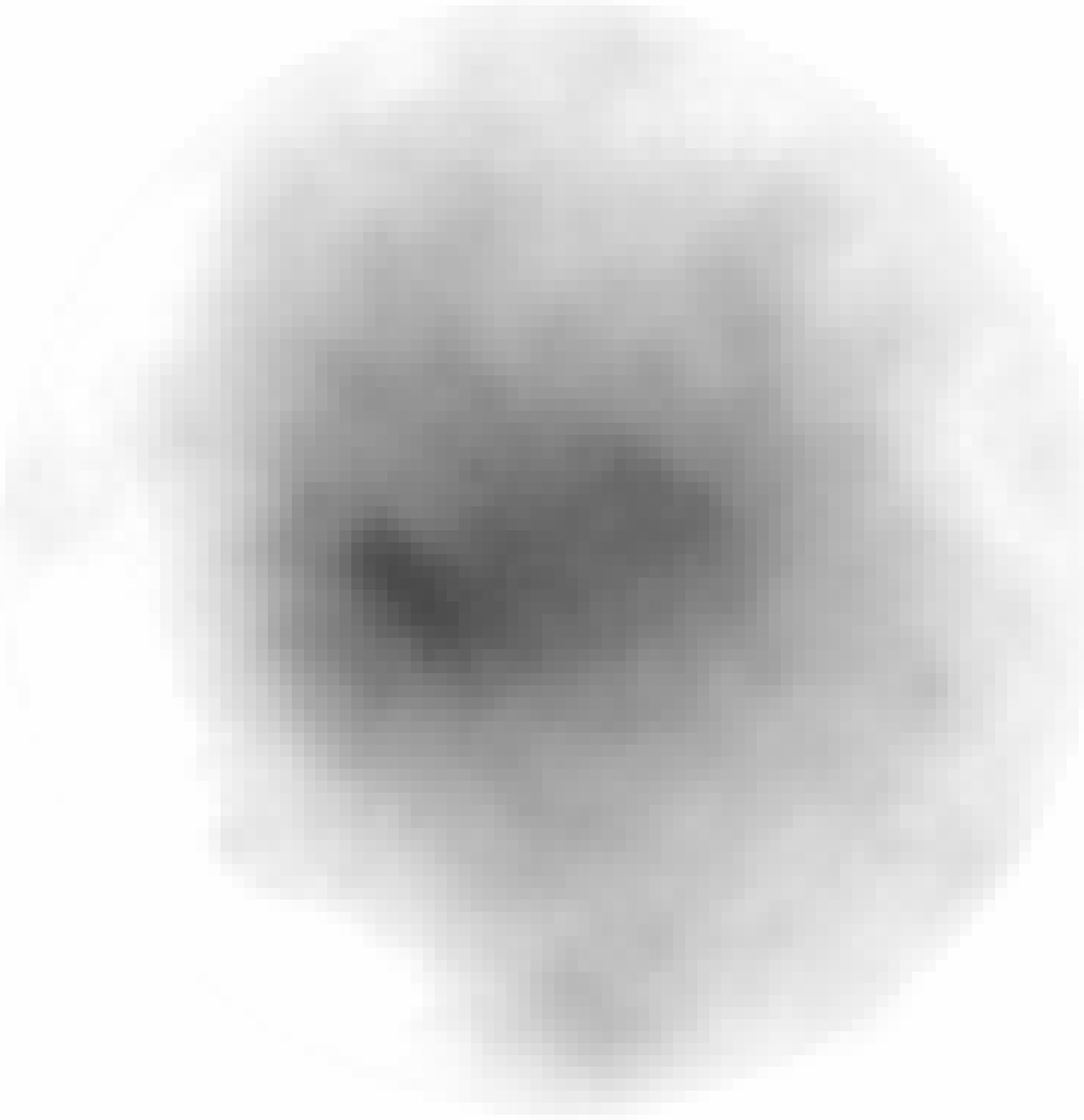}
\includegraphics[width=0.17\linewidth]{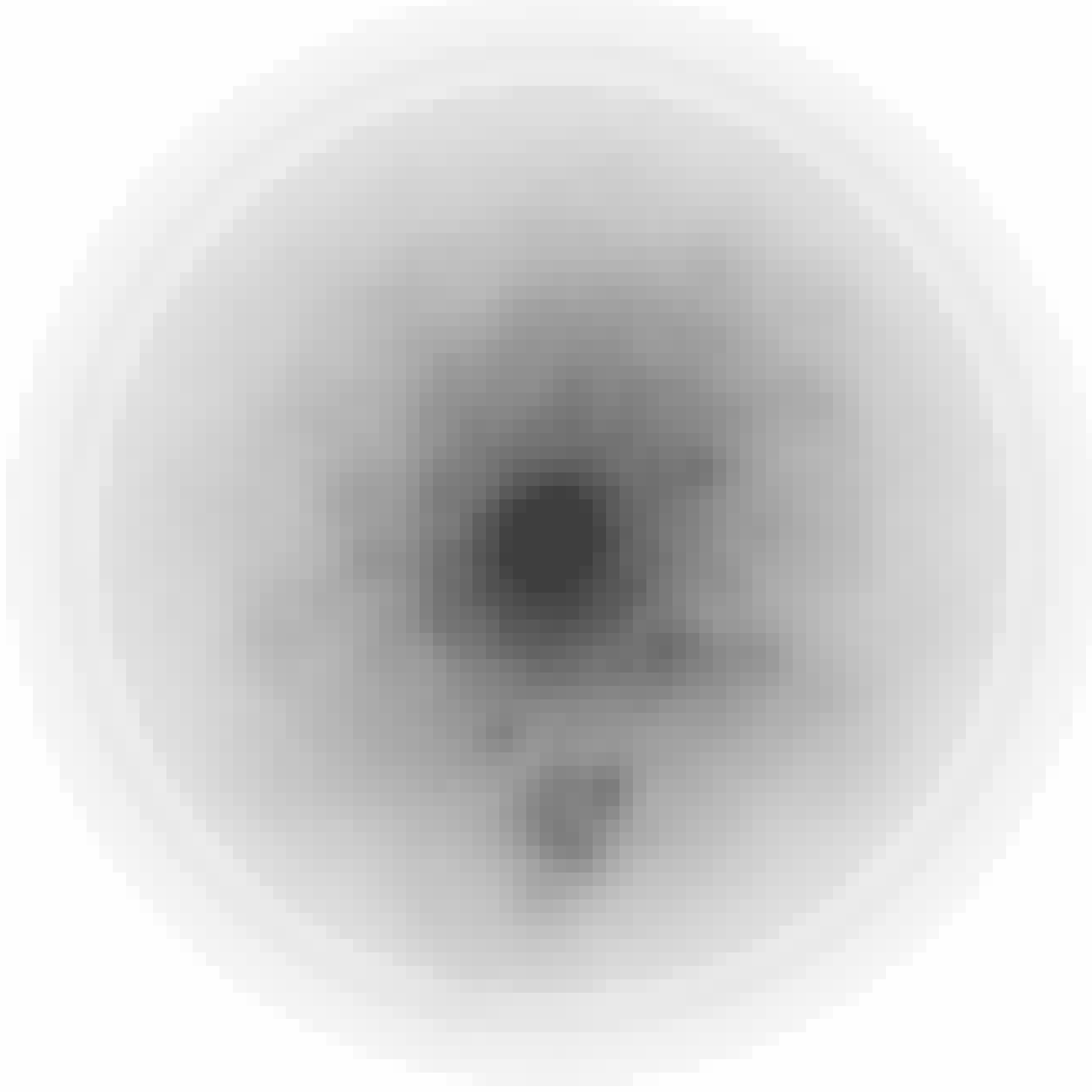}
\includegraphics[width=0.17\linewidth]{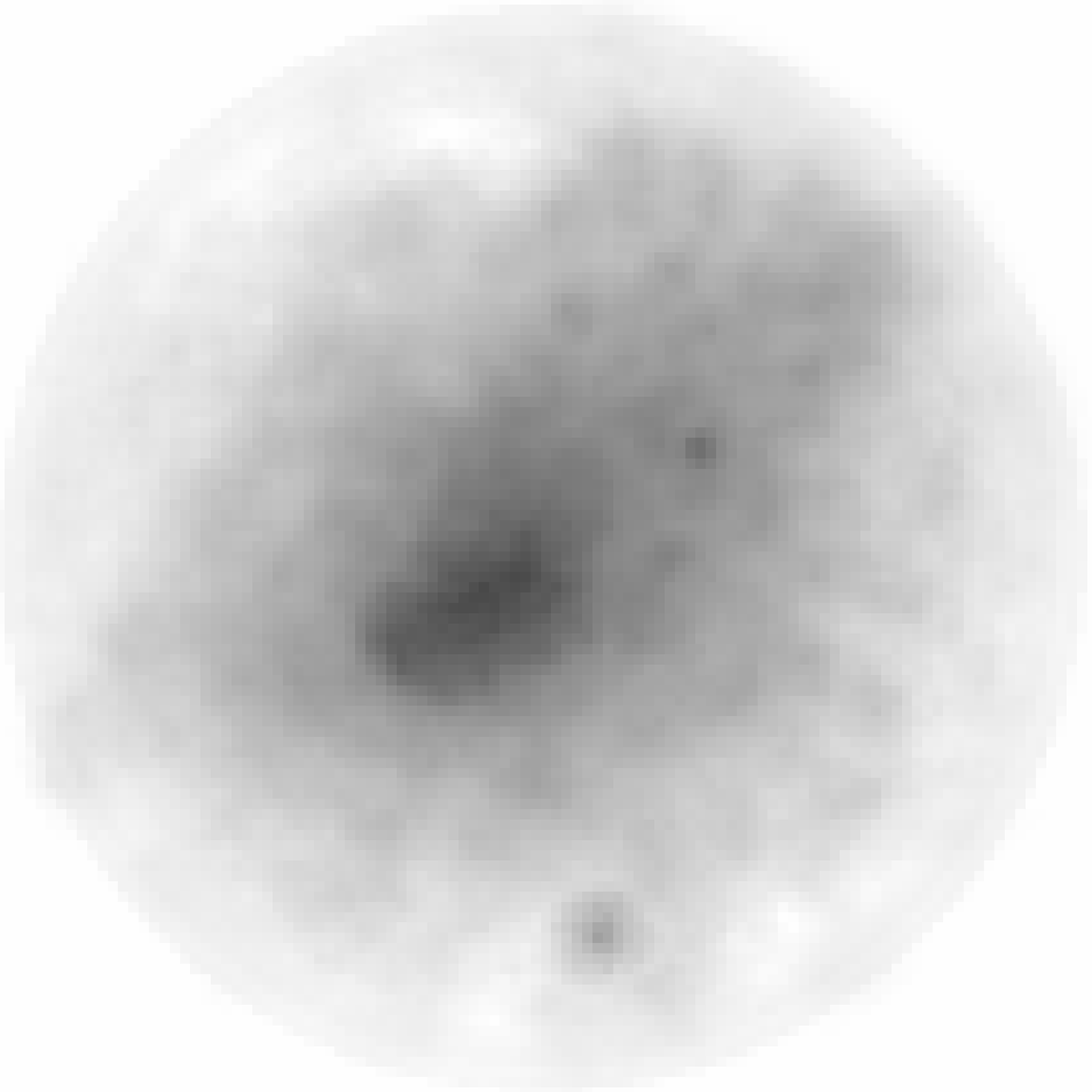}
\includegraphics[width=0.17\linewidth]{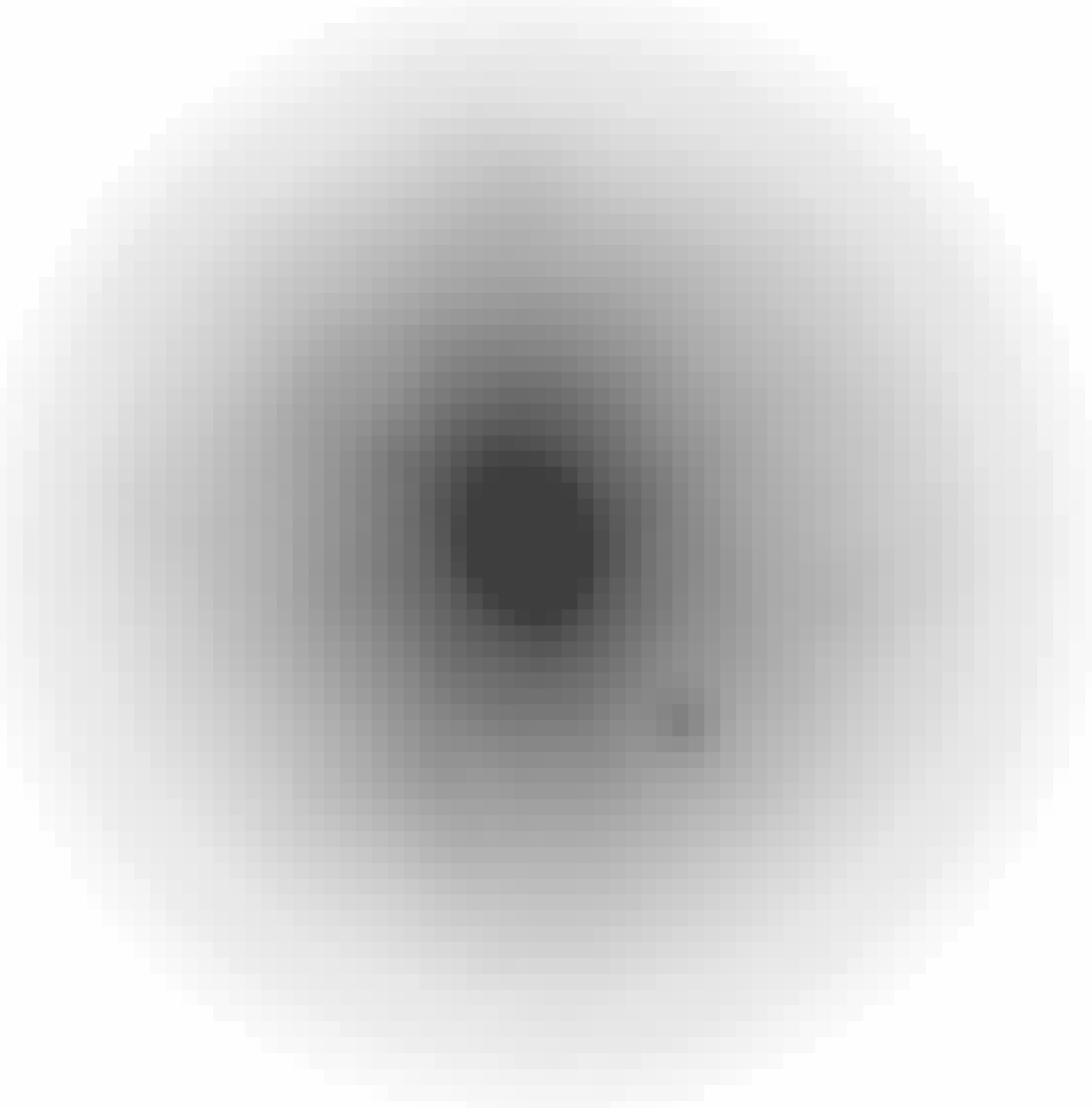}
\par                                           
\includegraphics[width=0.17\linewidth]{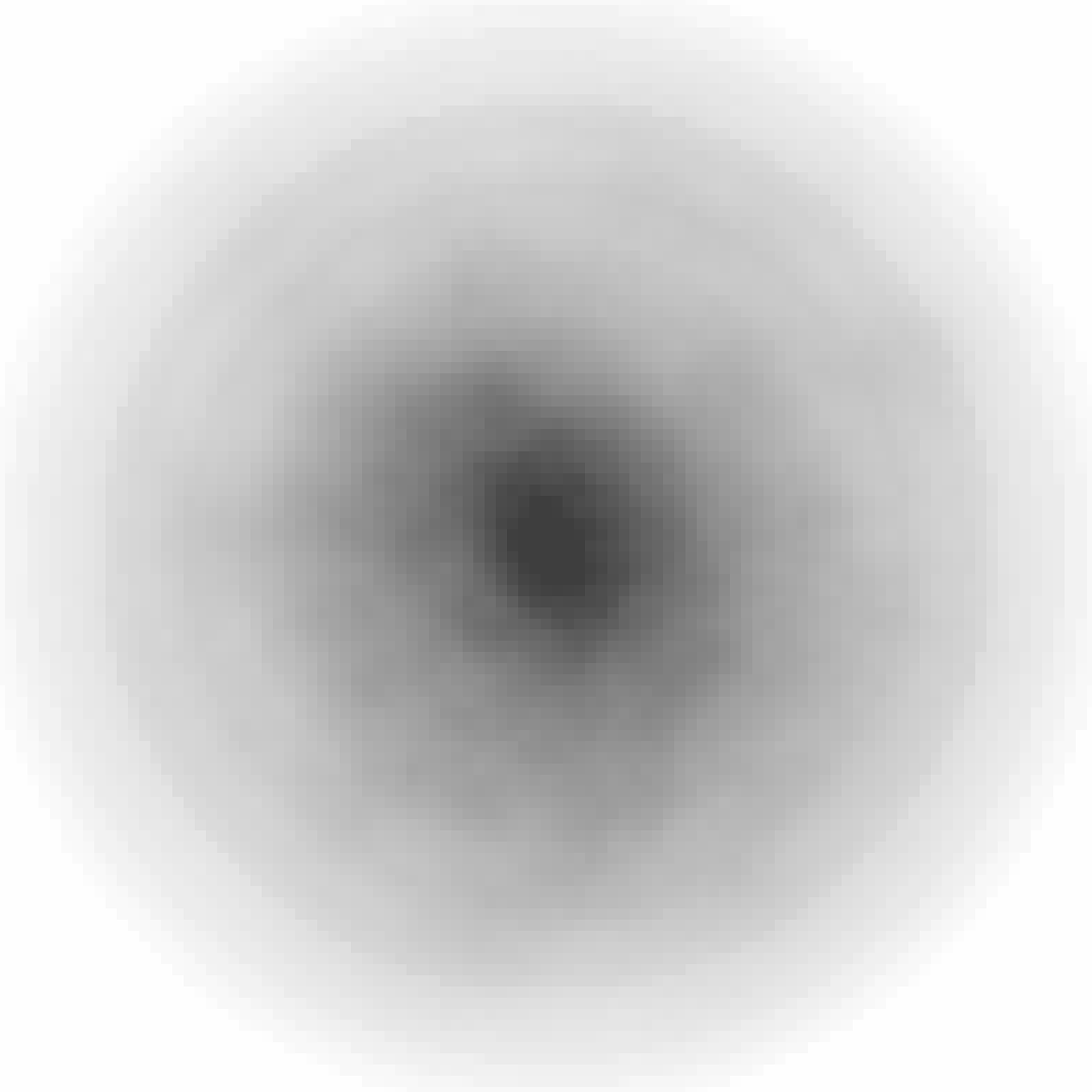}
\includegraphics[width=0.17\linewidth]{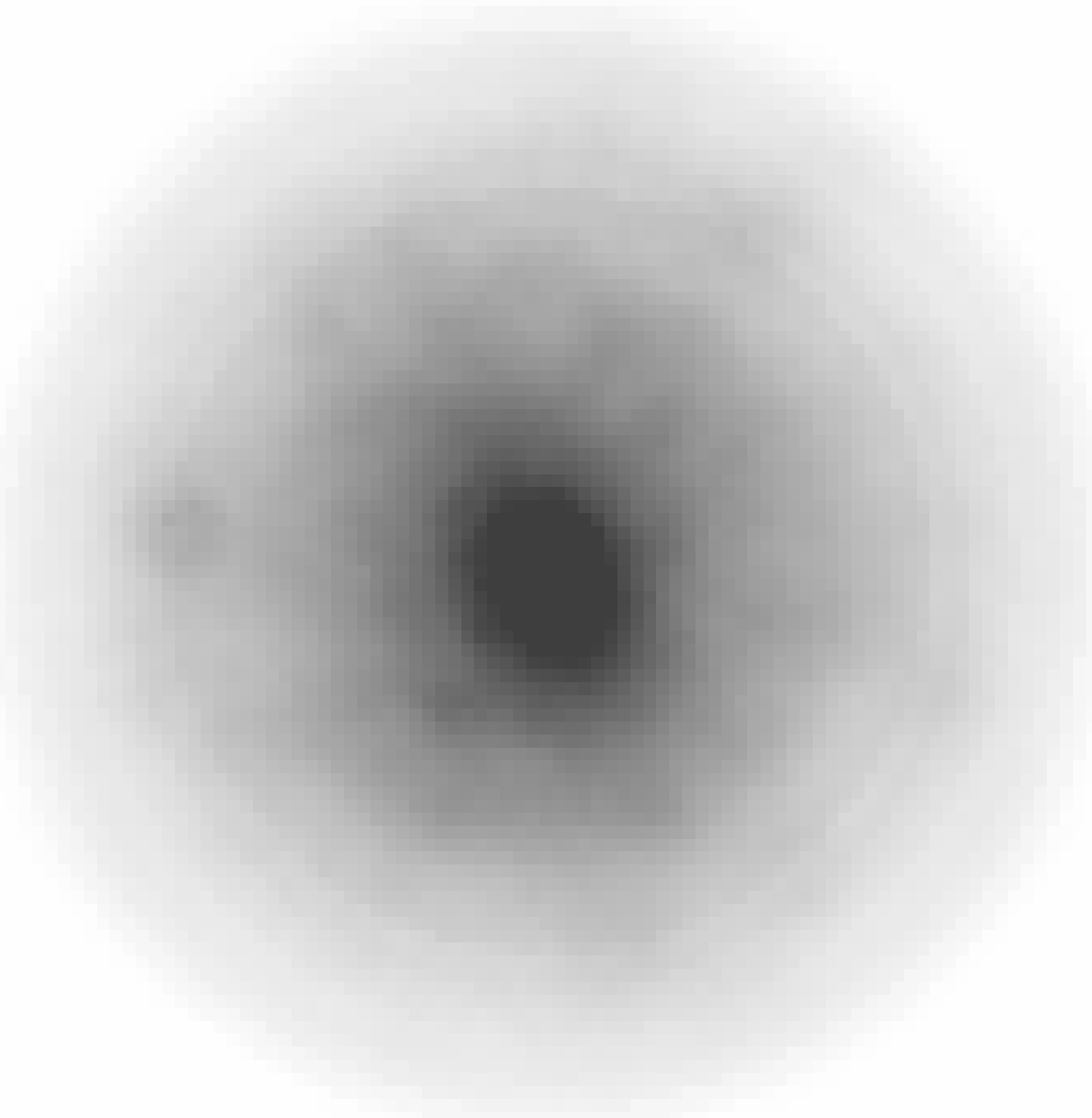}
\includegraphics[width=0.17\linewidth]{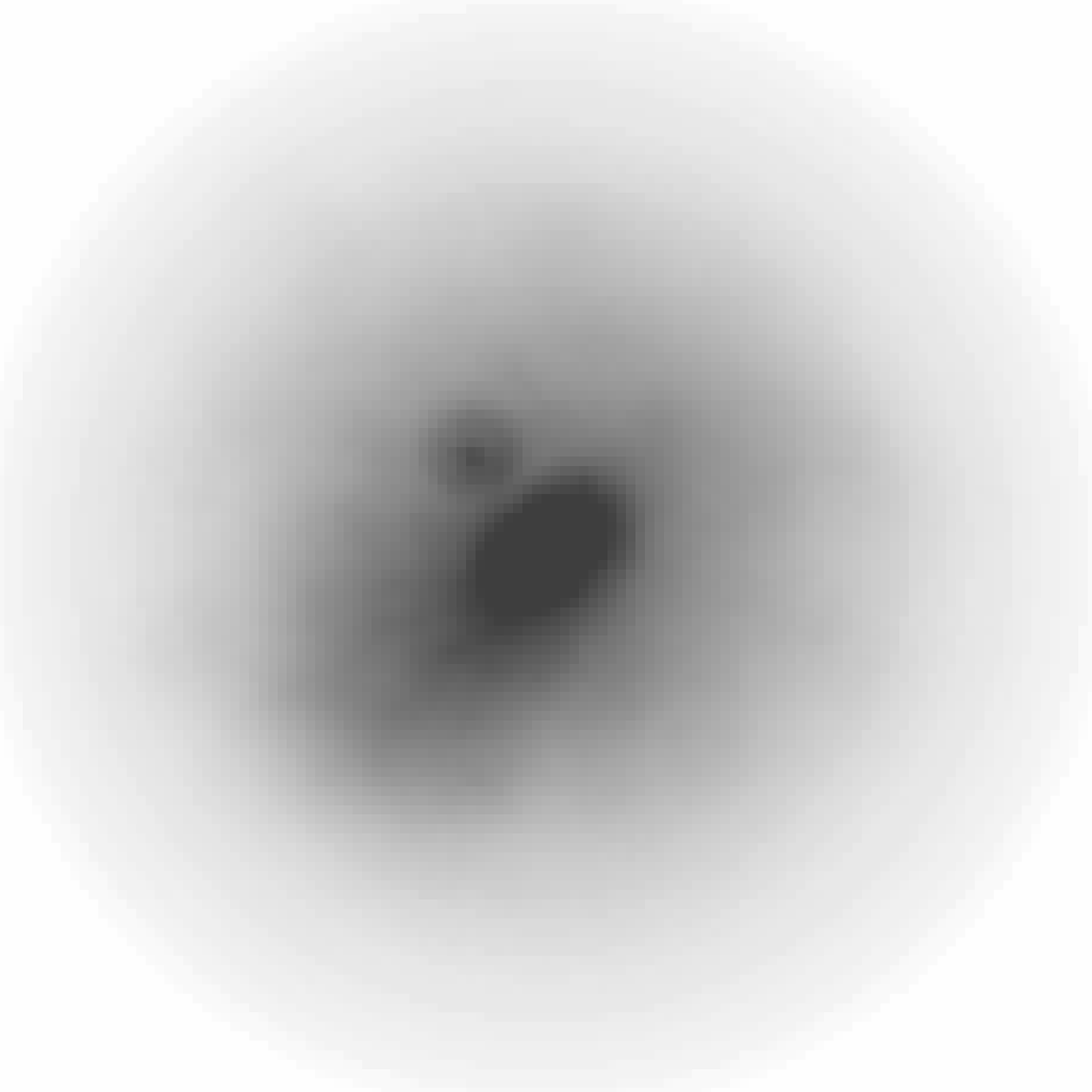}
\includegraphics[width=0.17\linewidth]{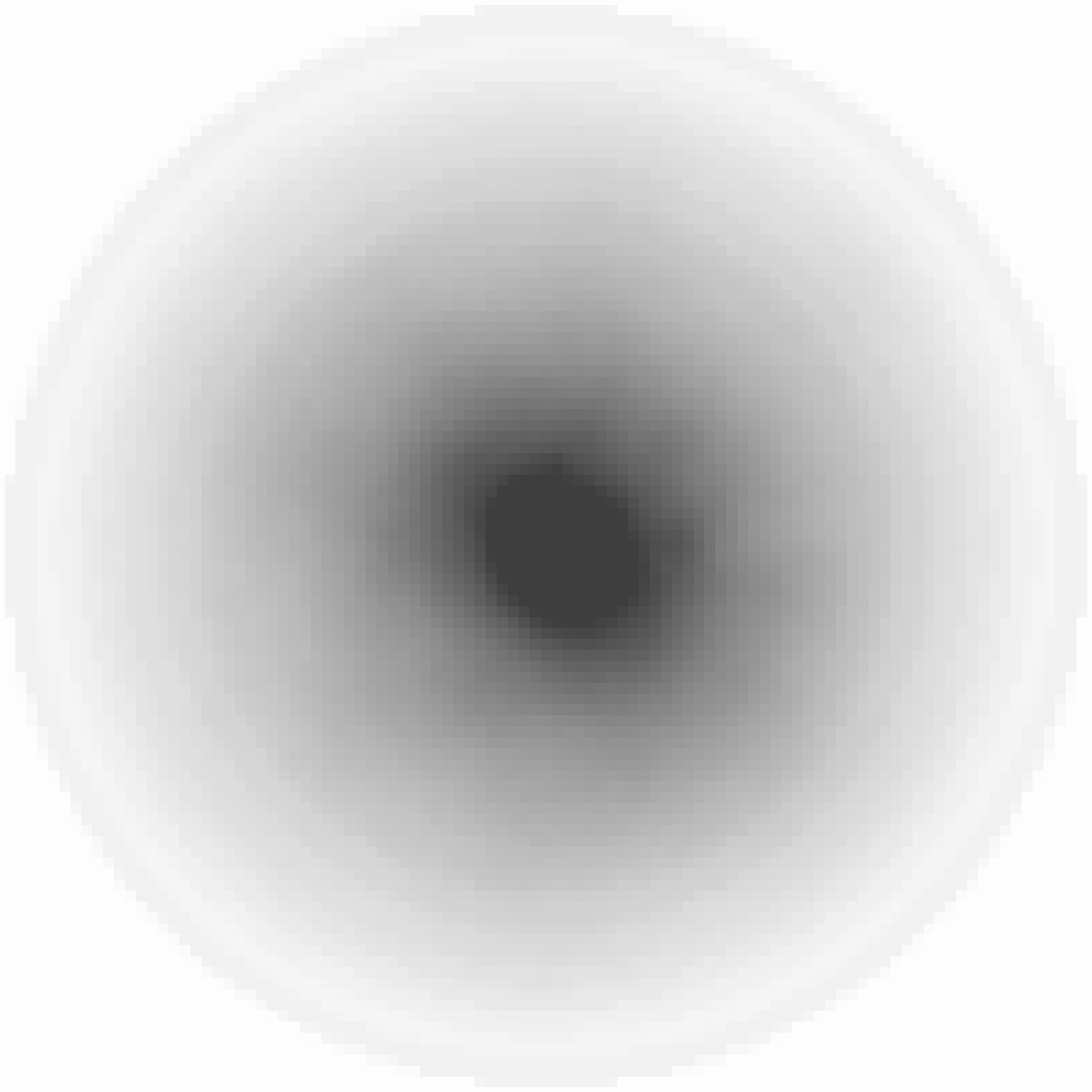}
\caption{Sample images of low-$z$ clusters used in the simulations
  described in \S~\ref{sec:areasys}. From left to right and top to
  bottom, the adaptively smoothed images are for A2634, A0576, A2589,
  EXO0422, A2657, A3376, A2063, A2052, A0754, A0085, A3667, A1795,
  A0401, A2029, A2142, and A0478. The images are scaled to the same
  physical size.} 
  \label{fig:realcl_images}
\end{figure*}

All the cases considered in Fig.\ref{fig:darea} still assume that the
cluster emission follows elliptical $\beta$-models. What if we consider
a more realistic range of cluster structures, from peaked cooling cores
to strong mergers? To check this, we used a complete flux-limited sample
of 38 clusters ($f_x>1.4\times10^{-11}$~\ergcm) in the redshift range
$0.03<z<0.1$ from the HIFLUGCS catalog \citep{2002ApJ...567..716R}. 
The minimum luminosity in this sample of low-redshift clusters
approximately corresponds to the \400d{} sensitivity limit at $z=0.35$. 
These clusters show a wide range of morphologies and represent an
unbiased (with respect to structure) snapshot of the local population
(Fig.\ref{fig:realcl_images}). Template images created from \emph{ROSAT}
and \emph{Chandra} observations of these clusters (shown in
Fig.\ref{fig:realcl_images}) were used in the simulations instead of the
elliptical $\beta$-models. Each cluster was put at $z=0.35$, 0.45, 0.55,
and $0.80$. We scaled the templates in flux and angular size according
to the distance to these redshifts but kept the X-ray luminosity and
physical size of each cluster constant. Constant luminosity corresponds
to a weakly X-ray luminosity function, approximately as observed
\citep{mullis04}.  Constant size corresponds approximately to a
non-evolving scale radius in the Navarro, Frenk \& White model
\citep{1997ApJ...490..493N} of the total mass density profile, as is
indeed expected \citep{2001MNRAS.321..559B}. 

Each cluster leaves a track in the $\Psel(f)$ plot when its redshift is
varied. The combined results for all clusters are shown by points in
Fig.\ref{fig:selprob}. There is a very good agreement with the
calculations using $\beta$-model clusters from our reference model. We
conclude that the deviations of the cluster X-ray morphologies from the
$\beta$-model do not play a significant role in the \400d{} survey
selection functions. 

\subsection{Cluster $\log N - \log S$.} 
\label{sec:logn-logs}

\begin{figure}
  \vspace*{-1mm}
  \centering 
  \includegraphics[width=0.95\linewidth,bb=20 170 567 685]{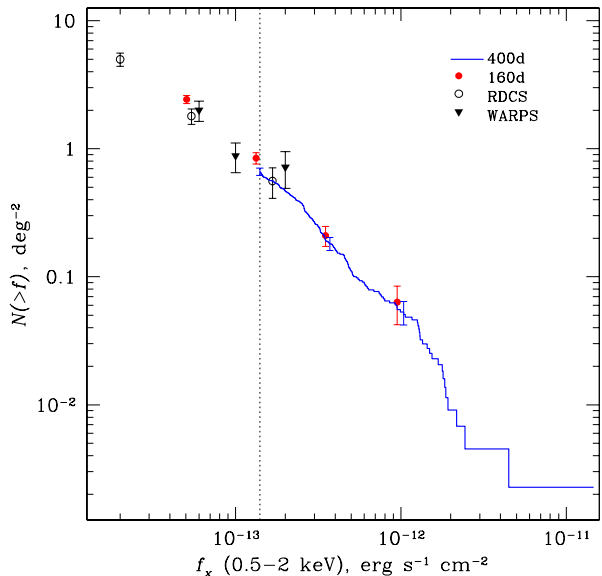}
  \caption{The $\log N - \log S$ function for the \400d{} cluster sample
    (solid histogram; error bars indicate statistical uncertainties at
    several flux thresholds). The results from earlier surveys (160d,
    \citealt{vikhl98a}; RDCS, \citealt{rosati98}; WARPS,
    \citealt{jones98}) are shown by points with error bars.} 
  \label{fig:lnls}
\end{figure}

If the fluxes of all clusters were measured precisely, the $\log N -
\log S$ could be computed using $\Psel(f)$ from
eq.(\ref{eq:prob:400d:f}),
\begin{equation}\label{eq:logn-logs:naive}
  N(>f) = \sum_{f_i>f} (A\,\Psel(f_i))^{-1}
\end{equation}
where $f_i$ is the X-ray flux of individual detected clusters and $A$ is
the geometric survey area. However, because of the flux measurement
errors, the $\log N - \log S$ estimated by eq.(\ref{eq:logn-logs:naive})
will be biased \citep[Eddington bias,][]{eddington40}. 
\citet{2003ApJ...584.1016K} discuss how to reconstruct the true $\log N
- \log S$ function by fitting an analytic model to the distribution of
measured object fluxes. However, it is also useful to reconstruct the
true $\log N - \log S$ non-parametrically. A possible approach
\citep{vikhl98a} is to define the effective sky coverage as a ratio of
differential $\log N - \log S$ for detected and input sources from a
realistic input population (e.g.\ a power law with Euclidean slope). 
This ratio is usually insensitive to the exact form of the input
population and therefore can be used for non-parametric $\log N - \log
S$ reconstruction. The corresponding calculation follows from
eq.(\ref{eq:nobs}):

\begin{equation}\label{eq:area(meas.flux)}
  \Aeff(f_m) = A \frac{\iiint P_m(f_m|f,r_c)\, P_d(f,r_c)
  \,n(f,r_c,z) \, d r_c \, df\, dz}{\iint n(f_m,r_c,z)\,dr_c\,dz},
\end{equation}
where $n(f,r_c,z)$ is the reference cluster population model
(Appendix\,\ref{sec:cluster:popul:models}). The cluster $\log N-\log S$
is then estimated as
\begin{equation}\label{eq:logn-logs:correct}
  N(>f) = \sum_{f_i>f} (\Aeff(f_i))^{-1}. 
\end{equation}
The results are shown in Fig.\ref{fig:lnls} in comparison with several
earlier surveys.  There is very good agreement in the overlapping flux
range. Note a marginal deficit of very bright clusters,
$f_x>3\times10^{-12}$~\ergcm, in our sample relative to a power law
extrapolation from fainter fluxes. This might be related to the fact
that many of the high-flux clusters were previously known and used as
targets for \emph{ROSAT} pointings.  However, this deficit is marginal. 
The observed $\log N - \log S$ distribution is in fact consistent with a
single power law throughout our flux range, $N(>S) = K\,
(S/1.4\times10^{-13})^{-\gamma}$ where $\gamma=-1.25\pm0.07$ and
$K=0.66\pm0.03$~per square degree.

\subsection{Survey Volume and X-ray Luminosity Function}
\label{sec:lf}

The area calculations discussed in \S\S\,\ref{sec:Psel}
and \ref{sec:logn-logs} are straightforwardly generalized for computations
of the search volume. For example, the comoving search volume for clusters
with \emph{true} luminosity $L$ in the redshift interval $z_1-z_2$ is
\begin{equation}\label{eq:V(L)}
  V(L) = A \int_{z_1}^{z_2} \Psel(f,z)\,\frac{dV}{dz}\, dz,
\end{equation}
where $dV/dz$ is the cosmological comoving volume per redshift interval
and $\Psel(f,z)$ is given by eq.(\ref{eq:prob:400d:f,z}). The flux and
luminosity in eq.(\ref{eq:V(L)}) are related through
eq.(\ref{eq:flux:lum}) in Appendix\,\ref{sec:cluster:popul:models}.  In
Fig.\ref{fig:volume}, we show the volume covered by the \400d{} survey
out to redshift $z$, for three representative X-ray luminosities,
$L_x=3\times10^{43}$, $10^{44}$, and $3\times10^{44}$~\ergs. Note that
the volume for luminous clusters, $L_x\gtrsim10^{44}$~\ergs at high
redshifts exceeds the volume typically covered by the catalogs based on
the \emph{ROSAT} All-Sky Survey.

The effective volume as a function of \emph{measured} X-ray luminosity,
including first-order corrections for the Eddington bias (see
\S\,\ref{sec:logn-logs}), is given by expressions
\begin{equation}
  \Aeff(f_m,z) = A\,\frac{\iint P_m(f_m|f,r_c)\, P_d(f,r_c)
  \,n(f,r_c,z) \, d r_c \, df}{\int n(f_m,r_c,z)\,dr_c},
\end{equation}

\begin{equation}\label{eq:vol:Lm}
  \Veff(L_m, z_1, z_2) = \int_{z_1}^{z_2} \Aeff(f_m,z)\,\frac{dV}{dz}\, dz. 
\end{equation}
This effective volume can be used for non-parametric estimation of the
X-ray luminosity function from the \400d{} data, shown in
Fig.~\ref{fig:lf1} and~\ref{fig:lf2} for the $z<0.3$ and $z>0.3$
subsamples, respectively. For comparison, we also show the low-$z$ XLF
from the REFLEX sample \citep{boeringer02}\footnote{Other measurements
  based on the \emph{ROSAT} All-Sky Survey, such as those from the BCS
  \citep{ebeling97} or RASS1 Bright \citep{degrandi99} catalogs, are
  consistent with the REFLEX result within the statistical
  uncertainties. We have chosen REFLEX because the XLF parameters are
  reported for the cosmology adopted here.} which is in very good
agreement with our results, thus providing an independent proof that our
flux measurements, calculations of the sky coverage etc., are accurate. 

\begin{figure}
  \vspace*{-1mm}
  \centering 
  \includegraphics[width=0.95\linewidth,bb=20 170 567 685]{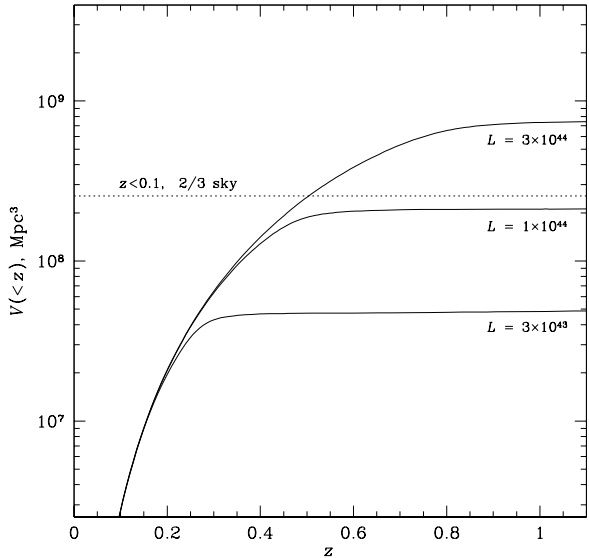} 
  \vspace*{-1mm}
  \caption{The volume covered by the \400d{} survey within redshift $z$,
    for three X-ray luminosities, $L_x=3\times10^{43}$, $10^{44}$, and
    $3\times10^{44}$~\ergs. The volumes are calculated from
    eq.(\ref{eq:V(L)}) where $z_1=0$. For reference, the horizontal line
    shows the typical volume covered at low redshifts by the
    \emph{ROSAT} All-Sky Survey cluster samples (excluding the region
    near the Galactic plane).} 
  \label{fig:volume}
\end{figure}

At $z>0.3$ (Fig.\,\ref{fig:lf2}), our XLF clearly shows negative
evolution at high $L_X$ compared with the low-redshift XLF. For example,
the local XLF predicts that the \400d{} should contain 116 clusters with
$L_x>10^{44}$~\ergs at $z>0.3$ while we found only 47 such objects
($\sim 7\sigma$ significance). The evolution is weaker for
low-luminosity clusters; the total number of the $L_x<10^{44}$~\ergs,
$z>0.3$ clusters in our sample is 17 while the non-evolving XLF
predicts~29.5.  Our XLF results are fully consistent with the earlier
studies by \citet{1992ApJ...386..408H}, \citet{rosati98},
\citet{vikhl98b}, \citet{jones98}, \citet{2001ApJ...553L.105G},
\citet{mullis04} but the evolution is measured with a much higher
statistical significance because of the larger survey area. 

To provide a quantitative characterization of the XLF evolution, we
follow the approach of \citet{rosati02} and \citet{mullis04} to fit the
data with the evolving \citet{1976ApJ...203..297S} model, $dN/dL =
n\,L_*^{-1}\,(L/L_*)^{-\alpha}\,\exp(-L/L_*)$, where the normalization
$n$ and characteristic luminosity $L_*$ are power-law functions of
$(1+z)$,
\begin{equation}
  n = n_0\,(1+z)^A \quad L_* = L_{*,0} (1+z)^B. 
\end{equation}
The best-fit parameters obtained with the Maximum Likelihood approach
\cite{1979ApJ...228..939C} are $A=0.05\pm0.82$ and $B=-1.43\pm0.39$.

\section{\emph{Chandra} observations of 400d clusters}
\label{sec:chandra:obs}

A complete sample of high-redshift clusters from the \400d{} survey was
observed by \emph{Chandra} (PIs L.~Van Speybroeck, S.~S.~Murray,
A.~Vikhlinin). Here, we use preliminary \emph{Chandra} results to
cross-check our X-ray flux estimates from the \emph{ROSAT} data. Due to
sufficiently long exposures, \emph{Chandra} traced the cluster surface
brightness to larger radii and thus provided accurate total fluxes
without the need to rely on $\beta$-model fits. 
Figure~\ref{fig:chandra:rosat} shows a comparison of \emph{Chandra}
fluxes with the \emph{ROSAT} estimates. The horizontal line shows the
flux limit of the \400d{} catalog. By design, all \emph{ROSAT} fluxes
are forced to be above this line, which leads to a significant bias at
low $f_x$. Using the statistical calibrations described in
\S\,\ref{sec:sim}, we can predict this bias.  Specifically, the average
measured flux for clusters with true flux $f$ is (cf.{}
eq.\ref{eq:prob:400d} and~\ref{eq:prob:400d:f,z})
\begin{equation}\label{eq:flux:bias:400d}
  \langle f_m \rangle = \frac
   {\int_{\fmin}^{\infty}
  f_m\, d f_m \int P_m(f_m|f,r_c)\,P_d(f,r_c) \, n(f,r_c) \, d r_c}
   {\int_{\fmin}^{\infty}
  d f_m \int P_m(f_m|f,r_c)\,P_d(f,r_c) \, n(f,r_c) \, d r_c}
\end{equation}
where $\fmin=1.4\times10^{-13}$~\ergcm. The predicted flux bias computed
from eq.(\ref{eq:flux:bias:400d}) is shown by the thick gray line in
Fig.\ref{fig:chandra:rosat}. It agrees very well with the actually
observed bias at $f\lesssim 2.5\times10^{-13}$~\ergcm. At higher fluxes,
where one expects no selection biases, the \emph{Chandra} and
\emph{ROSAT} fluxes agree to better than 10\%: the average observed
ratio is $f_{\rm ROSAT}/f_{\rm Chandra} = 1.02 \pm 0.04$.  The
comparison of our \emph{ROSAT}-derived fluxes with the accurate
\emph{Chandra} measurements thus demonstrates the validity of the
\400d{} statistical calibration and our cluster flux measurements
procedure. 

\begin{figure}
  \vspace*{-1mm}
  \centering 
  \includegraphics[width=0.95\linewidth,bb=20 170 567 685]{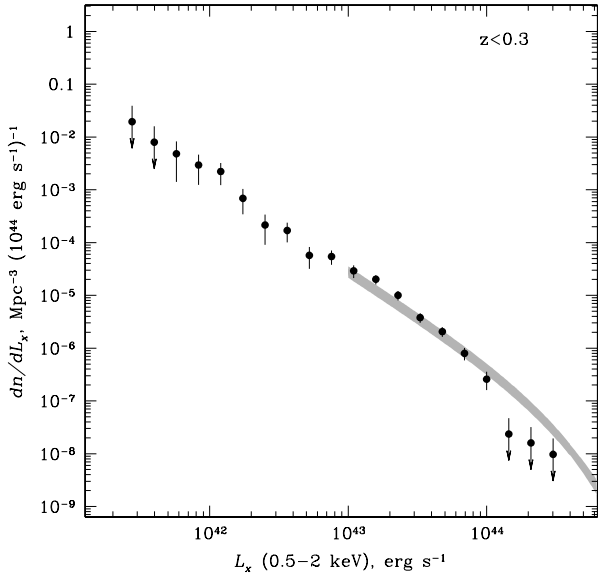}
  \vspace*{0.4mm}
  \caption{X-ray luminosity function for the \400d{} clusters at
    $z<0.3$. The grey line shows the low-redshift XLF measured in the
    REFLEX survey. The width of the line corresponds to the
    uncertainties of the REFLEX XLF.} 
  \label{fig:lf1}
\end{figure}

\section{Summary}
\label{sec:summ} 

We present a catalog of galaxy clusters detected in a new, 400~square
degrees~\emph{ROSAT} PSPC survey. The survey uses the central
17.5\arcmin{} region of 1610 individual pointings to high Galactic
latitude targets, essentially all \emph{ROSAT} PSPC data suitable for
detection of high-redshift clusters.  The X-ray analysis algorithm is
adopted from that in the 160d survey \cite{vikhl98a} with minimal
modifications. The \400d{} catalog includes 266 optically confirmed
galaxy clusters, groups and individual elliptical galaxies with flux
$f>1.4\times10^{-13}$erg$\,$s$^{-1}\,$cm$^{-2}$ in 0.5--2~keV energy
band. This sample is selected out of 287 candidate extended X-ray
sources; the success rate of the X-ray selection is therefore very high,
93\%. Redshifts of all clusters have been measured through optical
spectroscopy (see Hornstrup et al., in preparation). 

The statistical properties of the \400d{} sample have been carefully
calibrated by extensive Monte-Carlo simulations. We provide the
essential quantities, such as the cluster detection probability as a
function of flux and size, and describe how to use them for calculations
of the sky coverage or search volume in the given redshift interval. We
also study the sensitivity of the statistical calibration to details of
the cluster population models (e.g., the presence of the central X-ray
brightness cusps caused by the radiative cooling). These analyses show
that systematic uncertainties in the final area and volume calculations
are within 5\%, smaller than the Poisson uncertainties in our sample.

\begin{figure}[t]
  \vspace*{-1mm}
  \centering \includegraphics[width=0.95\linewidth,bb=20 170 567 685]{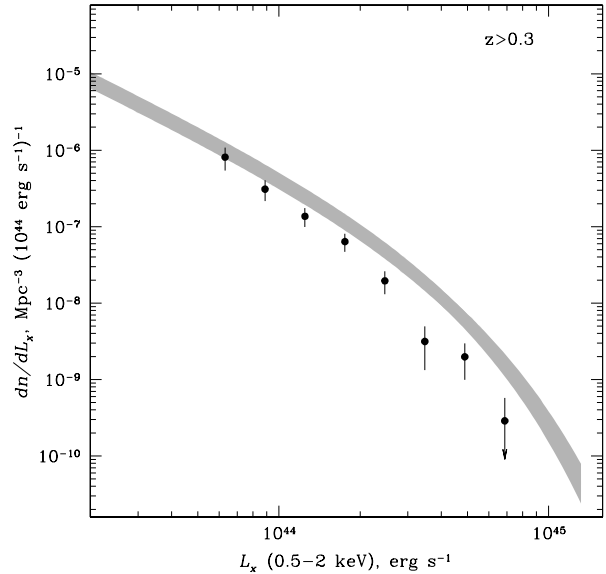}
  \vspace*{0.4mm}
  \caption{Same as Fig.\ref{fig:lf1} but for the \400d{} clusters at
    $z>0.3$.} 
  \label{fig:lf2}
\end{figure}

Our low-$z$ X-ray luminosity function agrees very well with the results
from the \emph{ROSAT} All-Sky Survey. At $z\gtrsim0.3$, the X-ray
luminosity function shows negative evolution significant at $\sim
7\,\sigma$. 

High-redshift clusters from the \400d{} catalog have been observed by
\emph{Chandra}. We will use these data to derive cosmological
constraints (Vikhlinin et al., in preparation). Relevant to the present
work is the comparison of the \emph{ROSAT}-derived fluxes with the
accurate values provided by the \emph{Chandra} data. This comparison
demonstrates the validity of the \400d{} statistical calibration and our
cluster flux measurements procedure. 

Machine-readable tables of our cluster catalog and associated
calibration data are also published on the WWW pages
\texttt{http://hea.iki.rssi.ru/400d} and
\texttt{http://hea-www.harvard.edu/400d}. 

\begin{figure}[t]
  \vspace*{-1mm}
  \centering \includegraphics[width=0.95\linewidth,bb=20 170 567 685]{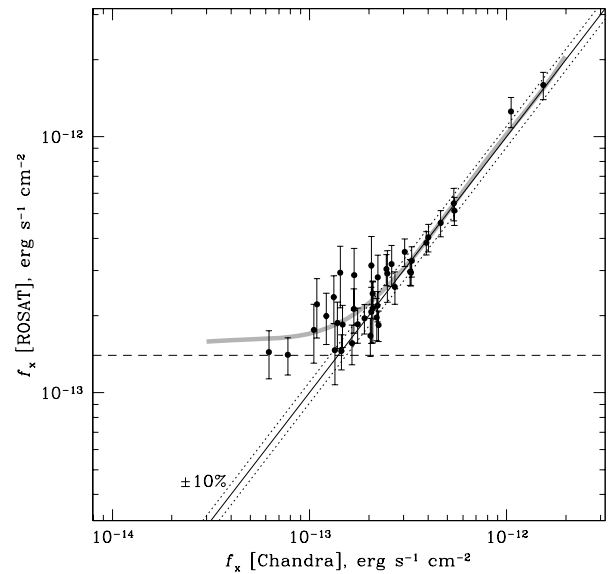}
  \caption{Comparison of the \400d{} cluster flux measurements with the
    accurate values derived from deep \emph{Chandra} observations. The
    horizontal dashed lines indicates the flux limit of the \400d{}
    catalog. The thick gray line shows the expected average measured flux
    as a function of true flux (eq.~\ref{eq:flux:bias:400d}).} 
  \label{fig:chandra:rosat}
\end{figure}

\acknowledgements We acknowledge the contribution of C.~Mullis,
B.~R.~McNamara, W.~Forman, C.~Jones, J.~P.~Henry, and I.~Gioia to the
``parent'' project, the 160d~\emph{ROSAT} survey. We also thank
P.~Berlind, M.~Calkins, M.~Westover, J.~E.~Gonzalez, and G.~Hertling for
help with the optical observing and data reduction.  We acknowledge
generous support from the Time Allocation Committees of various optical
telescopes including RTT-150 operated by the National Observatory of
Turkey, Kazan State University, and IKI and Nordic Optical Telescope
operated jointly by Denmark, Finland, Iceland, Norway, and Sweden, in
the Spanish Observatorio del Roque de los Muchachos of the Instituto de
Astrofisica de Canarias. Financial support was provided by NASA (grant
NAG5-9217 and contract NAS8-39073), and grants from the Russian Academy
of Sciences, Government, and Basic Research Foundation (02-02-16619,
05-02-16540, NSH-2083.2003.2, NSH-1100.2006.2, MK-4064.2005.2). HQ was
partially supported by FONDAP Centro de Astrofisica. 

\bibliography{p400d}

\begin{appendix}

\section{Summary of notations}

For quick reference, we provide a summary of notations used throughout the
paper. 

$r_c$ --- core-radius of the $\beta$-model; in calculations of the
survey area and volumes, it is assumed to be in units of angular size;

$f$, $f_x$ --- true cluster flux in the 0.5--2 keV band;

$f_m$ --- measured flux in the 0.5--2 keV band;

$P_d(f,r_c)$ --- the probability for the cluster to be detected and
identified as extended X-ray source (\S\,\ref{sec:eff});

$P_m(f_m|f,r_c)$ --- the probability for the detected cluster to have
measured flux $f_m$ (\S\,\ref{sec:me});

$\Psel$ --- the probability that the cluster is detected and has $f_m$
above the \400d{} catalog selection threshold (\S\,\ref{sec:Psel});

$n(f,r_c,z)$ --- number density of clusters at redshift $z$ as a
function of true observed flux and angular size;

$A$ --- geometric area of the survey (intersection of the 18.5\arcmin{}
circles centered on individual pointings, minus the target regions);

$\Aeff(f_m)$ --- effective sky coverage as a function of measured flux
(\S\,\ref{sec:logn-logs});

$V(L)$ --- comoving search volume for clusters with true luminosity $L$
(\S\,\ref{sec:lf});

$\Veff(L_m)$ --- effective comoving search volume as a function of
measured luminosity $L_m$ (\S\,\ref{sec:lf});

\section{Reference model of the cluster population}
\label{sec:cluster:popul:models}

The calculation of essential quantities such as the survey area or volume
involves averaging of the detection probabilities over the expected
distribution of cluster sizes and fluxes
(\S\,\ref{sec:areas:and:volumes}). For this, we use the reference model
which assumes a non-evolving population of $\beta$-model clusters with
the distribution of structural parameters and the X-ray luminosity
function fixed by detailed observations of the low-redshift objects, as
detailed below. 

\begin{figure}
  \centering 
  \includegraphics[width=0.5\linewidth,bb=20 170 567 685]{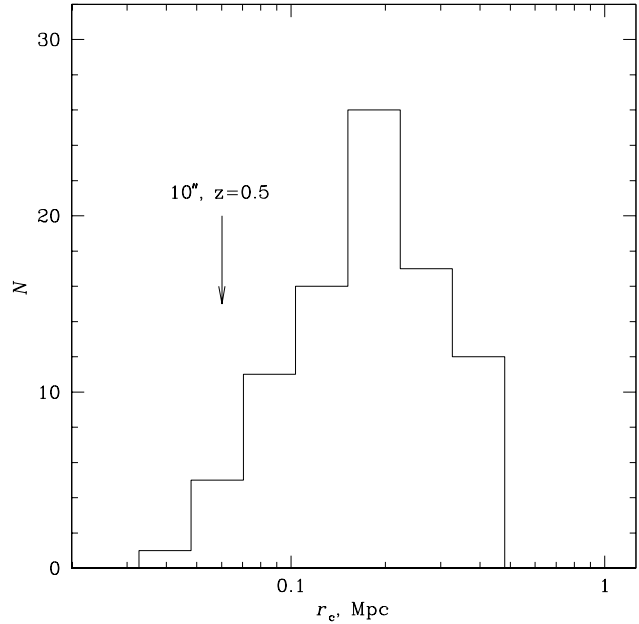}
  \caption{The distribution of cluster core radii in the \citet{jf99}
    sample. The core radii have been corrected for the average trend
    with $L_x$. The arrow indicates the proper distance that corresponds
    to 10\arcsec{} at $z=0.5$ (approximately the angular size below
    which the discrimination of extended and point sources becomes
    ineffective).} 
  \label{fig:dsr}
\end{figure}

The distribution of cluster core radii is taken from the \emph{Einstein}
sample of \cite{jf99}. The use of Jones \& Forman sample is justified
by the fact that nearby clusters at $z\approx0.1$ were observed by
\emph{Einstein} approximately with the same proper-size resolution as
\emph{ROSAT} distant clusters at $z\approx0.5$. The average core radius
in the Jones \& Forman sample shows a trend with the cluster luminosity
which can be approximated as $r_c \propto L_x^{0.5}$ for
$L_x<10^{44}$~\ergs\ and constant average $r_c$ at higher luminosities. 
We corrected individual $r_c$ measurements in the Jones \& Forman sample
by this trend to obtain the master distribution of $r_c$ shown in
Fig.~\ref{fig:dsr}. The distribution for any luminosity is obtained by
scaling the master distribution by the trend of average $r_c$ with
$L_x$. The distribution of $\beta$-parameters is also adopted from
\cite{jf99}, with one modification. The Jones \& Forman distribution
contains objects with $\beta<0.6$ and even $\beta<0.5$. Such low values
are unrealistic (e.g., the total flux diverges for $\beta\le0.5$). They
are artifacts of fitting a single $\beta$-model to the cluster with cool
cores and are in fact not found if the cores are excluded from the fit
\citep{vikhl99a}.  For our work, the relevant slope is that at large
radii (it affects the total flux estimates). Therefore, we truncated the
Jones \& Forman distribution at $\beta<0.6$. Finally, our $\beta$-models
are elliptical with the distribution of axis ratios from
\cite{mohr95}. 

Our adopted X-ray luminosity function is the \cite{1976ApJ...203..297S}
fit to the REFLEX survey data \citep{boeringer02},
\begin{equation}
  dN/dL = C\, L^{-\alpha}\, \exp(-L/L_*)
\end{equation}
with $\alpha=1.69$ and $L_*=2.57\times 10^{44}$~\ergs{} \citep[we use
the][conversion of $L_*$ to the 0.5--2~keV band]{mullis04}. The REFLEX
XLF is consistent with other \emph{ROSAT} All-Sky Survey measurements
\citep{ebeling97,degrandi99}. We use the REFLEX results because they are
reported for our adopted cosmology. 

To compute the survey area and search volume, we need the distribution
of clusters as a function of observed flux, not the rest-frame
luminosity (see \S~\ref{sec:Psel} and~\ref{sec:lf}). This can be
obtained from the luminosity function through an obvious relation,
\begin{equation}\label{eq:flux:lum}
  dN/df = \mathrm{const} \times dN/dL, \qquad  
  f = \frac{L}{4\pi\,d_L(z)^2}\,K(z),
\end{equation}
where $d_L(z)$ is the cosmological luminosity distance and $K(z)$ is the
$K$-correction factor that describes redshifting of the source spectrum. 
The $K$-correction can be easily computed for any given source spectrum
\citep[see, e.g.,][specifically for the case of the cluster X-ray
spectra]{jones98}. The $K$-correction in principle depends on the
temperature and, more weakly, on the metallicity of the ICM. The
temperature can be estimated from the $L_x-T$ correlation with the
accuracy sufficient for this purpose. We use a non-evolving $L_x-T$
relation that follows the \citet{markevitch98} measurements for the
high-$T$ clusters and the \citet{1998PASJ...50..187F} data for low-mass
clusters and groups. It is now established that the $L_x-T$ relation in
fact evolves \citep{2002ApJ...578L.107V}. However, this does not affect
the luminosity-to-flux conversion because the evolution is not strong
and the $K$-correction is a weak function of temperature in the high-$T$
regime \citep[see, e.g., Fig.7 in][]{jones98}.

\end{appendix}

{
\LongTables
\begin{deluxetable}{lccccccl}
  \tablecaption{Cluster catalog\label{tab:cat}}
  \tablehead{
    \colhead{Num.} &
    \colhead{R.A.} &
    \colhead{Dec.} &
    \colhead{$f_x$,} &
    \colhead{$z$} &
    \colhead{$z$ ref.} &
    \colhead{$L_X$,} &
    Note\\
    \colhead{}  & \multicolumn{2}{c}{(J2000)} &
    \colhead{$10^{-13}$~cgs} & \colhead{} & \colhead{} &
    \colhead{erg s$^{-1}$} \\
    (1) &
    \colhead{(2)} &
    \colhead{(3)} &
    \colhead{(4)} &
    \colhead{(5)} &
    \colhead{(6)} &
    \colhead{(7)} &
    (8)
  }
  \startdata
1 & 00:29:50.6 & $+$13:30:14 & \phn3.14 $\pm$ 0.78 & 0.251\phn & \phn~ & $5.38 \times 10^{43}$ & ~ \\
2 & 00:30:33.6 & $+$26:18:16 & \phn2.44 $\pm$ 0.29 & 0.500\phn & \phn1 & $1.85 \times 10^{44}$ & VMF~001 \\
3 & 00:45:18.8 & $-$29:24:02 & \phn3.41 $\pm$ 0.49 & 0.257\phn & \phn~ & $6.12 \times 10^{43}$ & ~ \\
4 & 00:50:59.3 & $-$09:29:14 & \phn3.66 $\pm$ 0.50 & 0.199\phn & \phn1 & $3.81 \times 10^{43}$ & VMF~003   \\
5 & 00:56:03.5 & $-$37:32:43 & 14.92 $\pm$ 1.53 & 0.163\phn & \phn2 & $9.79 \times 10^{43}$ & ~ \\
6 & 00:56:55.0 & $-$22:13:51 & \phn2.48 $\pm$ 0.45 & 0.116\phn & \phn1 & $8.29 \times 10^{42}$ & VMF~005 \\
7 & 00:57:24.7 & $-$26:16:50 & 18.60 $\pm$ 2.23 & 0.113\phn & \phn3 & $5.66 \times 10^{43}$ & VMF~007, A~0122 \\
8 & 01:06:59.0 & $+$32:09:30 & \phn4.53 $\pm$ 0.52 & 0.112\phn & \phn4 & $1.37 \times 10^{43}$ & \S\,\ref{sec:notes} \\
9 & 01:13:11.0 & $+$28:28:16 & \phn3.99 $\pm$ 0.97 & 0.262\phn & \phn~ & $7.41 \times 10^{43}$ & ~ \\
10 & 01:16:39.7 & $-$03:30:11 & \phn3.22 $\pm$ 0.46 & 0.0810 & \phn5 & $5.02 \times 10^{42}$ & ~ \\
11 & 01:22:36.0 & $-$28:32:05 & \phn2.69 $\pm$ 0.61 & 0.256\phn & \phn1 & $4.82 \times 10^{43}$ & VMF~010, A~S0154 \\
12 & 01:24:48.1 & $+$09:32:29 & \phn3.87 $\pm$ 0.46 & 0.0079 & \phn6 & $5.29 \times 10^{40}$ & NGC~0524 \\
13 & 01:26:54.5 & $+$19:12:42 & \phn1.49 $\pm$ 0.43 & 0.0427 & \phn7 & $6.23 \times 10^{41}$ & IC~0115, A~0195 \\
14 & 01:32:54.6 & $-$42:59:46 & \phn3.25 $\pm$ 0.81 & 0.0876 & \phn1 & $5.96 \times 10^{42}$ & VMF~014 \\
15 & 01:39:53.7 & $+$18:10:07 & \phn2.73 $\pm$ 0.36 & 0.176\phn & \phn3 & $2.20 \times 10^{43}$ & VMF~017, A~0227 \\
16 & 01:41:32.3 & $-$30:34:42 & \phn3.14 $\pm$ 0.94 & 0.442\phn & \phn~ & $1.81 \times 10^{44}$ & \S\,\ref{sec:notes} \\
17 & 01:42:50.6 & $+$20:25:13 & \phn2.61 $\pm$ 0.52 & 0.271\phn & \phn1 & $5.25 \times 10^{43}$ & VMF~018 \\
18 & 01:52:41.3 & $-$13:58:13 & \phn1.84 $\pm$ 0.25 & 0.833\phn & \phn8 & $4.19 \times 10^{44}$ & ~ \\
19 & 01:54:12.5 & $-$59:37:33 & \phn1.45 $\pm$ 0.36 & 0.360\phn & \phn1 & $5.55 \times 10^{43}$ & VMF~020  \\
20 & 01:59:18.2 & $+$00:30:09 & \phn3.27 $\pm$ 0.45 & 0.386\phn & \phn1 & $1.40 \times 10^{44}$ & VMF~021 \\
21 & 02:06:49.9 & $-$13:09:09 & \phn2.61 $\pm$ 0.49 & 0.321\phn & \phn1 & $7.59 \times 10^{43}$ & VMF~023 \\
22 & 02:09:52.8 & $-$51:16:19 & \phn1.72 $\pm$ 0.66 & 0.206\phn & \phn~ & $1.95 \times 10^{43}$ & ~ \\
23 & 02:16:33.7 & $-$17:47:27 & \phn1.40 $\pm$ 0.23 & 0.578\phn & \phn8 & $1.50 \times 10^{44}$ & ~ \\
24 & 02:23:28.2 & $-$08:52:12 & \phn2.53 $\pm$ 0.34 & 0.163\phn & \phn9 & $1.73 \times 10^{43}$ & ~ \\
25 & 02:28:13.5 & $-$10:05:45 & \phn2.43 $\pm$ 0.37 & 0.149\phn & \phn1 & $1.38 \times 10^{43}$ & VMF~026 \\
26 & 02:28:22.6 & $+$23:25:23 & \phn2.59 $\pm$ 0.56 & 0.305\phn & \phn~ & $6.79 \times 10^{43}$ & ~ \\
27 & 02:30:26.6 & $+$18:36:22 & \phn2.21 $\pm$ 0.58 & 0.799\phn & \phn~ & $4.59 \times 10^{44}$ & ~ \\
28 & 02:37:59.6 & $-$52:24:47 & \phn7.33 $\pm$ 0.79 & 0.136\phn & 10 & $3.33 \times 10^{43}$ & VMF~028, A~3038 \\
29 & 02:45:45.7 & $+$09:36:36 & \phn6.33 $\pm$ 2.05 & 0.147\phn & \phn~ & $3.41 \times 10^{43}$ & ~ \\
30 & 02:50:03.6 & $+$19:07:51 & \phn2.59 $\pm$ 0.39 & 0.122\phn & \phn8 & $9.57 \times 10^{42}$ & ~ \\
31 & 02:51:17.8 & $-$20:55:46 & \phn2.62 $\pm$ 0.55 & 0.325\phn & \phn~ & $7.84 \times 10^{43}$ & ~ \\
32 & 02:59:33.8 & $+$00:13:45 & \phn3.24 $\pm$ 0.56 & 0.194\phn & \phn1 & $3.19 \times 10^{43}$ & VMF~031 \\
33 & 03:02:21.3 & $-$04:23:29 & 15.89 $\pm$ 1.95 & 0.350\phn & \phn~ & $5.29 \times 10^{44}$ & ~ \\
34 & 03:04:24.7 & $-$07:02:13 & \phn2.59 $\pm$ 0.67 & 0.135\phn & 11 & $1.19 \times 10^{43}$ & ~ \\
35 & 03:06:28.7 & $-$09:43:50 & 10.40 $\pm$ 1.24 & 0.0342 & 12 & $2.71 \times 10^{42}$ & IC~1880 \\
36 & 03:07:04.7 & $-$06:28:51 & \phn5.99 $\pm$ 0.82 & 0.347\phn & 11 & $2.01 \times 10^{44}$ & ~ \\
37 & 03:18:33.4 & $-$03:02:56 & \phn4.60 $\pm$ 0.54 & 0.370\phn & \phn9 & $1.79 \times 10^{44}$ & ~ \\
38 & 03:20:18.1 & $-$42:59:13 & \phn2.99 $\pm$ 0.67 & 0.158\phn & \phn~ & $1.92 \times 10^{43}$ & ~ \\
39 & 03:20:37.8 & $-$43:11:52 & \phn2.62 $\pm$ 0.67 & 0.149\phn & \phn~ & $1.47 \times 10^{43}$ & A~S0343 \\
40 & 03:22:59.4 & $-$13:38:15 & \phn2.06 $\pm$ 0.68 & 0.334\phn & \phn~ & $6.62 \times 10^{43}$ & ~ \\
41 & 03:23:59.5 & $-$19:16:34 & \phn3.26 $\pm$ 0.74 & 0.332\phn & \phn~ & $1.01 \times 10^{44}$ & ~ \\
42 & 03:27:54.5 & $+$02:33:47 & \phn9.41 $\pm$ 1.00 & 0.0302 & 14 & $1.91 \times 10^{42}$ & UGC~02748 \\
43 & 03:28:36.1 & $-$21:40:04 & \phn2.14 $\pm$ 0.57 & 0.590\phn & \phn~ & $2.33 \times 10^{44}$ & ~ \\
44 & 03:32:13.5 & $-$29:10:39 & \phn4.83 $\pm$ 0.71 & 0.150\phn & 15 & $2.75 \times 10^{43}$ & ~ \\
45 & 03:33:10.2 & $-$24:56:41 & \phn2.36 $\pm$ 0.50 & 0.475\phn & \phn~ & $1.61 \times 10^{44}$ & ~ \\
46 & 03:34:03.7 & $-$39:00:49 & \phn6.43 $\pm$ 0.71 & 0.0623 & \phn3 & $3.86 \times 10^{42}$ & A~3135 \\
47 & 03:36:49.4 & $-$28:04:53 & \phn9.46 $\pm$ 1.67 & 0.105\phn & \phn3 & $2.49 \times 10^{43}$ & A~3141 \\
48 & 03:38:11.8 & $-$22:56:24 & \phn1.73 $\pm$ 0.23 & 0.173\phn & \phn~ & $1.35 \times 10^{43}$ & ~ \\
49 & 03:39:24.3 & $-$33:13:09 & \phn3.46 $\pm$ 0.88 & 0.269\phn & \phn~ & $6.83 \times 10^{43}$ & A~3150 \\
50 & 03:40:27.2 & $-$28:40:20 & 17.77 $\pm$ 3.44 & 0.0680 & \phn3 & $1.90 \times 10^{43}$ & A~3151 \\
51 & 03:40:51.6 & $-$28:23:10 & \phn3.18 $\pm$ 0.58 & 0.346\phn & \phn~ & $1.08 \times 10^{44}$ & ~ \\
52 & 03:48:22.4 & $-$33:28:33 & 10.38 $\pm$ 1.48 & 0.165\phn & 16 & $7.09 \times 10^{43}$ & A~3169 \\
53 & 03:50:43.9 & $-$38:01:25 & \phn2.88 $\pm$ 0.78 & 0.363\phn & \phn~ & $1.09 \times 10^{44}$ & \S\,\ref{sec:notes} \\
54 & 03:54:34.6 & $-$42:33:33 & \phn2.05 $\pm$ 0.55 & 0.224\phn & \phn~ & $2.77 \times 10^{43}$ & ~ \\
55 & 03:54:35.4 & $-$37:45:20 & 19.72 $\pm$ 2.59 & 0.251\phn & \phn~ & $3.22 \times 10^{44}$ & A~3184 \\
56 & 03:55:29.9 & $-$36:34:03 & 10.84 $\pm$ 1.83 & 0.320\phn & 17 & $3.01 \times 10^{44}$ & A~S0400, MS 0353.6-3642 \\
57 & 03:55:59.3 & $-$37:41:46 & \phn2.92 $\pm$ 0.67 & 0.473\phn & \phn~ & $1.95 \times 10^{44}$ & ~ \\
58 & 04:05:24.3 & $-$41:00:15 & \phn1.54 $\pm$ 0.37 & 0.686\phn & \phn~ & $2.35 \times 10^{44}$ & ~ \\
59 & 04:17:25.8 & $-$45:12:28 & \phn4.88 $\pm$ 1.05 & 0.213\phn & 15 & $5.81 \times 10^{43}$ & A~3240 \\
60 & 04:21:03.5 & $-$46:29:34 & \phn3.51 $\pm$ 0.84 & 0.131\phn & \phn~ & $1.50 \times 10^{43}$ & A~3247 \\
61 & 04:22:28.8 & $-$50:09:01 & \phn4.00 $\pm$ 1.18 & 0.0901 & \phn~ & $7.77 \times 10^{42}$ & ~ \\
62 & 04:28:42.4 & $-$38:05:47 & \phn2.08 $\pm$ 0.55 & 0.154\phn & \phn3 & $1.27 \times 10^{43}$ & VMF~036, A~3259 \\
63 & 04:46:38.0 & $-$04:21:02 & \phn7.71 $\pm$ 0.88 & 0.177\phn & \phn~ & $6.12 \times 10^{43}$ & ~ \\
64 & 04:58:55.1 & $-$00:29:21 & 24.84 $\pm$ 2.61 & 0.0150 & 18 & $5.73 \times 10^{41}$ & NGC~1713 \\
65 & 05:05:58.2 & $-$28:26:02 & \phn1.41 $\pm$ 0.21 & 0.131\phn & \phn1 & $6.12 \times 10^{42}$ & VMF~038 \\
66 & 05:06:04.0 & $-$28:40:50 & \phn1.97 $\pm$ 0.41 & 0.136\phn & \phn1 & $9.18 \times 10^{42}$ & VMF~039 \\
67 & 05:09:43.4 & $-$08:36:41 & \phn1.98 $\pm$ 0.26 & 0.125\phn & \phn~ & $7.72 \times 10^{42}$ & ~ \\
68 & 05:21:10.5 & $-$25:30:36 & \phn1.76 $\pm$ 0.45 & 0.581\phn & \phn1 & $1.88 \times 10^{44}$ & VMF~040 \\
69 & 05:22:13.8 & $-$36:24:49 & \phn1.84 $\pm$ 0.34 & 0.472\phn & \phn1 & $1.25 \times 10^{44}$ & VMF~041 \\
70 & 05:28:40.1 & $-$32:51:29 & \phn1.99 $\pm$ 0.25 & 0.273\phn & \phn1 & $4.12 \times 10^{43}$ & VMF~042 \\
71 & 05:32:41.8 & $-$46:14:17 & \phn4.11 $\pm$ 0.43 & 0.135\phn & \phn1 & $1.85 \times 10^{43}$ & VMF~044 \\
72 & 05:33:53.1 & $-$57:46:45 & \phn1.70 $\pm$ 0.62 & 0.297\phn & \phn1 & $4.26 \times 10^{43}$ & VMF~045 \\
73 & 05:42:50.8 & $-$41:00:05 & \phn2.19 $\pm$ 0.29 & 0.642\phn & \phn~ & $2.85 \times 10^{44}$ & ~ \\
74 & 05:44:13.3 & $-$25:55:40 & \phn1.55 $\pm$ 0.33 & 0.260\phn & \phn~ & $2.91 \times 10^{43}$ & foreground A~0548 \\
75 & 06:10:32.0 & $-$48:48:26 & \phn2.49 $\pm$ 0.28 & 0.243\phn & \phn~ & $3.99 \times 10^{43}$ & ~ \\
76 & 06:34:34.1 & $-$62:26:46 & \phn2.77 $\pm$ 0.76 & 0.270\phn & \phn~ & $5.58 \times 10^{43}$ & ~ \\
77 & 06:35:28.4 & $-$62:34:06 & \phn3.92 $\pm$ 0.84 & 0.157\phn & \phn~ & $2.46 \times 10^{43}$ & a part of A~3398 \\
78 & 07:20:17.7 & $+$71:32:11 & \phn1.48 $\pm$ 0.22 & 0.268\phn & 19 & $2.97 \times 10^{43}$ & ~ \\
79 & 07:20:53.7 & $+$71:08:57 & \phn1.82 $\pm$ 0.24 & 0.230\phn & 19 & $2.62 \times 10^{43}$ & ~ \\
80 & 08:09:41.0 & $+$28:11:58 & \phn5.49 $\pm$ 0.80 & 0.399\phn & \phn~ & $2.49 \times 10^{44}$ & \S\,\ref{sec:notes} \\
81 & 08:10:24.2 & $+$42:16:19 & 23.89 $\pm$ 2.63 & 0.0640 & \phn1 & $2.24 \times 10^{43}$ & VMF~047 \\
82 & 08:19:54.7 & $+$56:34:39 & \phn3.08 $\pm$ 0.52 & 0.260\phn & \phn1 & $5.68 \times 10^{43}$ & VMF~050 \\
83 & 08:20:26.6 & $+$56:45:27 & \phn2.29 $\pm$ 0.48 & 0.0429 & \phn1 & $9.63 \times 10^{41}$ & VMF~051 \\
84 & 08:38:31.3 & $+$19:48:17 & \phn5.25 $\pm$ 1.34 & 0.123\phn & \phn~ & $1.96 \times 10^{43}$ & ~ \\
85 & 08:41:07.4 & $+$64:22:41 & \phn2.91 $\pm$ 0.32 & 0.343\phn & \phn1 & $9.73 \times 10^{43}$ & VMF~056  \\
86 & 08:49:11.4 & $+$37:31:23 & \phn1.46 $\pm$ 0.27 & 0.240\phn & \phn1 & $2.32 \times 10^{43}$ & VMF~062, a part of A~0708 \\
87 & 08:52:32.9 & $+$16:18:07 & \phn3.71 $\pm$ 0.68 & 0.0980 & \phn1 & $8.60 \times 10^{42}$ & VMF~063 \\
88 & 08:53:13.4 & $+$57:59:44 & \phn1.99 $\pm$ 0.45 & 0.475\phn & \phn1 & $1.37 \times 10^{44}$ & VMF~064 \\
89 & 09:00:04.7 & $+$39:20:24 & \phn3.48 $\pm$ 0.68 & 0.0951 & \phn~ & $7.57 \times 10^{42}$ & ~ \\
90 & 09:07:20.0 & $+$16:39:25 & 14.87 $\pm$ 1.66 & 0.0756 & 20 & $1.98 \times 10^{43}$ & VMF~068, A~0744 \\
91 & 09:10:16.1 & $+$60:12:17 & \phn1.42 $\pm$ 0.35 & 0.181\phn & \phn~ & $1.23 \times 10^{43}$ & a part of A~0742 \\
92 & 09:21:13.2 & $+$45:28:44 & \phn2.39 $\pm$ 0.43 & 0.315\phn & \phn1 & $6.75 \times 10^{43}$ & VMF~070 \\
93 & 09:26:36.6 & $+$12:42:59 & \phn1.67 $\pm$ 0.28 & 0.489\phn & \phn1 & $1.23 \times 10^{44}$ & VMF~071 \\
94 & 09:43:32.4 & $+$16:40:02 & \phn2.31 $\pm$ 0.36 & 0.256\phn & \phn1 & $4.15 \times 10^{43}$ & VMF~073 \\
95 & 09:43:45.0 & $+$16:44:13 & \phn2.12 $\pm$ 0.47 & 0.180\phn & \phn1 & $1.79 \times 10^{43}$ & VMF~074 \\
96 & 09:50:07.7 & $+$70:33:58 & \phn3.20 $\pm$ 0.54 & 0.210\phn & \phn~ & $3.72 \times 10^{43}$ & ~ \\
97 & 09:53:36.0 & $+$70:54:29 & \phn9.07 $\pm$ 1.32 & 0.185\phn & \phn~ & $7.89 \times 10^{43}$ & A~0875 \\
98 & 09:53:44.6 & $+$69:47:29 & \phn1.60 $\pm$ 0.20 & 0.214\phn & \phn~ & $1.98 \times 10^{43}$ & ~ \\
99 & 09:56:02.8 & $+$41:07:08 & \phn1.56 $\pm$ 0.28 & 0.587\phn & \phn1 & $1.71 \times 10^{44}$ & VMF~079 \\
100 & 09:58:13.0 & $+$55:16:06 & \phn4.82 $\pm$ 0.88 & 0.214\phn & \phn1 & $5.81 \times 10^{43}$ & VMF~081, A~0899 \\
101 & 09:58:19.3 & $+$47:02:17 & \phn2.82 $\pm$ 0.62 & 0.390\phn & 21 & $1.25 \times 10^{44}$ & ~ \\
102 & 10:02:07.7 & $+$68:58:48 & \phn1.97 $\pm$ 0.37 & 0.500\phn & \phn~ & $1.51 \times 10^{44}$ & \S\,\ref{sec:notes} \\
103 & 10:03:04.5 & $+$32:53:36 & \phn3.55 $\pm$ 0.44 & 0.416\phn & 22 & $1.79 \times 10^{44}$ & ~ \\
104 & 10:03:06.6 & $-$19:25:47 & \phn2.77 $\pm$ 0.46 & 0.243\phn & \phn~ & $4.44 \times 10^{43}$ & ~ \\
105 & 10:07:08.4 & $-$20:31:26 & \phn1.43 $\pm$ 0.44 & 0.105\phn & \phn~ & $3.88 \times 10^{42}$ & ~ \\
106 & 10:10:15.8 & $+$54:30:12 & \phn2.11 $\pm$ 0.30 & 0.0450 & \phn1 & $9.80 \times 10^{41}$ & VMF~084 \\
107 & 10:11:25.4 & $+$54:50:06 & \phn2.00 $\pm$ 0.52 & 0.294\phn & \phn1 & $4.86 \times 10^{43}$ & VMF~086 \\
108 & 10:13:27.8 & $-$01:36:42 & \phn2.28 $\pm$ 0.55 & 0.276\phn & \phn~ & $4.82 \times 10^{43}$ & ~ \\
109 & 10:13:36.9 & $+$49:33:05 & \phn4.40 $\pm$ 1.05 & 0.133\phn & \phn1 & $1.95 \times 10^{43}$ & VMF~087 \\
110 & 10:18:00.9 & $+$21:54:35 & \phn2.32 $\pm$ 0.42 & 0.240\phn & \phn~ & $3.65 \times 10^{43}$ & ~ \\
111 & 10:27:10.7 & $+$39:08:06 & \phn4.63 $\pm$ 0.56 & 0.338\phn & \phn~ & $1.48 \times 10^{44}$ & ~ \\
112 & 10:33:51.9 & $+$57:03:11 & \phn1.45 $\pm$ 0.37 & 0.0463 & \phn1 & $7.18 \times 10^{41}$ & VMF~089 \\
113 & 10:36:11.3 & $+$57:13:31 & \phn1.88 $\pm$ 0.41 & 0.203\phn & \phn1 & $2.08 \times 10^{43}$ & VMF~090 \\
114 & 10:38:01.8 & $+$41:46:38 & \phn2.71 $\pm$ 0.42 & 0.125\phn & 23 & $1.06 \times 10^{43}$ & A~1056 \\
115 & 10:39:31.3 & $+$39:47:38 & \phn1.92 $\pm$ 0.46 & 0.0926 & \phn~ & $4.00 \times 10^{42}$ & ~ \\
116 & 10:42:24.3 & $-$00:08:16 & \phn3.44 $\pm$ 0.47 & 0.139\phn & 24 & $1.67 \times 10^{43}$ & ~ \\
117 & 10:48:00.6 & $-$11:24:11 & \phn1.85 $\pm$ 0.37 & 0.0650 & \phn1 & $1.84 \times 10^{42}$ & VMF~091 \\
118 & 10:58:12.6 & $+$01:36:57 & 13.00 $\pm$ 1.82 & 0.0385 & 25 & $4.31 \times 10^{42}$ & VMF~095, a part of A~1139 \\
119 & 11:10:04.5 & $-$29:57:06 & \phn1.71 $\pm$ 0.31 & 0.200\phn & \phn~ & $1.82 \times 10^{43}$ & ~ \\
120 & 11:16:54.7 & $+$18:03:20 & \phn6.23 $\pm$ 0.79 & 0.0032 & 26 & $1.37 \times 10^{40}$ & NGC~3607 \\
121 & 11:17:30.1 & $+$17:44:45 & \phn1.44 $\pm$ 0.31 & 0.547\phn & \phn1 & $1.36 \times 10^{44}$ & VMF~098 \\
122 & 11:20:07.6 & $+$43:18:07 & \phn2.97 $\pm$ 0.34 & 0.600\phn & \phn9 & $3.30 \times 10^{44}$ & ~ \\
123 & 11:20:58.3 & $+$23:26:34 & \phn2.12 $\pm$ 0.42 & 0.562\phn & \phn1 & $2.08 \times 10^{44}$ & VMF~100 \\
124 & 11:23:10.6 & $+$14:09:40 & \phn1.82 $\pm$ 0.42 & 0.340\phn & \phn1 & $6.07 \times 10^{43}$ & VMF~101 \\
125 & 11:24:36.6 & $+$41:55:55 & \phn4.02 $\pm$ 0.93 & 0.195\phn & \phn1 & $3.97 \times 10^{43}$ & VMF~103 \\
126 & 11:27:45.4 & $+$43:09:47 & \phn1.83 $\pm$ 0.53 & 0.181\phn & 17 & $1.58 \times 10^{43}$ & MS~1125.3+4324 \\
127 & 11:28:54.2 & $+$42:52:00 & \phn1.69 $\pm$ 0.32 & 0.411\phn & \phn~ & $8.55 \times 10^{43}$ & ~ \\
128 & 11:35:54.4 & $+$21:31:04 & \phn1.78 $\pm$ 0.38 & 0.133\phn & \phn1 & $7.93 \times 10^{42}$ & VMF~104 \\
129 & 11:38:43.2 & $+$03:15:33 & \phn1.59 $\pm$ 0.34 & 0.127\phn & \phn1 & $6.46 \times 10^{42}$ & VMF~105 \\
130 & 11:42:04.5 & $+$21:45:00 & \phn4.56 $\pm$ 1.29 & 0.131\phn & \phn1 & $1.95 \times 10^{43}$ & VMF~106 \\
131 & 11:42:06.3 & $+$10:08:52 & \phn4.71 $\pm$ 0.65 & 0.119\phn & \phn3 & $1.63 \times 10^{43}$ & A~1354 \\
132 & 11:42:16.6 & $+$10:27:02 & \phn3.26 $\pm$ 0.46 & 0.117\phn & \phn~ & $1.10 \times 10^{43}$ & foreground A~135, \S\,\ref{sec:notes} \\
133 & 11:46:26.9 & $+$28:54:19 & \phn3.92 $\pm$ 0.52 & 0.149\phn & \phn1 & $2.19 \times 10^{43}$ & VMF~107 \\
134 & 11:52:35.7 & $+$37:32:46 & \phn3.49 $\pm$ 0.62 & 0.230\phn & \phn~ & $4.95 \times 10^{43}$ & ~ \\
135 & 11:59:51.2 & $+$55:31:56 & \phn7.42 $\pm$ 0.76 & 0.0808 & 11 & $1.14 \times 10^{43}$ & VMF~110, MS~1157.3+5548 \\
136 & 12:00:07.7 & $+$68:09:07 & \phn3.67 $\pm$ 0.73 & 0.265\phn & 11 & $7.01 \times 10^{43}$ & foreground A~1432 \\
137 & 12:00:49.5 & $-$03:27:30 & \phn1.85 $\pm$ 0.27 & 0.396\phn & \phn1 & $8.59 \times 10^{43}$ & VMF~111 \\
138 & 12:01:04.7 & $+$12:09:31 & \phn4.77 $\pm$ 1.15 & 0.304\phn & \phn~ & $1.21 \times 10^{44}$ & ~ \\
139 & 12:02:13.7 & $+$57:51:53 & \phn1.47 $\pm$ 0.39 & 0.677\phn & \phn~ & $2.19 \times 10^{44}$ & ~ \\
140 & 12:06:33.5 & $-$07:44:24 & 12.88 $\pm$ 1.52 & 0.0680 & \phn1 & $1.38 \times 10^{43}$ & VMF~114 \\
141 & 12:11:16.0 & $+$39:11:41 & \phn3.18 $\pm$ 0.37 & 0.340\phn & 17 & $1.04 \times 10^{44}$ & VMF~115, MS~1208.7+3928 \\
142 & 12:12:19.2 & $+$27:33:14 & 12.54 $\pm$ 1.69 & 0.353\phn & \phn~ & $4.28 \times 10^{44}$ & a part of A~1489 \\
143 & 12:12:59.0 & $+$27:27:13 & \phn7.93 $\pm$ 1.09 & 0.179\phn & \phn~ & $6.42 \times 10^{43}$ & a part of A~1489 \\
144 & 12:13:34.4 & $+$02:53:57 & \phn1.43 $\pm$ 0.29 & 0.409\phn & \phn1 & $7.24 \times 10^{43}$ & VMF~116 \\
145 & 12:16:19.8 & $+$26:33:21 & \phn1.54 $\pm$ 0.40 & 0.428\phn & \phn1 & $8.52 \times 10^{43}$ & VMF~117 \\
146 & 12:17:43.7 & $+$47:29:14 & \phn6.09 $\pm$ 0.64 & 0.270\phn & \phn~ & $1.19 \times 10^{44}$ & ~ \\
147 & 12:17:48.5 & $+$22:55:16 & \phn3.00 $\pm$ 0.59 & 0.140\phn & \phn~ & $1.48 \times 10^{43}$ & ~ \\
148 & 12:20:17.4 & $+$75:22:12 & \phn9.53 $\pm$ 1.00 & 0.0059 & 14 & $7.07 \times 10^{40}$ & NGC~4291 \\
149 & 12:21:25.0 & $+$49:18:07 & \phn2.07 $\pm$ 0.51 & 0.700\phn & \phn1 & $3.25 \times 10^{44}$ & VMF~119 \\
150 & 12:22:01.9 & $+$27:09:19 & \phn1.87 $\pm$ 0.39 & 0.472\phn & \phn~ & $1.27 \times 10^{44}$ & ~ \\
151 & 12:22:16.2 & $+$25:59:40 & \phn1.89 $\pm$ 0.32 & 0.160\phn & \phn~ & $1.26 \times 10^{43}$ & ~ \\
152 & 12:24:17.5 & $+$75:31:39 & \phn1.44 $\pm$ 0.21 & 0.0056 & 14 & $9.77 \times 10^{39}$ & NGC~4386 \\
153 & 12:26:22.2 & $+$62:24:38 & \phn1.43 $\pm$ 0.43 & 0.398\phn & 11 & $6.79 \times 10^{43}$ & ~ \\
154 & 12:26:57.7 & $+$33:32:50 & \phn2.94 $\pm$ 0.34 & 0.888\phn & 28 & $7.52 \times 10^{44}$ & ~ \\
155 & 12:27:14.1 & $+$08:58:15 & \phn4.14 $\pm$ 0.53 & 0.0873 & 29 & $7.54 \times 10^{42}$ & ~ \\
156 & 12:30:14.1 & $+$23:26:04 & \phn3.13 $\pm$ 0.73 & 0.221\phn & \phn~ & $4.06 \times 10^{43}$ & ~ \\
157 & 12:31:45.6 & $+$41:37:11 & \phn2.82 $\pm$ 0.34 & 0.176\phn & \phn~ & $2.26 \times 10^{43}$ & ~ \\
158 & 12:35:06.4 & $+$41:17:44 & \phn3.01 $\pm$ 0.83 & 0.189\phn & \phn~ & $2.81 \times 10^{43}$ & A~1565 \\
159 & 12:36:28.6 & $+$12:24:21 & \phn2.45 $\pm$ 0.38 & 0.0667 & 31 & $2.56 \times 10^{42}$ & IC 3574 \\
160 & 12:36:56.8 & $+$25:50:27 & \phn2.52 $\pm$ 0.38 & 0.175\phn & \phn~ & $2.01 \times 10^{43}$ & ~ \\
161 & 12:48:36.4 & $-$05:48:01 & \phn3.47 $\pm$ 0.47 & 0.0041 & 25 & $1.28 \times 10^{40}$ & NGC~4697 \\
162 & 12:52:04.7 & $-$29:20:51 & \phn2.15 $\pm$ 0.40 & 0.188\phn & \phn1 & $2.00 \times 10^{43}$ & VMF~124 \\
163 & 12:53:04.7 & $+$62:48:10 & \phn2.32 $\pm$ 0.45 & 0.235\phn & \phn~ & $3.45 \times 10^{43}$ & A~1636 \\
164 & 12:59:51.0 & $+$31:20:48 & \phn5.68 $\pm$ 1.53 & 0.0523 & \phn5 & $3.56 \times 10^{42}$ & ~ \\
165 & 13:01:43.4 & $+$10:59:35 & \phn2.80 $\pm$ 0.71 & 0.231\phn & \phn1 & $4.00 \times 10^{43}$ & VMF~130 \\
166 & 13:08:32.9 & $+$53:42:15 & \phn1.69 $\pm$ 0.26 & 0.330\phn & \phn9 & $5.32 \times 10^{43}$ & ~ \\
167 & 13:11:12.7 & $+$32:28:58 & \phn4.67 $\pm$ 0.56 & 0.245\phn & \phn1 & $7.49 \times 10^{43}$ & VMF~132 \\
168 & 13:12:19.4 & $+$39:00:58 & \phn2.59 $\pm$ 0.38 & 0.404\phn & \phn~ & $1.24 \times 10^{44}$ & ~ \\
169 & 13:13:39.1 & $-$32:50:41 & \phn2.55 $\pm$ 0.35 & 0.0518 & \phn~ & $1.58 \times 10^{42}$ & ~ \\
170 & 13:29:27.9 & $+$11:43:23 & 12.73 $\pm$ 2.18 & 0.0228 & \phn4 & $1.12 \times 10^{42}$ & VMF~136, NGC~5171, NGC~5176 \\
171 & 13:29:49.4 & $-$33:10:23 & \phn1.80 $\pm$ 0.30 & 0.0511 & \phn~ & $1.09 \times 10^{42}$ & ~ \\
172 & 13:31:31.0 & $+$62:38:24 & \phn2.19 $\pm$ 0.42 & 0.219\phn & \phn~ & $2.82 \times 10^{43}$ & ~ \\
173 & 13:34:20.3 & $+$50:31:05 & \phn1.85 $\pm$ 0.29 & 0.620\phn & \phn9 & $2.26 \times 10^{44}$ & ~ \\
174 & 13:38:05.7 & $-$29:44:22 & \phn4.11 $\pm$ 0.49 & 0.189\phn & 17 & $3.81 \times 10^{43}$ & MS~1335.2-2928 \\
175 & 13:38:50.2 & $+$38:51:18 & \phn5.98 $\pm$ 0.77 & 0.246\phn & 33 & $9.61 \times 10^{43}$ & 3C 28, \S\,\ref{sec:notes} \\
176 & 13:40:33.5 & $+$40:17:46 & \phn1.61 $\pm$ 0.29 & 0.171\phn & \phn1 & $1.23 \times 10^{43}$ & VMF~144 \\
177 & 13:40:54.0 & $+$39:58:28 & \phn3.47 $\pm$ 0.66 & 0.169\phn & \phn1 & $2.55 \times 10^{43}$ & VMF~145, A~1774 \\
178 & 13:41:52.0 & $+$26:22:49 & 80.95 $\pm$ 8.30 & 0.0755 & \phn3 & $1.06 \times 10^{44}$ & VMF~146, A~1775 \\
179 & 13:43:27.9 & $+$55:46:55 & \phn1.99 $\pm$ 0.24 & 0.0673 & 11 & $2.13 \times 10^{42}$ & VMF~150, a part of A~1783 \\
180 & 13:49:00.2 & $+$49:18:33 & \phn5.52 $\pm$ 0.96 & 0.167\phn & 22 & $3.88 \times 10^{43}$ & a part of A~1804 \\
181 & 13:54:16.7 & $-$02:21:46 & \phn1.46 $\pm$ 0.23 & 0.546\phn & \phn1 & $1.37 \times 10^{44}$ & VMF~151 \\
182 & 13:57:19.4 & $+$62:32:42 & \phn1.95 $\pm$ 0.26 & 0.525\phn & \phn~ & $1.67 \times 10^{44}$ & ~ \\
183 & 14:06:55.0 & $+$28:34:16 & \phn2.56 $\pm$ 0.31 & 0.118\phn & \phn1 & $8.82 \times 10^{42}$ & VMF~154 \\
184 & 14:10:13.4 & $+$59:42:38 & \phn3.35 $\pm$ 0.69 & 0.316\phn & 11 & $9.37 \times 10^{43}$ & VMF~155 \\
185 & 14:10:15.9 & $+$59:38:27 & \phn2.01 $\pm$ 0.68 & 0.319\phn & 11 & $5.85 \times 10^{43}$ & VMF~156 \\
186 & 14:16:26.8 & $+$23:15:31 & 13.05 $\pm$ 1.38 & 0.138\phn & \phn9 & $6.09 \times 10^{43}$ & ~ \\
187 & 14:16:28.1 & $+$44:46:38 & \phn4.04 $\pm$ 0.50 & 0.400\phn & \phn1 & $1.86 \times 10^{44}$ & VMF~158 \\
188 & 14:18:31.2 & $+$25:10:47 & \phn7.54 $\pm$ 0.78 & 0.290\phn & \phn1 & $1.71 \times 10^{44}$ & VMF~159 \\
189 & 14:27:58.2 & $+$26:30:23 & \phn1.46 $\pm$ 0.23 & 0.0324 & 32 & $3.45 \times 10^{41}$ & IC 4436 \\
190 & 14:34:17.3 & $-$32:29:04 & 12.63 $\pm$ 2.07 & 0.241\phn & \phn~ & $1.92 \times 10^{44}$ & A~S0766 \\
191 & 14:36:58.4 & $+$55:07:45 & \phn2.63 $\pm$ 0.54 & 0.125\phn & \phn~ & $1.03 \times 10^{43}$ & ~ \\
192 & 14:38:50.6 & $+$64:23:39 & \phn2.62 $\pm$ 0.42 & 0.146\phn & \phn1 & $1.43 \times 10^{43}$ & VMF~164 \\
193 & 14:48:36.2 & $-$27:49:10 & \phn5.11 $\pm$ 1.08 & 0.175\phn & \phn~ & $3.99 \times 10^{43}$ & may be a part of A~3609 \\
194 & 14:51:17.4 & $+$18:41:00 & \phn2.73 $\pm$ 0.56 & 0.0439 & 36 & $1.20 \times 10^{42}$ & IC 1062 \\
195 & 15:01:18.3 & $-$08:30:33 & 14.00 $\pm$ 1.75 & 0.108\phn & \phn~ & $3.88 \times 10^{43}$ & ~ \\
196 & 15:04:39.1 & $-$14:53:58 & \phn2.79 $\pm$ 0.76 & 0.284\phn & \phn~ & $6.24 \times 10^{43}$ & ~ \\
197 & 15:15:33.0 & $+$43:46:35 & \phn3.45 $\pm$ 0.68 & 0.137\phn & \phn1 & $1.62 \times 10^{43}$ & VMF~168         \\
198 & 15:24:40.3 & $+$09:57:35 & \phn3.04 $\pm$ 0.42 & 0.516\phn & \phn1 & $2.45 \times 10^{44}$ & VMF~170 \\
199 & 15:33:17.1 & $+$31:08:55 & 18.17 $\pm$ 2.96 & 0.0673 & \phn3 & $1.90 \times 10^{43}$ & A~2092 \\
200 & 15:37:44.6 & $+$12:00:21 & \phn2.67 $\pm$ 0.74 & 0.134\phn & \phn1 & $1.19 \times 10^{43}$ & VMF~171 \\
201 & 15:52:12.3 & $+$20:13:42 & \phn4.97 $\pm$ 0.61 & 0.136\phn & \phn1 & $2.29 \times 10^{43}$ & VMF~175 \\
202 & 16:14:11.5 & $+$34:25:25 & \phn2.30 $\pm$ 0.45 & 0.269\phn & \phn~ & $4.61 \times 10^{43}$ & ~ \\
203 & 16:29:46.1 & $+$21:23:55 & \phn2.53 $\pm$ 0.53 & 0.184\phn & \phn1 & $2.24 \times 10^{43}$ & VMF~178     \\
204 & 16:30:14.7 & $+$24:34:49 & 17.79 $\pm$ 2.54 & 0.0655 & \phn1 & $1.75 \times 10^{43}$ & VMF~179 \\
205 & 16:31:04.9 & $+$21:21:54 & \phn3.01 $\pm$ 0.60 & 0.0980 & \phn1 & $6.99 \times 10^{42}$ & VMF~180 \\
206 & 16:39:55.5 & $+$53:47:55 & 13.07 $\pm$ 1.43 & 0.111\phn & \phn1 & $3.84 \times 10^{43}$ & VMF~182, A~2220  \\
207 & 16:41:11.0 & $+$82:32:26 & \phn8.03 $\pm$ 1.16 & 0.206\phn & \phn1 & $8.79 \times 10^{43}$ & VMF~183 \\
208 & 16:41:52.3 & $+$40:01:29 & \phn2.94 $\pm$ 0.80 & 0.464\phn & \phn1 & $1.88 \times 10^{44}$ & VMF~184 \\
209 & 16:58:33.9 & $+$34:30:08 & \phn3.37 $\pm$ 0.68 & 0.330\phn & \phn1 & $1.03 \times 10^{44}$ & VMF~187      \\
210 & 17:00:42.7 & $+$64:12:58 & \phn4.56 $\pm$ 0.48 & 0.225\phn & \phn3 & $6.10 \times 10^{43}$ & VMF~189, A~2246 \\
211 & 17:01:22.6 & $+$64:14:09 & \phn3.85 $\pm$ 0.40 & 0.453\phn & \phn1 & $2.32 \times 10^{44}$ & VMF~190 \\
212 & 17:22:53.9 & $+$41:05:30 & \phn2.87 $\pm$ 0.63 & 0.309\phn & \phn1 & $7.69 \times 10^{43}$ & VMF~193 \\
213 & 17:29:00.6 & $+$74:40:40 & \phn1.74 $\pm$ 0.51 & 0.213\phn & \phn1 & $2.13 \times 10^{43}$ & VMF~194    \\
214 & 17:46:27.4 & $+$68:48:57 & \phn2.23 $\pm$ 0.34 & 0.217\phn & \phn1 & $2.81 \times 10^{43}$ & VMF~195 \\
215 & 17:51:08.6 & $+$65:31:59 & \phn3.41 $\pm$ 0.71 & 0.0428 & 32 & $1.42 \times 10^{42}$ & NGC~6505 \\
216 & 17:51:38.6 & $+$67:19:19 & \phn4.81 $\pm$ 1.42 & 0.0933 & 40 & $1.00 \times 10^{43}$ & ~ \\
217 & 18:06:55.8 & $+$65:37:11 & \phn4.93 $\pm$ 1.00 & 0.263\phn & 40 & $9.15 \times 10^{43}$ & ~ \\
218 & 18:07:28.1 & $+$69:46:22 & \phn1.41 $\pm$ 0.39 & 0.0941 & \phn~ & $3.04 \times 10^{42}$ & ~ \\
219 & 18:19:10.0 & $+$69:09:39 & \phn2.58 $\pm$ 0.89 & 0.205\phn & \phn~ & $2.88 \times 10^{43}$ & ~ \\
220 & 18:23:43.5 & $+$56:58:26 & \phn2.59 $\pm$ 0.41 & 0.105\phn & \phn~ & $7.03 \times 10^{42}$ & ~ \\
221 & 18:46:42.7 & $-$74:31:59 & \phn4.53 $\pm$ 0.75 & 0.141\phn & \phn~ & $2.25 \times 10^{43}$ & ~ \\
222 & 20:03:13.3 & $-$32:47:31 & \phn2.53 $\pm$ 0.61 & 0.256\phn & \phn~ & $4.54 \times 10^{43}$ & ~ \\
223 & 20:03:28.4 & $-$55:56:47 & \phn4.77 $\pm$ 0.60 & 0.0148 & 41 & $2.29 \times 10^{41}$ & VMF~196, A~S0840 \\
224 & 20:11:53.2 & $-$35:55:40 & \phn2.13 $\pm$ 0.52 & 0.172\phn & \phn~ & $1.65 \times 10^{43}$ & ~ \\
225 & 20:51:38.6 & $-$57:04:24 & \phn1.43 $\pm$ 0.24 & 0.0599 & \phn~ & $1.20 \times 10^{42}$ & ~ \\
226 & 21:14:19.8 & $-$68:00:56 & \phn2.57 $\pm$ 0.30 & 0.130\phn & \phn1 & $1.09 \times 10^{43}$ & VMF~201 \\
227 & 21:37:06.9 & $+$00:26:48 & \phn2.77 $\pm$ 0.63 & 0.0509 & \phn1 & $1.65 \times 10^{42}$ & VMF~202 \\
228 & 21:37:51.2 & $-$42:51:08 & \phn5.08 $\pm$ 0.66 & 0.185\phn & \phn~ & $4.46 \times 10^{43}$ & A~3791 \\
229 & 22:13:30.4 & $-$16:56:05 & \phn1.83 $\pm$ 0.36 & 0.297\phn & \phn1 & $4.57 \times 10^{43}$ & VMF~207 \\
230 & 22:20:09.1 & $-$52:28:01 & 18.80 $\pm$ 2.22 & 0.102\phn & 42 & $4.64 \times 10^{43}$ & A~3864 \\
231 & 22:22:13.9 & $-$52:35:13 & \phn2.61 $\pm$ 0.41 & 0.174\phn & \phn~ & $2.05 \times 10^{43}$ & may be a part of A~3870 \\
232 & 22:39:24.4 & $-$05:46:57 & \phn2.22 $\pm$ 0.32 & 0.242\phn & \phn8 & $3.54 \times 10^{43}$ & VMF~208, a part of A~2465 \\
233 & 22:39:38.8 & $-$05:43:13 & \phn3.24 $\pm$ 0.46 & 0.243\phn & \phn8 & $5.15 \times 10^{43}$ & VMF~210, a part of A~2465 \\
234 & 22:47:29.5 & $+$03:37:08 & \phn2.30 $\pm$ 0.50 & 0.200\phn & \phn1 & $2.43 \times 10^{43}$ & VMF~211 \\
235 & 22:58:07.1 & $+$20:55:06 & \phn5.06 $\pm$ 0.60 & 0.288\phn & 17 & $1.14 \times 10^{44}$ & VMF~213, MS~2255.7+2039 \\
236 & 23:05:25.7 & $-$35:45:43 & \phn1.55 $\pm$ 0.29 & 0.201\phn & \phn1 & $1.67 \times 10^{43}$ & VMF~214 \\
237 & 23:18:05.5 & $-$42:35:34 & \phn1.56 $\pm$ 0.30 & 0.209\phn & \phn1 & $1.83 \times 10^{43}$ & VMF~216 \\
238 & 23:19:34.2 & $+$12:26:12 & \phn3.83 $\pm$ 0.49 & 0.126\phn & \phn1 & $1.49 \times 10^{43}$ & VMF~217 \\
239 & 23:23:14.8 & $+$18:11:23 & \phn5.04 $\pm$ 0.99 & 0.154\phn & \phn~ & $3.02 \times 10^{43}$ & ~ \\
240 & 23:25:39.0 & $-$54:43:57 & \phn2.26 $\pm$ 0.68 & 0.102\phn & \phn1 & $5.78 \times 10^{42}$ & VMF~218 \\
241 & 23:48:53.4 & $-$31:17:12 & \phn3.25 $\pm$ 0.44 & 0.184\phn & \phn1 & $2.85 \times 10^{43}$ & VMF~221, A~4043 \\
242 & 23:55:11.5 & $-$15:00:29 & \phn2.66 $\pm$ 0.65 & 0.0857 & \phn1 & $4.69 \times 10^{42}$ & VMF~223 
\enddata
\tablecomments{
Cluster redshift references are:
 1 --- \cite{vikhl98a,mullis03};
 2 --- \cite{cappi98};
 3 --- \cite{strublerood99};
 4 --- \cite{mahdavigeller04};
 5 --- \cite{rines03};
 6 --- \cite{2000A&AS..145..263S};
 7 --- \cite{huchra99};
 8 --- \cite{perlman02};
 9 --- \cite{romer00};
 10 --- \cite{1990A&A...234...60N};
 11 --- \cite{abazajian05};
 12 --- \cite{dacosta98};
 13 --- \cite{aco89};
 14 --- \cite{devaucouleurs91};
 15 --- \cite{colless01};
 16 --- \cite{batuski99};
 17 --- \cite{stocke91};
 18 --- \cite{wegner03};
 19 --- \cite{mason00};
 20 --- \cite{1978ApJ...223..740P};
 21 --- \cite{molthagen97};
 22 --- \cite{boeringer00};
 23 --- \cite{miller02};
 24 --- \cite{appenzeller98};
 25 --- \cite{2000MNRAS.313..469S};
 26 --- \cite{2005MNRAS.356.1440D};
 27 --- \cite{1985AJ.....90.1681B};
 28 --- \cite{ebeling01b};
 29 --- \cite{burke03};
 30 --- \cite{2002AJ....123.2990B};
 31 --- \cite{binggeli93};
 32 --- \cite{falco99};
 33 --- \cite{hb91};
 34 --- \cite{keel96};
 35 --- \cite{1990ApJS...72..433H};
 36 --- \cite{wegner99};
 37 --- \cite{miller04};
 38 --- \cite{owen95};
 39 --- \cite{kochanek01};
 40 --- \cite{gioia03};
 41 --- \cite{1991ApJS...75..935D};
 42 --- \cite{Katgert98};
If redshift reference is missing, the redshift is our measurement
\citep{hornstrup05}.
The codes in column (8) are VMF --- 160d cluster \citep{vikhl98a,mullis03}; 
A --- Abell cluster \citep{aco89}; MS --- EMSS \citep{stocke91}.
}
\end{deluxetable}

}

\begin{turnpage}
  \begin{deluxetable}{lcccccclcll}
  \tablecaption{Clusters at \emph{ROSAT} target $z$\label{tab:tz}}
  \tablehead{
    \colhead{Num.} &
    \colhead{R.A.} &
    \colhead{Dec.} &
    \colhead{$f_x$,} &
    \colhead{$z$} &
    \colhead{$z$ ref.} &
    \colhead{$L_X$,} &
    Note &
    \multicolumn{3}{l}{\emph{ROSAT} target:}\\
    \colhead{}  & \multicolumn{2}{c}{(J2000)} &
    \colhead{$10^{-13}$~cgs} & & &
    \colhead{erg s$^{-1}$} &  &
    \colhead{$z$} & \colhead{Type\phm{AA}} & Name\\
    (1) &
    \colhead{(2)} &
    \colhead{(3)} &
    \colhead{(4)} &
    \colhead{(5)} &
    \colhead{(6)} &
    \colhead{(7)} &
    (8) &
    \colhead{(9)} &
    \colhead{(10)} &
    (11)
  }
  \startdata
1 & 03:22:20.5 & $-$49:18:42 & \phn4.04 $\pm$ 0.66 & 0.0670 & 13 & $4.23 \times 10^{42}$ & VMF~032, A~S0346    & 0.0710 & AGN & 2MASX~J03231532$-$4931064\\
2 & 03:31:59.4 & $-$45:24:24 & \phn1.63 $\pm$ 0.36 & 0.0671 & \phn~ & $1.73 \times 10^{42}$ & ~  & 0.0665 &  & {\raggedright Horologium Supercluster \citep{lucey83}}\\
3 & 08:38:10.4 & $+$25:06:02 & 18.48 $\pm$ 3.02 & 0.0286 & \phn4 & $3.34 \times 10^{42}$ & CGCG 120-014  & 0.0286 & AGN & NGC~2622, MKN~1218\\
4 & 10:20:45.3 & $+$38:31:28 & \phn5.16 $\pm$ 0.65 & 0.0530 & \phn~ & $3.32 \times 10^{42}$ & ~  & 0.0491 & GClstr & RX~J1022.1$+$3830\\
5 & 10:24:19.1 & $+$68:04:56 & \phn5.46 $\pm$ 0.62 & 0.201\phn & \phn9 & $5.72 \times 10^{43}$ & A~0981  & 0.2031 & GClstr & A~0998\\
6 & 12:04:04.0 & $+$28:07:08 & 10.25 $\pm$ 1.09 & 0.167\phn & 17 & $7.17 \times 10^{43}$ & VMF~112, MS~1201.5+2824  & 0.1653 & AGN & GQ~Com\\
7 & 12:36:59.5 & $+$63:11:18 & 19.75 $\pm$ 2.22 & 0.302\phn & \phn3 & $4.78 \times 10^{44}$ & A~1576  & 0.2970 & BL~Lac & HB89~1235$+$632\\
8 & 13:29:39.3 & $+$29:46:05 & \phn2.25 $\pm$ 0.34 & 0.0473 & 32 & $1.16 \times 10^{42}$ & CGCG 161-067  & 0.0470 & AGN & VCV2001~J132908.8$+$295023\\
9 & 13:59:34.0 & $+$62:19:01 & \phn4.93 $\pm$ 0.54 & 0.332\phn & \phn~ & $1.52 \times 10^{44}$ & ~  & 0.3280 & GClstr & ZwCl~1358.1$+$6245\\
10 & 14:25:04.5 & $+$37:58:16 & \phn2.04 $\pm$ 0.41 & 0.163\phn & \phn~ & $1.39 \times 10^{43}$ & ~  & 0.1712 & GClstr & A~1914\\
11 & 14:44:06.4 & $+$63:44:54 & \phn1.74 $\pm$ 0.38 & 0.298\phn & 35 & $4.38 \times 10^{43}$ & VMF~165, A~1969  & 0.2990 & BL~Lac & MS~1443.5$+$6349\\
12 & 15:00:02.6 & $+$22:34:05 & \phn1.46 $\pm$ 0.42 & 0.230\phn & \phn1 & $2.12 \times 10^{43}$ & VMF~166  & 0.2350 & BL~Lac & FBQS~J150101.8$+$223806\\
13 & 15:14:22.0 & $+$36:36:22 & \phn5.14 $\pm$ 0.65 & 0.372\phn & 17 & $2.01 \times 10^{44}$ & MS~1512.4+3647  & 0.3707 & QSO & HB89~1512$+$370\\
14 & 15:36:35.3 & $+$01:33:20 & \phn7.14 $\pm$ 1.83 & 0.309\phn & \phn~ & $1.87 \times 10^{44}$ & ~  & 0.3120 & BL~Lac & RBS~1517\\
15 & 15:40:10.1 & $+$66:11:18 & \phn3.29 $\pm$ 0.59 & 0.245\phn & 37 & $5.34 \times 10^{43}$ & ~  & 0.2465 & GClstr & A~2125\\
16 & 15:41:10.7 & $+$66:26:30 & \phn2.91 $\pm$ 0.39 & 0.245\phn & \phn9 & $4.72 \times 10^{43}$ & ~  & 0.2465 & GClstr & A~2125\\
17 & 15:47:20.4 & $+$20:57:01 & \phn2.54 $\pm$ 0.88 & 0.266\phn & \phn1 & $4.93 \times 10^{43}$ & VMF~174  & 0.2643 & QSO & PG~1545$+$210\\
18 & 16:20:21.6 & $+$17:23:19 & \phn2.08 $\pm$ 0.39 & 0.112\phn & \phn1 & $6.44 \times 10^{42}$ & VMF~177  & 0.1124 & QSO & MRK~0877, PG1617$+$175\\
19 & 16:27:00.1 & $+$55:28:18 & 21.06 $\pm$ 2.44 & 0.130\phn & \phn3 & $8.61 \times 10^{43}$ & A~2201  & 0.1330 & QSO & SBS~1626$+$554\\
20 & 16:59:02.1 & $+$32:29:15 & \phn3.37 $\pm$ 0.48 & 0.0621 & 38 & $3.03 \times 10^{42}$ & ~  & 0.0635 & GClstr & A~2241\\
21 & 17:01:27.9 & $+$34:00:35 & \phn3.72 $\pm$ 0.74 & 0.0960 & 11 & $8.25 \times 10^{42}$ & ~  & 0.0968 & GClstr & A~2244\\
22 & 17:22:15.4 & $+$30:42:45 & 12.22 $\pm$ 2.04 & 0.0467 & 39 & $6.02 \times 10^{42}$ & CGCG 170-018  & 0.0430 & AGN & MRK506\\
23 & 18:43:30.6 & $+$79:49:57 & \phn3.15 $\pm$ 0.42 & 0.0510 & 17 & $1.88 \times 10^{42}$ & MS~1846.9+7947  & 0.0561 & AGN & 3C~390.3\\
24 & 21:39:04.8 & $-$23:33:08 & \phn5.18 $\pm$ 0.69 & 0.320\phn & \phn~ & $1.47 \times 10^{44}$ & ~  & 0.3130 & GClstr & MS~2137.3$-$2353
\enddata
\tablecomments{See notes to Table~\ref{tab:cat}}
\end{deluxetable}

\end{turnpage}

\begin{deluxetable}{ccccccll}
  \tablecaption{Other extended X-ray sources\label{tab:notcl}}
  \tablehead{
    \colhead{R.A.} &
    \colhead{Dec.} &
    \colhead{$f_x$,} &
    \colhead{$z$} &
    \colhead{$z$ ref.} &
    \colhead{$L_X$,} &
    Type &
    Note \\
    \multicolumn{2}{c}{(J2000)} &
    \colhead{$10^{-13}$~cgs} & \colhead{} & \colhead{} &
    \colhead{erg s$^{-1}$} \\
    \colhead{(1)} &
    \colhead{(2)} &
    \colhead{(3)} &
    \colhead{(4)} &
    \colhead{(5)} &
    \colhead{(6)} &
    (7) &
    (8)
  }
  \startdata
09:55:18.0 & $+$69:50:57 & \phn2.48 $\pm$ 0.32 & \nodata &   & \nodata & M82 wind & ~ \\
12:19:09.9 & $+$47:05:40 & \phn1.41 $\pm$ 0.18 & 0.0009 & 14 & $2.49 \times 10^{38}$ & dE & UGC 07356 \\
12:21:56.1 & $+$04:29:09 & \phn4.38 $\pm$ 0.71 & 0.0052 & 27 & $2.58 \times 10^{40}$ & SAB, Sy2 & MESSIER~061 \\
12:30:01.0 & $+$13:38:41 & \phn4.62 $\pm$ 0.71 & 0.0045 & 30 & $2.04 \times 10^{40}$ & SB0, Sy2 & NGC~4477 \\
14:13:15.1 & $-$03:12:36 & 26.05 $\pm$ 2.80 & 0.0062 & 34 & $2.15 \times 10^{41}$ & Sa, Sy2 & NGC~5506 
\enddata
\tablecomments{See notes to Table~\ref{tab:cat}}
\end{deluxetable}

\begin{deluxetable}{cccl}
  \tablecaption{False detections\label{tab:false}}
  \tablehead{
    \colhead{R.A.} &
    \colhead{Dec.} &
    \colhead{$f_x$,} &
    Note \\
    \multicolumn{2}{c}{(J2000)} &
    \colhead{$10^{-13}$~cgs}  \\
    \colhead{(1)} &
    \colhead{(2)} &
    \colhead{(3)} &
    (4)
  }
  \startdata
03:22:26.4 & $-$21:08:34 & \phn9.15 $\pm$ 1.49 & \S\,\ref{sec:notes} \\
03:55:46.9 & $-$39:25:23 & \phn2.30 $\pm$ 0.41 & ~ \\
05:22:37.1 & $-$36:28:53 & \phn1.41 $\pm$ 0.41 & ~ \\
06:54:15.9 & $+$69:05:52 & \phn1.44 $\pm$ 0.46 & ~ \\
10:07:09.9 & $+$35:02:47 & \phn1.54 $\pm$ 0.24 & ~ \\
11:39:58.6 & $+$31:53:04 & \phn1.45 $\pm$ 0.36 & ~ \\
12:30:35.6 & $+$76:59:34 & \phn1.55 $\pm$ 0.49 & ~ \\
12:43:50.8 & $+$02:44:12 & \phn3.22 $\pm$ 0.76 & ~ \\
13:37:54.2 & $+$38:54:01 & \phn1.43 $\pm$ 0.41 & VMF~143  \\
14:18:45.6 & $+$06:44:02 & \phn1.64 $\pm$ 0.24 & VMF~160 \\
14:28:28.0 & $+$01:06:31 & \phn1.41 $\pm$ 0.29 & ~ \\
15:00:51.3 & $+$22:44:56 & \phn1.78 $\pm$ 0.35 & VMF~16, \S\,\ref{sec:notes} \\
17:58:26.0 & $+$65:30:58 & \phn3.93 $\pm$ 0.53 & ~ \\
19:58:02.6 & $-$57:35:39 & \phn1.98 $\pm$ 0.65 & ~ \\
20:05:13.5 & $-$56:12:57 & \phn3.51 $\pm$ 0.49 & VMF~198 \\
22:01:54.9 & $-$51:24:04 & \phn1.53 $\pm$ 0.37 & ~ 
\enddata
\end{deluxetable}

\end{document}